\documentclass[12pt, draftclsnofoot, onecolumn]{IEEEtran}
\usepackage{amssymb}
\usepackage[cmex10]{amsmath}
\usepackage{stfloats}
\usepackage{graphicx}
\usepackage{subfigure}
\usepackage{tabularx}
\usepackage{epsfig,epsf,color,balance,cite}
\usepackage{verbatim}
\usepackage{url}
\usepackage{bm}
\usepackage{mathrsfs}
\usepackage{amsthm}
\DeclareMathAlphabet\mathbfcal{OMS}{cmsy}{b}{n}

\newtheorem{theorem}{Theorem}

\newtheorem{lemma}{Lemma}
\newtheorem{remark}{Remark}
\usepackage{algorithm}
\usepackage{algpseudocode}
\usepackage{graphicx}
\usepackage{epstopdf}
\IEEEoverridecommandlockouts

\begin{document}

\title{MAC Wiretap Channels with Confidential and Open Messages: Improved Achievable Region and Low-complexity Precoder Design}

\author{
	\IEEEauthorblockN{Hao Xu, \emph{Member, IEEE}\IEEEauthorrefmark{0},
		Kai-Kit Wong, \emph{Fellow, IEEE}\IEEEauthorrefmark{0},
		and
		Giuseppe Caire, \emph{Fellow, IEEE}\IEEEauthorrefmark{0}
	}
	\thanks{
		This work was supported by the European Union's Horizon 2020 Research and Innovation Programme under Marie Skłodowska-Curie Grant No. 101024636 and the Alexander von Humboldt Foundation.
		
		H. Xu and K.-K. Wong are with the Department of Electronic and Electrical Engineering, University College London, London WC1E 7JE, UK (e-mail: hao.xu@ucl.ac.uk; kai-kit.wong@ucl.ac.uk).
		
		G. Caire is with the Faculty of Electrical Engineering and Computer Science at the Technical University of Berlin, 10587 Berlin, Germany (e-mail: caire@tu-berlin.de).
	}
}

\maketitle

\begin{abstract}
This paper investigates the achievable region and precoder design for multiple access wiretap (MAC-WT) channels, where each user transmits both secret and open (i.e., non-confidential) messages.
All these messages are intended for the legitimate receiver (or Bob for brevity) and the eavesdropper (Eve) is interested only in the secret messages of all users.
By allowing users with zero secret message rate to act as conventional MAC channel users with no wiretapping, we show that the achievable region of the discrete memoryless (DM) MAC-WT channel given in \cite{xu2022achievable} can be enlarged.
In \cite{xu2022achievable}, the achievability was proven by considering the two-user case, making it possible to prove a key auxiliary lemma by directly using the Fourier-Motzkin elimination procedure.
However, this approach does not generalize to the case with any number of users.
In this paper, we provide a new region that generally enlarges that in \cite{xu2022achievable} and provide general achievability proof.
\end{abstract}

\IEEEpeerreviewmaketitle

\section{Introduction}
\label{section1}

To meet the tremendous demand for wireless communications, the future mobile systems will incorporate many different network topologies and large numbers of devices that may access and leave at any time, making it difficult to generate and manage cryptographic keys. 
In addition, advances in quantum computing make systems based on classical computational cryptography intrinsically insecure. 
Hence, the conventional cryptographic encryption methods, which rely on secret keys and assumptions of limited computational ability at eavesdroppers (Eves), are no longer sufficient to guarantee secrecy in the future mobile networks.
To address these issues, advanced signal processing techniques developed for embedding security directly in the physical layer have emerged and triggered considerable research interest in recent years \cite{yang2015safeguarding, 7762075}.
Different from the cryptographic encryption methods employed in the application layer, physical layer security techniques exploit the random propagation properties of radio channels and advanced signal processing techniques to prevent Eves from wiretapping.
Furthermore, physical layer security guarantees privacy in an information theoretic sense, i.e., without relying on Eves' limited computational limitations.
Ever since the early seminal works \cite{shannon1949communication, wyner1975wire, leung1978gaussian, csiszar1978broadcast}, the research of wiretap channels or physical layer security has evolved into various network topologies over the past decades, e.g., multiple access (MAC) wiretap channels \cite{ekrem2008secrecy, 4036106, 4455769, 5961828, nafea2019generalizing, tekin2008gaussian, tekin2008general, 9174164, xu2022achievable}, broadcast channels \cite{5730586, 5550390, 5605348, 6584931, 9133130}, interference channels \cite{4529283, 4595013, 5752448, 6006610, 7060726, 7313047}, relay-aided channels \cite{955145, 4608977, 5352243, 6601774, 7105936, 7355564, 7551149}, etc.
This paper mainly studies the physical layer security problem for MAC wiretap (MAC-WT) channels.
Hence, we introduce MAC-WT related works in the following.

We begin the review from works that studied MAC-WT systems with two transmitters \cite{ekrem2008secrecy, 4036106, 4455769, 5961828, nafea2019generalizing}.
Reference \cite{ekrem2008secrecy} considered a discrete memoryless (DM) MAC-WT channel with a weaker Eve which has access to a degraded version of the main channel, and developed an outer bound for the secrecy capacity region.
In \cite{4036106}, one user wishes to communicate confidential messages to a common receiver while the other one is permitted to eavesdrop.
Upper (converse) and lower (achievable) bounds for this communication situation were investigated.
A similar system was considered in \cite{4455769} where, differently, each user attempts to transmit both common and confidential information to the destination, and sees the other user as an Eve. 
Reference \cite{5961828} extended the work in \cite{4036106} and \cite{4455769} to a fading cognitive MAC channel.
In \cite{nafea2019generalizing}, the MAC-WT system with a DM main channel and different wiretapping scenarios was studied.

In \cite{xu2022achievable} and \cite{tekin2008gaussian, tekin2008general, 9174164}, the more general MAC-WT channel with any number of users was investigated.
Specifically, \cite{tekin2008gaussian} provided achievable rate regions for a Gaussian single-input single-output (SISO) MAC-WT channel with a weaker Eve under different secrecy constraints.
The work was then extended by \cite{tekin2008general} to a non-degraded MAC-WT channel where, besides confidential information, each user has also an open (i.e., non-confidential) message intended for Bob.
An achievable rate region for both secret and open message rates was provided in \cite{tekin2008general}. 
However, as we showed in \cite{xu2022achievable} and \cite{9174164}, there is a problem with the coding scheme in \cite{tekin2008general}, because of which rate tuples that are actually non-achievable are mistakenly claimed to be achievable.
In this sense, \cite{xu2022achievable} and \cite{9174164} provide a general achievable rate region for the MAC-WT scenario with confidential and open messages while \cite{tekin2008general} does not.
Note that in contrast to the works which considered only confidential information, by simultaneously transmitting secret and open messages, the spectral efficiency of the system can be greatly increased \cite{xu2022achievable}, \cite{9174164}.  

Based on the information-theoretic results, many references have studied the resource allocation problems for MAC-WT channels \cite{tekin2008general, hao2018resource, 8895802, lee2017precoder, xu2022achievable}.
The sum secrecy rate of a Gaussian SISO MAC-WT channel was maximized by power control in \cite{tekin2008general}, and it was shown that in the optimal case, only a subset of the strong users will transmit using the maximum power while the other users will keep inactive.
\cite{hao2018resource} and \cite{8895802} maximized the sum secrecy rate of a single-input multi-output (SIMO) device-to-device (D2D) underlaid network.
However, it was assumed in these two papers that both Bob and Eve adopted independent decoding, i.e., treating interference from the other users as independent additive noise.
Notice that designing a system under the assumption that Eve uses a suboptimal detection scheme may be risky, since the system secrecy may break down if Eve actually applies an enhanced receiver.
Reference \cite{lee2017precoder} considered the sum secrecy rate maximization problem of a Gaussian multiple-input multiple-output (MIMO) MAC-WT system, where both Bob and Eve applied the optimal joint decoding scheme.
However, a special power constraint was set to the signal vector covariance matrices, which limited the secrecy performance of the network.
In \cite{xu2022achievable}, the general power constraint was considered and it was shown that the system secrecy performance could be greatly improved compared with \cite{lee2017precoder}.

In this paper, we continue the study of the information-theoretic secrecy problem for a general MAC-WT channel in \cite{xu2022achievable}.
To make full use of the channel resources, besides the confidential message, each user also has an open message for Bob.
Eve aims to wiretap the confidential information of all users.
The main contributions of this paper are summarized below.

$\bullet$ In \cite{xu2022achievable}, we provided an achievable region for the DM   MAC-WT channel in \cite[Lemma~$1$]{xu2022achievable} and proved the achievability by considering a simple two-user case.
Though the general proof can be developed by following similar steps, the key auxiliary result \cite[Lemma~$7$]{xu2022achievable} does not appear to be directly extended to the case with arbitrary number of users $K$.
An extension of \cite[Lemma~$7$]{xu2022achievable} (see Theorem~\ref{lemma_FM_gene_K2} in this paper) is also of great importance for the general achievability proof in this paper.
To prove Theorem~\ref{lemma_FM_gene_K2}, we first give the general form of \cite[Lemma~$7$]{xu2022achievable} with any $K$ in Lemma~\ref{theorem_FM}.
For a simple system with a few users, this lemma can be proven by directly using the Fourier-Motzkin procedure \cite[Appendix D]{el2011network}.
However, as $K$ grows, this direct proof becomes unmanageable due to the excessively large number of inequalities.
We proved the $K = 3$ case by direct elimination in note \cite{xu2022note}, which shows how the number of inequalities quickly explodes with $K$. 
Besides the extremely high complexity, another disadvantage of the direct elimination strategy is that it works only if the number of users is given.
Obviously, this makes the strategy inappropriate for the proof of Lemma~\ref{theorem_FM} since it is a general result for any $K$.
In Appendix~\ref{Prove_theorem_FM}, we circumvent this problem and provide a general proof of Lemma~\ref{theorem_FM}.
Then, using Lemma~\ref{theorem_FM}, the general proof of Theorem~\ref{lemma_FM_gene_K2} follows.

$\bullet$ In \cite{xu2022achievable}, `garbage' messages were introduced to all users to ensure perfect secrecy.
This is important if a user's secrecy rate is positive.
However, when a user has zero secrecy rate, two options are possible: 1) still introducing `garbage' messages such that Eve cannot decode all of its information; 2) acting as a conventional MAC channel user with no wiretapping and just transmitting its open message.
We show in this paper that there is a trade-off between these two options, i.e., both of them have advantages and disadvantages.
Using the first option, the user's signal is noise and non-decodable for Eve and thus helps reduce Eve's wiretapping capability.
But as we will show, this option introduces many more constraints in determining the corresponding achievable region.
With the second option, since it is possible for Eve to decode the user's open message and then cancel it from the received signal, Eve may have a stronger wiretapping capability.
However, this option causes fewer constraints to the user's open message rate, which is of course good for determining the achievable region.
In this paper, we consider all possible cases and provide in Theorem~\ref{lemma_DM_exten} an achievable region for the DM MAC-WT channel, which improves that given in \cite[Lemma~$1$]{xu2022achievable}.
Note that different from \cite{xu2022achievable} which proved the achievability for a simple two-user case, we provide the general proof of Theorem~\ref{lemma_DM_exten} for arbitrary $K$.
Considering a special case where all users have only confidential information, we get directly from Theorem~\ref{lemma_DM_exten} an achievable region for the classical DM MAC-WT channel with no open messages.
The achievable region for such a channel has already been studied in \cite{ekrem2008secrecy} and \cite{tekin2008gaussian}.
But we show in Lemma~\ref{achi_secrecy_only2} that our result improves these provided by \cite{ekrem2008secrecy} and \cite{tekin2008gaussian}.

Notations: we use calligraphic capital letters to denote sets, $|\cdot|$ to denote the cardinality of a set, `` $\setminus$ " to represent the set subtraction operation, and ${\cal X}_1 \times {\cal X}_2$ for the Cartesian product of the sets ${\cal X}_1$ and ${\cal X}_2$.
We use line over a calligraphic letter to indicate it is the complement of a set, e.g., ${\overline S} = {\cal K} \setminus {\cal S}$ if ${\cal S} \subseteq {\cal K}$.
We use calligraphic subscript to denote the set of elements whose indexes take values from the subscript set, e.g., ${\cal X}_{\cal K} = \{{\cal X}_k, \forall k \in {\cal K}\}$, $X_{\cal K} = \{X_k, \forall k \in {\cal K}\}$, $\bm X_{\cal K} = \{\bm X_k, \forall k \in {\cal K}\}$, etc.
$\mathbb N$ is the set of natural numbers, and $\mathbb R$ and $\mathbb C$ are the real and complex spaces, respectively. 
Boldface upper (lower) case letters are used to denote matrices (vectors). 
A similar convention but with boldface upper-case letters is used for random vectors.
${\bm I}_B$ stands for the $B \times B$ dimensional identity matrix and $\bm 0$ denotes the all-zero vector or matrix.
Superscript $(\cdot)^H$ denotes the conjugated-transpose operation, ${\mathbb E}\left[\cdot\right]$ denotes the expectation operation, and $[\cdot]^+ \triangleq \max (\cdot,0)$.
The logarithm function $\log$ is base $2$.


\section{DM MAC-WT Channel}
\label{DM_model}

In this section, we introduce the general DM MAC-WT channel model and define metrics based on which coding schemes can be designed to guarantee perfect secrecy.
Then, we provide an extension of \cite[Lemma~$7$]{xu2022achievable}, which plays quite an important role for the achievability proof.

\subsection{DM MAC-WT Channel Model}
\label{DM_MAC_channel_model}

\begin{figure}
	\centering
	\includegraphics[scale=0.80]{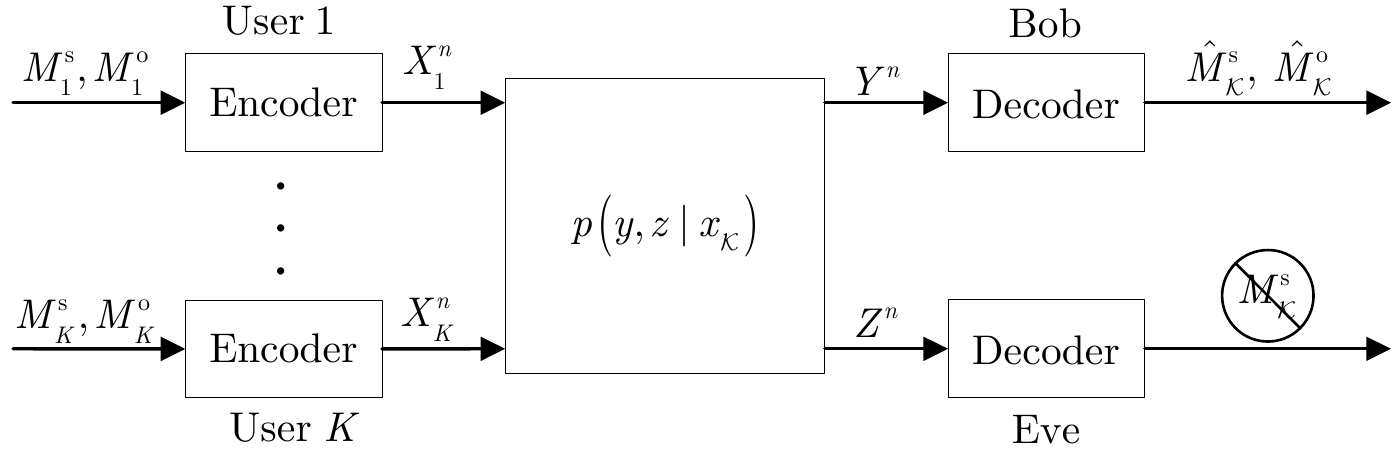}
	\caption{Block diagram of a DM MAC-WT channel.}
	\label{Fig1}
\end{figure}

As shown in Fig.~\ref{Fig1}, we consider a DM MAC-WT channel with $K$ users, a legitimate receiver (or Bob for brevity), and an eavesdropper (Eve).
Let ${\cal K} = \{1, \cdots, K\}$ denote the set of all users.
The DM MAC-WT system can then be denoted by $\left({\cal X}_{\cal K}, p(y,z|x_{\cal K}), {\cal Y}, {\cal Z}\right)$ (in short $p(y,z|x_{\cal K})$), where ${\cal X}_k$, $\cal Y$, and $\cal Z$ are finite alphabets, $x_k \in {\cal X}_k$ is the channel inputs from user $k$, and $y \in {\cal Y}$ and $z \in {\cal Z}$ are respectively channel outputs at Bob and Eve. 
For brevity, we use the short-hand notation $p(x_k)$ to indicate $P_{X_k}(x_k)$. 
Analogous short-hand notations are clear from the context. 

Each user $k \in {\cal K}$ transmits a secret message $M_k^{\text s}$ and an open message $M_k^{\text o}$ to Bob.
Eve attempts to overhear all the secret messages but is not interested in the open messages.
User $k$ encodes its information into a codeword $X_k^n$, and then transmits it over the channel with transition probability $p(y,z|x_{\cal K})$.
Upon receiving the sequence $Y^n$, Bob decodes the messages of all users.
To avoid leakage of confidential information to Eve, the secret messages of all users, i.e., $M_k^{\text s}, k \in {\cal K}$, should be protected.
Let $R_k^{\text s}$ and $R_k^{\text o}$ denote the rate of user $k$'s secret and open messages. 
Then, a $\left( 2^{n R_1^{\text s}}, 2^{n R_1^{\text o}}, \cdots, 2^{n R_K^{\text s}}, 2^{n R_K^{\text o}}, n \right)$ secrecy code for the considered DM MAC-WT channel consists of
\begin{itemize}
	\item Secret and open message sets: ${\cal M}_k^{\text s} = \left[1:2^{n R_k^{\text s}}\right]$ and ${\cal M}_k^{\text o} \!=\! \left[1:2^{n R_k^{\text o}}\right], \forall k \!\in\! {\cal K}$.
	Messages $M_k^{\text s}$ and $M_k^{\text o}$ are uniformly distributed over the corresponding sets ${\cal M}_k^{\text s}$ and ${\cal M}_k^{\text o}$. 
	\item $K$ randomized encoders: the encoder of user $k$ maps the message pair $(m_k^{\text s}, m_k^{\text o}) \in {\cal M}_k^{\text s} \times {\cal M}_k^{\text o}$ to a codeword $x_k^n$.
	\item A decoder at Bob which maps the received noisy sequence $y^n$ to the message estimate $\left( {\hat m}_k^{\text s}, {\hat m}_k^{\text o} \right) \in {\cal M}_k^{\text s} \times {\cal M}_k^{\text o}, \forall k \in {\cal K}$.
\end{itemize}

To ensure that all messages can be perfectly decoded at Bob and all confidential information can be perfectly protected from Eve, we define two metrics, i.e., the average probability of error
\begin{equation}\label{Pe}
P_{\text e} = {\text {Pr}} \left\{ \left( {\hat M}_{\cal K}^{\text s}, {\hat M}_{\cal K}^{\text o} \right) \neq \left( M_{\cal K}^{\text s}, M_{\cal K}^{\text o} \right) \right\},
\end{equation}
for Bob and the information leakage rate for Eve. 
For the MAC-WT channel, a widely used information leakage rate is $\frac{1}{n} I (M_{\cal K}^{\text s}; Z^n)$ and it is required that this rate vanishes as $n$ goes to infinity \cite{xu2022achievable, ekrem2008secrecy, tekin2008gaussian, tekin2008general}.
In this paper, we use a different leakage rate, based on which perfect secrecy can be realized and the achievable rate region in existing literatures can be improved.

We divide the user set ${\cal K}$ into two parts, i.e., ${\cal K}' \subseteq {\cal K}$ with $R_k^{\text s} \geq 0, \forall k \in {\cal K}'$, and ${\overline {{\cal K}'}} = {\cal K} \setminus {\cal K}'$ with $R_k^{\text s} = 0, \forall k \in {\overline {{\cal K}'}}$.
Here we do not require $R_k^{\text s} > 0$ if $k \in {\cal K}'$ such that users in ${\overline {{\cal K}'}}$ can also be reclassified to ${\cal K}'$.
To protect the confidential information of users in ${\cal K}'$, the coding scheme has to at least ensure that Eve cannot directly decode their messages using the normal MAC decoding scheme even if Eve has codebooks of all users.
As for user $k \in {\overline {{\cal K}'}}$, since $R_k^{\text s} = 0$, there is no such requirement and it is thus possible for Eve to get its open message $M_k^{\text o}$.
For a given ${\cal K}' \subseteq {\cal K}$, since $R_k^{\text s} = 0, \forall k \in {\overline {{\cal K}'}}$ and it is possible that Eve gets $M_k^{\text o}, \forall k \in {\overline {{\cal K}'}}$, we have
\begin{align}\label{leak_rate_1}
\frac{1}{n} I (M_{\cal K}^{\text s}; Z^n) & = \frac{1}{n} I (M_{{\cal K}'}^{\text s}; Z^n) \nonumber\\
& \leq \frac{1}{n} I (M_{{\cal K}'}^{\text s}; M_{\overline {{\cal K}'}}^{\text o}, Z^n) \nonumber\\
& = \frac{1}{n} I (M_{{\cal K}'}^{\text s}; Z^n| M_{\overline {{\cal K}'}}^{\text o}),
\end{align}
where the last step holds since the messages of all users are independent.
If $\frac{1}{n} I (M_{\cal K}^{\text s}; Z^n)$ is used to evaluate the secrecy level, a hidden assumption is that Eve cannot decode the messages of all users, including $M_k^{\text o}, \forall k \in {\overline {{\cal K}'}}$, since otherwise due to (\ref{leak_rate_1}), Eve may be able to extract the confidential information from $(M_{\overline {{\cal K}'}}^{\text o}, Z^n)$ even if $\frac{1}{n} I (M_{{\cal K}'}^{\text s}; Z^n) \rightarrow 0$.
Hence, we argue that the secrecy requirement based solely on $\frac{1}{n} I (M_{\cal K}^{\text s}; Z^n)$ implicitly forces any achievable coding scheme to prevent Eve to decode the open messages of users in ${\overline {{\cal K}'}}$. 
This requirement may be too restrictive and thus may limit the achievable region.  
As an alternative, we propose to use
\begin{equation}\label{leakage_rate}
R_{{\text E}, {\cal K}'} = \frac{1}{n} I (M_{{\cal K}'}^{\text s}; Z^n| M_{\overline {{\cal K}'}}^{\text o}),
\end{equation}
to evaluate the system's secrecy level.
For perfect secrecy, we require $R_{{\text E}, {\cal K}'} \rightarrow 0$ and take the union over all possible choices of ${\cal K}'$.
Obviously, by using (\ref{leakage_rate}), it is no longer necessary to design codes that prevent Eve from decoding $M_k^{\text o}, \forall k \in {\overline {{\cal K}'}}$.

Note that here we are not saying $\frac{1}{n} I (M_{\cal K}^{\text s}; Z^n)$ is always a bad metric in designing the codes and determining the achievable region.
Actually, as we will show later, by preventing Eve from decoding $M_k^{\text o}, \forall k \in {\overline {{\cal K}'}}$, the signal of users in ${\overline {{\cal K}'}}$ performs as noise to Eve, and may thus limit its wiretapping capability.
What we want to emphasize is that more possibilities should be taken into account.
In this paper, we use (\ref{leakage_rate}) to evaluate the secrecy level for all possible ${\cal K}' \subseteq {\cal K}$.
As stated above, users in ${\overline {{\cal K}'}}$ can be reclassified to ${\cal K}'$.
If ${\cal K}' = {\cal K}$, i.e., ${\overline {{\cal K}'}} = \phi$, it is obvious that (\ref{leak_rate_1}) holds with equality, indicating that the case using metric $\frac{1}{n} I (M_{\cal K}^{\text s}; Z^n)$ is also considered.

With the metrics defined in (\ref{Pe}) and (\ref{leakage_rate}), a rate tuple $(R_1^{\text s}, R_1^{\text o},\cdots, R_K^{\text s}, R_K^{\text o})$ is said to be achievable if for any $\delta > 0$, there exist a subset ${\cal K}' \subseteq {\cal K}$ and a sequence of 
$\left( 2^{n R_1^{\text s}}, 2^{n R_1^{\text o}}, \cdots, 2^{n R_K^{\text s}}, 2^{n R_K^{\text o}}, n \right)$ codes such that $R_k^{\text s} = 0, \forall k \in {\overline {{\cal K}'}}$ and
\begin{align}\label{RE}
& \lim_{n \rightarrow \infty} P_{\text e} \leq \delta, \nonumber\\
& \lim_{n \rightarrow \infty} R_{{\text E}, {\cal K}'} \leq \delta.
\end{align}

\subsection{Extension of \cite[Lemma~$7$]{xu2022achievable}}
\label{extension_FM}

In wiretap channels, to protect or hide the confidential information, `garbage' messages are usually introduced to the users \cite{xu2022achievable, 9174164, el2011network}.
We also do so in this paper.
It is thus important to study the existence condition of the `garbage' message rate  (see a two-user example in \cite[Lemma~$7$]{xu2022achievable}).
With the condition satisfied, `garbage' messages can be introduced at a feasible rate and perfect secrecy can be proven.
We give the condition for the considered DM MAC-WT channel in the following theorem, which plays quite an important role for the achievability proof in this paper.
\begin{theorem}\label{lemma_FM_gene_K2}
	Let $(X_{\cal K}, Y, Z) \sim \prod_{k=1}^K p(x_k) p(y,z| x_{\cal K})$.
	For any ${\cal K}' \subseteq {\cal K}$, if $I(X_{\cal S}; Y| X_{\overline {\cal S}}, X_{\overline {{\cal K}'}}) \geq I(X_{\cal S}; Z| X_{\overline {{\cal K}'}}), \forall {\cal S} \subseteq {\cal K}'$, then, for any rate tuple
	$(R_1^{\text s}, R_1^{\text o},\cdots, R_K^{\text s}, R_K^{\text o})$ satisfying
	\begin{equation}\label{region_DM0}
	\left\{\!\!\!
	\begin{array}{ll}
	R_k^{\text s} = 0, ~\forall~ k \in {\overline {{\cal K}'}}, \\
	\sum\limits_{k \in \cal S} R_k^{\text s} + \sum\limits_{k \in {\cal S} \setminus {\cal S}'} R_k^{\text o} + \sum\limits_{k \in {\cal T}} R_k^{\text o} \leq I(X_{\cal S}, X_{\cal T}; Y| X_{\overline {\cal S}}, X_{\overline {\cal T}}) - I(X_{{\cal S}'}; Z| X_{\overline {{\cal K}'}}), \\
	\quad\quad\quad\quad\quad\quad\quad\quad\quad\quad\quad\quad \forall~ {\cal S} \subseteq {\cal K}',~ {\cal S}' \subseteq {\cal S},~ {\cal T} \subseteq {\overline {{\cal K}'}},
	\end{array} \right.
	\end{equation}
	there exist $R_k^{\text g}, \forall k \in {\cal K}'$ such that
	\begin{equation}\label{region_FM2}
	\left\{\!\!\!
	\begin{array}{ll}
	R_k^{\text g} \geq 0, ~\forall~ k \in {\cal K}', \\
	\sum\limits_{k \in {\cal S}} (R_k^{\text s} + R_k^{\text o} + R_k^{\text g}) + \sum\limits_{k \in {\cal T}} R_k^{\text o} \leq I(X_{\cal S}, X_{\cal T}; Y| X_{\overline {\cal S}}, X_{\overline {\cal T}}), ~\forall~ {\cal S} \subseteq {\cal K}',~ {\cal T} \subseteq {\overline {{\cal K}'}}, \\
	\sum\limits_{k \in {\cal S}} (R_k^{\text o} + R_k^{\text g}) \geq I(X_{\cal S}; Z| X_{\overline {{\cal K}'}}), ~\forall~ {\cal S} \subseteq {\cal K}',
	\end{array} \right.
	\end{equation}
	where ${\overline {{\cal K}'}} = {\cal K} \setminus {\cal K}'$, ${\overline S} = {\cal K}' \setminus {\cal S}$, ${\overline {\cal T}} = {\overline {{\cal K}'}} \setminus {\cal T}$, and $R_k^{\text g}$ is the rate of the `garbage' message added by user $k$ to protect its confidential information
\end{theorem}
\itshape \textbf{Proof:} \upshape
We give in the following Lemma~\ref{theorem_FM}, which is a special case of Theorem~\ref{lemma_FM_gene_K2} for ${\cal K}' = {\cal K}$ and directly extends \cite[Lemma~$7$]{xu2022achievable} to the case with any $K$. 
Lemma~\ref{theorem_FM} is proven in Appendix~\ref{Prove_theorem_FM}. 
Then, using this lemma, Theorem~\ref{lemma_FM_gene_K2} is proven in Appendix~\ref{Prove_lemma_FM_gene_K2}.
\hfill $\Box$

In contrast to \cite[Lemma~$7$]{xu2022achievable}, which considers only two users and can thus be easily proven, Theorem~\ref{lemma_FM_gene_K2} not only extends the result to the general case with any $K$, but also gives the existence conditions of $R_k^{\text g}, \forall k \in {\cal K}'$ for all possible subsets ${\cal K}' \subseteq {\cal K}$, which is quite important for the achievability proof in this paper.
To apply the coding scheme provided in the next section, it is also necessary to know how to find the feasible $R_k^{\text g}, \forall k \in {\cal K}'$.
We show that this is easy.
For a given ${\cal K}' \subseteq {\cal K}$ and any rate tuple $(R_1^{\text s}, R_1^{\text o},\cdots, R_K^{\text s}, R_K^{\text o})$ satisfying (\ref{region_DM0}), the linear inequalities in (\ref{region_FM2}) define a polytope as a feasible region of $R_k^{\text g}, \forall k \in {\cal K}'$, which should be non-empty since otherwise the region defined by (\ref{region_DM0}) is empty.
Then, we may apply Dantzig's simplex algorithm to obtain $R_k^{\text g}, \forall k \in {\cal K}'$ \cite{gass2003linear}.

For the case ${\cal K}' = {\cal K}$, the basic result obtained from Theorem~\ref{lemma_FM_gene_K2} is collected in the following lemma.
\begin{lemma}\label{theorem_FM}
	Let $(X_{\cal K}, Y, Z) \sim \prod_{k=1}^K p(x_k) p(y,z| x_{\cal K})$.
	If $I(X_{\cal S}; Y| X_{\overline {\cal S}}) \geq I(X_{\cal S}; Z), \forall {\cal S} \subseteq {\cal K}$, for any rate tuple $(R_1^{\text s}, R_1^{\text o},\cdots, R_K^{\text s}, R_K^{\text o})$ satisfying 
	\begin{align}\label{rate_region0}
	\sum_{k \in \cal S} R_k^{\text s} + \sum_{k \in {\cal S} \setminus {\cal S}'} R_k^{\text o} \leq I(X_{\cal S}; Y| X_{\overline {\cal S}}) - I(X_{{\cal S}'}; Z), ~\forall~ {\cal S} \subseteq {\cal K},~ {\cal S}' \subseteq {\cal S},
	\end{align}
	there exist $R_k^{\text g}, \forall k \in {\cal K}$ such that
	\begin{equation}\label{region_FM1}
	\left\{\!\!\!
	\begin{array}{ll}
	R_k^{\text g} \geq 0, ~\forall~ k \in {\cal K}, \\
	\sum\limits_{k \in {\cal S}} (R_k^{\text s} + R_k^{\text o} + R_k^{\text g}) \leq I(X_{\cal S}; Y| X_{\overline {\cal S}}), ~\forall~ {\cal S} \subseteq {\cal K}, \\
	\sum\limits_{k \in {\cal S}} (R_k^{\text o} + R_k^{\text g}) \geq I(X_{\cal S}; Z), ~\forall~ {\cal S} \subseteq {\cal K},
	\end{array} \right.
	\end{equation}
	where ${\overline S} = {\cal K} \setminus {\cal S}$.
\end{lemma}
\itshape \textbf{Proof:} \upshape
See Appendix~\ref{Prove_theorem_FM}.
\hfill $\Box$

\begin{remark}\label{remark_FM}
	As already remarked before, Lemma~\ref{theorem_FM} is a direct extension of \cite[Lemma~$7$]{xu2022achievable} from a two-user case to the general $K$.
	For the simple system with a small $K$, e.g., $K = 1$ or $K = 2$, as stated in \cite{xu2022achievable}, Lemma~\ref{theorem_FM} can be proven by eliminating $R_k^{\text g}$ in (\ref{region_FM1}) using the Fourier-Motzkin procedure \cite[Appendix D]{el2011network} and showing that (\ref{rate_region0}) is the projection of (\ref{region_FM1}) onto the hyperplane $\{ R_k^{\text g} = 0, \forall k \in {\cal K}\}$.
	However, when $K$ increases, the number of inequalities resulted in the elimination procedure grows very quickly (doubly exponentially), making it quite difficult or even impractical to prove this lemma by following this brute-force way.
	Besides the great complexity, another problem is that the strategy works only if $K$ is given.
	Obviously, this makes the strategy inappropriate for the proof of Lemma~\ref{theorem_FM} since it is a general result for any $K$.
	
	In note \cite{xu2022note}, we consider a system with $K = 3$ and prove Lemma~\ref{theorem_FM} by eliminating $R_1^{\text g}$, $R_2^{\text g}$, and $R_3^{\text g}$ one by one.
	From (\ref{region_FM1}) we first get $4$ upper bounds and $5$ lower bounds on $R_1^{\text g}$.
	By pairing up these lower and upper bounds, we get $20$ inequalities, based on which $8$ upper bounds and $7$ lower bounds on $R_2^{\text g}$ are obtained.
	We then eliminate $R_2^{\text g}$ and get $56$ inequalities, in which most of them are redundant.
	Neglecting the redundant terms, we further get $9$ upper bounds and $7$ lower bounds on $R_3^{\text g}$, and $63$ inequalities by pairing them up.
	Neglecting the redundant terms, we show that the inequalities left construct (\ref{rate_region0}).
	This unpublished note is mentioned here and made public in \cite{xu2022note} to illustrate how difficult the brute-force Fourier-Motzkin elimination is, even in the simple case of $3$ users. 
\end{remark}

\section{Improved Achievable Region of the DM MAC-WT Channel}
\label{DM_region}

As stated in the introduction part, the information-theoretic secrecy problem for a MAC-WT channel with both secret and open messages has been studied by  \cite{tekin2008general} and \cite{xu2022achievable}. 
However, it was shown in \cite{xu2022achievable} that there is a problem with the coding scheme in \cite{tekin2008general}, because of which rate tuples that are non-achievable are mistakenly claimed to be achievable.
As a correction, \cite{xu2022achievable} provided an achievable region in \cite[Lemma~$1$]{xu2022achievable} and its proof for the two-user case.
In this section, we show that the achievable region provided in \cite[Lemma~$1$]{xu2022achievable} can be further improved and give the general proof for any $K$. 
Using the result, we also show that a new achievable secrecy rate region can be obtained for the conventional MAC-WT channel with only secret messages, which improves the regions given in the existing literature.

\subsection{Motivations of Improving \cite[Lemma~$1$]{xu2022achievable}}
\label{Motivations}

Here we explain why it is possible to improve \cite[Lemma~$1$]{xu2022achievable}.
For convenience, we rewrite \cite[Lemma~$1$]{xu2022achievable} below.
\begin{lemma}\label{theorem_region_DM}
	\cite[Lemma~$1$]{xu2022achievable} 
	Let $(X_{\cal K}, Y, Z) \sim \prod_{k=1}^K p(x_k) p(y,z| x_{\cal K})$. 
	Then, any rate tuple $(R_1^{\text s}, R_1^{\text o},$ $\cdots, R_K^{\text s}, R_K^{\text o})$ satisfying
	\begin{align}\label{region_DM}
	\sum_{k \in \cal S} R_k^{\text s} + \sum_{k \in {\cal S} \setminus {\cal S}'} R_k^{\text o} \leq \left[ I(X_{\cal S}; Y| X_{\overline {\cal S}}) - I(X_{{\cal S}'}; Z) \right]^+, ~\forall~ {\cal S} \subseteq {\cal K},~ {\cal S}' \subseteq {\cal S},
	\end{align}
	is achievable.
	Let ${\mathscr R} (X_{\cal K})$ denote the set of rate tuples satisfying (\ref{region_DM}).
	Then, the convex hull of the union of ${\mathscr R} (X_{\cal K})$ over all $\prod_{k=1}^K p(x_k)$ is an achievable rate region of the DM MAC-WT channel.
\end{lemma}

In Subsection~\ref{achi_region_DM}, we give Theorem~\ref{lemma_DM_exten}, which improves the region in Lemma~\ref{theorem_region_DM}, and provide the general proof for any $K$ in Appendix~\ref{prove_lemma_DM_exten}.
Now we consider two special cases of Lemma~\ref{theorem_region_DM} and provide some observations that evidence the fact that the region of Lemma~\ref{theorem_region_DM} can be improved. 

First, consider the case with $R_k^{\text o} = 0, \forall k \in {\cal K}$.
Then, (\ref{region_DM}) becomes
\begin{equation}\label{rate_region0_1}
\sum_{k \in \cal S} R_k^{\text s} \leq \left[ I(X_{\cal S}; Y| X_{\overline {\cal S}}) - I(X_{{\cal S}'}; Z) \right]^+, ~\forall~ {\cal S} \subseteq {\cal K}, ~ {\cal S}' \subseteq {\cal S},
\end{equation}
which can be further simplified as (\ref{region_secrecy_only}) since for any ${\cal S} \subseteq {\cal K}$, it is obvious that $I(X_{\cal S}; Z) \geq I(X_{{\cal S}'}; Z), \forall {\cal S}' \subseteq {\cal S}$.
We give an achievable secrecy rate region in the following lemma.
\begin{lemma}\label{achi_secrecy_only}
	Assume that $R_k^{\text o} = 0, \forall k \in {\cal K}$ and $(X_{\cal K}, Y, Z) \sim \prod_{k=1}^K p(x_k) p(y,z| x_{\cal K})$. 
	Any rate tuple $(R_1^{\text s}, \cdots, R_K^{\text s})$ satisfying
	\begin{equation}\label{region_secrecy_only}
	\sum_{k \in \cal S} R_k^{\text s} \leq \left[I(X_{\cal S}; Y| X_{\overline {\cal S}}) - I(X_{\cal S}; Z) \right]^+,~ \forall~ {\cal S} \subseteq {\cal K},
	\end{equation}
	is achievable.
	Let ${\mathscr R}^{\text s} (X_{\cal K})$ denote the set of rate tuples satisfying (\ref{region_secrecy_only}).
	Then, the convex hull of the union of ${\mathscr R}^{\text s} (X_{\cal K})$ over all $\prod_{k=1}^K p(x_k)$ is an achievable secrecy rate region of the DM MAC-WT channel with only secret messages.
\end{lemma}
Similar achievable secrecy rate regions for Gaussian MAC-WT channels can be found in \cite[Theorem~$1$]{tekin2008gaussian} (with two users), \cite[Theorem~$2$]{ekrem2008secrecy} (with a weaker Eve which has access to a degraded version of the main channel), and also \cite[Theorem~$1$]{tekin2008general} (by letting all open message rates be $0$).
Notice that we have shown in \cite[Remark~$1$ and Appendix~G]{xu2022achievable} that \cite[Theorem~$1$]{tekin2008general}, which bounds the secret and open message rates is in general incorrect. 
We show in Lemma~\ref{achi_secrecy_only2} in Subsection~\ref{achi_region_DM} that the region provided in Lemma~\ref{achi_secrecy_only} can be enlarged.
The results in \cite{tekin2008gaussian}, \cite{ekrem2008secrecy}, and \cite{tekin2008general} can thus also be improved.

Next we consider another special case with $R_k^{\text s} = 0, \forall k \in {\cal K}$, i.e., each user has only an open message for Bob.
Intuitively, we hope to get from Lemma~\ref{theorem_region_DM} the capacity region of the standard MAC channel with $K$ users and no wiretapping.
Unfortunately, we show that it is not the case.
With $R_k^{\text s} = 0, \forall k \in {\cal K}$, (\ref{region_DM}) becomes
\begin{equation}\label{rate_region0_3}
\sum_{k \in {\cal S} \setminus {\cal S}'} R_k^{\text o} \leq \left[I(X_{\cal S}; Y| X_{\overline {\cal S}}) - I(X_{{\cal S}'}; Z) \right]^+, ~\forall~ {\cal S} \subseteq {\cal K},~ {\cal S}' \subseteq {\cal S},
\end{equation}
which can be divided into two parts as follows
\begin{subequations}\label{rate_region0_3_12}
\begin{align}
& \sum_{k \in {\cal S}} R_k^{\text o} \leq I(X_{\cal S}; Y| X_{\overline {\cal S}}), ~\forall~ {\cal S} \subseteq {\cal K}, \label{rate_region0_3_1}\\
& \sum_{k \in {\cal S} \setminus {\cal S}'} R_k^{\text o} \leq \left[I(X_{\cal S}; Y| X_{\overline {\cal S}}) - I(X_{{\cal S}'}; Z) \right]^+, ~\forall~ {\cal S} \subseteq {\cal K},~ {\cal S}' \subseteq {\cal S},~ {{\cal S}'} \neq \phi. \label{rate_region0_3_2}
\end{align}
\end{subequations}
Obviously, (\ref{rate_region0_3_1}) over all $\prod_{k=1}^K p(x_k)$ constructs the capacity region of a conventional MAC channel with no wiretapping.
However, due to (\ref{rate_region0_3_2}), the region becomes smaller, indicating that the achievability results given in Lemma~\ref{theorem_region_DM} do not handle `well' the case where all (or some) users do not have confidential information.
This is because in \cite{xu2022achievable}, the coding scheme introduces `garbage' messages to all users and ensures $\sum_{k \in {\cal S}} (R_k^{\text o} + R_k^{\text g}) \geq I(X_{\cal S}; Z), \forall {\cal S} \subseteq {\cal K}$ to hide the confidential information.
This is important if $R_k^{\text s} > 0$.
But if $R_k^{\text s} = 0$, i.e., user $k$ has no confidential information, this may be unnecessary or even have a negative effect in determining the achievable region.
We will give the more detailed explanations in Subsection~\ref{achi_region_DM}.

\subsection{Improved Achievable Region}
\label{achi_region_DM}

By allowing users with zero secret message rate to act as conventional MAC channel users with no wiretapping, we obtain a better achievable region in the following Theorem~\ref{lemma_DM_exten}.
We show later that this theorem not only improves Lemma~\ref{theorem_region_DM}, by considering the special cases with only secret or open messages as we have done in Subsection~\ref{Motivations}, it also improves Lemma~\ref{achi_secrecy_only} and yields the standard MAC capacity region as a byproduct in a natural immediate way.
 
\begin{theorem}\label{lemma_DM_exten}
	Let $(X_{\cal K}, Y, Z) \sim \prod_{k=1}^K p(x_k) p(y,z| x_{\cal K})$. 
	For each given $ {\cal K}' \subseteq {\cal K}$, any rate tuple $(R_1^{\text s}, R_1^{\text o},\cdots, R_K^{\text s}, R_K^{\text o})$ satisfying
	\begin{equation}\label{region_DM_exten}
	\left\{
	\begin{array}{ll}
	R_k^{\text s} = 0, ~\forall~ k \in {\overline {{\cal K}'}}, \\
	\sum\limits_{k \in \cal S} R_k^{\text s} + \sum\limits_{k \in {\cal S} \setminus {\cal S}'} R_k^{\text o} + \sum\limits_{k \in {\cal T}} R_k^{\text o}	\leq \left[ I(X_{\cal S}, X_{\cal T}; Y| X_{\overline {\cal S}}, X_{\overline {\cal T}}) - I(X_{{\cal S}'}; Z| X_{\overline {{\cal K}'}}) \right]^+, \\
	\quad\quad\quad\quad\quad\quad\quad\quad\quad\quad\quad\quad \forall~ {\cal S} \subseteq {\cal K}',~ {\cal S}' \subseteq {\cal S},~ {\cal T} \subseteq {\overline {{\cal K}'}},
	\end{array} \right.
	\end{equation}
	is achievable, where $\overline {\cal S}$, $\overline {\cal T}$, and $\overline {{\cal K}'}$ are defined as in (\ref{region_FM2}).
	Let ${\mathscr R} (X_{\cal K}, {\cal K}')$ denote the set of rate tuples satisfying (\ref{region_DM_exten}).
	Then, the convex hull of the union of ${\mathscr R} (X_{\cal K}, {\cal K}')$ over all $\prod_{k=1}^K p(x_k)$ and $ {\cal K}' \subseteq {\cal K}$ is an achievable rate region of the DM MAC-WT channel.
\end{theorem}
\itshape \textbf{Proof:} \upshape
See Appendix~\ref{prove_lemma_DM_exten}.
\hfill $\Box$

Obviously, Lemma~\ref{theorem_region_DM} is a special case of Theorem~\ref{lemma_DM_exten} with ${\cal K}' = {\cal K}$.

\begin{remark}\label{remark_lemma_DM_exten}
	Note that the partition of ${\cal K}$ into ${\cal K}'$ and ${\overline {{\cal K}'}}$ is very important since it considers a trade-off, which influences the size of the achievable region.
	To clarify this, we talk about a specific user $j$ in ${\cal K}$, and assume that all the other users ${\cal K} \setminus \{j\}$ are in ${\cal K}'$ for convenience.
	If $R_j^{\text s} > 0$, it is obvious from the achievability proof in Appendix~\ref{prove_lemma_DM_exten} that $j$ should be included in ${\cal K}'$ such that its confidential information can be perfectly protected by the open and introduced `garbage' messages.
	Differently, if $R_j^{\text s} = 0$, there are two cases for $j$, i.e., $j \in {\cal K}'$ and $j \in {\overline {{\cal K}'}}$.
	We show that both these two cases have advantages and disadvantages in determining the corresponding achievable regions.
	For clarity, we only give some brief explanations in this remark.
	The more detailed analysis is provide in Appendix~\ref{specific_j}.
	
	In the first case with $j \in {\cal K}'$, ${\cal K}' = {\cal K}$ and ${\overline {{\cal K}'}} = \phi$.
    We ensure 
    \begin{equation}\label{j_in_ensure}
    \sum_{k \in {\cal S}} (R_k^{\text o} + R_k^{\text g}) \geq I(X_{\cal S}; Z), ~\forall~ {\cal S} \subseteq {\cal K},
    \end{equation}
    such that the information leakage rate vanishes, i.e.,
    \begin{equation}\label{j_in_leak}
    \lim_{n \rightarrow \infty} \frac{1}{n} I (M_{\cal K}^{\text s}; Z^n) \leq \delta.
    \end{equation}
    In the second case with $j \in {\overline {{\cal K}'}}$, ${\cal K}' = {\cal K} \setminus \{j\}$ and ${\overline {{\cal K}'}} = \{j\}$.
    We ensure 
    \begin{equation}\label{j_notin_ensure}
    \sum_{k \in {\cal S}} (R_k^{\text o} + R_k^{\text g}) \geq I(X_{\cal S}; Z| X_j), ~\forall~ {\cal S} \subseteq {\cal K} \setminus \{j\},
    \end{equation}
    such that
    \begin{equation}\label{j_notin_leak}
    \lim_{n \rightarrow \infty} \frac{1}{n} I (M_{{\cal K} \setminus \{j\}}^{\text s}; Z^n| M_j^{\text o}) \leq \delta.
    \end{equation}
    Note that (\ref{j_in_ensure}) considers the combination of all users, while (\ref{j_notin_ensure}) does not take user $j$ into account.
    Hence, it is not possible for Eve to decode the open message $M_j^{\text o}$ in the first case, but is possible in the second case.
    This explains why we consider tighter lower bounds (with $X_j$ as a known condition) in (\ref{j_notin_ensure}) and a stronger condition on the leakage rate in (\ref{j_notin_leak}).
    As shown in (\ref{FM_project4}), the tighter lower bounds in (\ref{j_notin_ensure}) cause tighter upper bounds to the rate sums in ${\mathscr R} (X_{\cal K}, {\cal K} \setminus \{j\})$.
    This shows the advantage of case one and disadvantage of case two. 
    
    On the other hand, since (\ref{j_in_ensure}) considers one more user $j$, it includes $2^{K-1}$ more inequalities than (\ref{j_notin_ensure}).
    Specifically, (\ref{j_in_ensure}) sets lower bounds to $\sum_{k \in {\cal S}} (R_k^{\text o} + R_k^{\text g}) + R_j^{\text o} + R_j^{\text g}, \forall {\cal S} \subseteq {\cal K} \setminus \{j\}$ while (\ref{j_notin_ensure}) does not.
    Then, it is known from the Fourier-Motzkin procedure given in Appendix~\ref{Prove_theorem_FM} and also shown in (\ref{FM_project2}) that we get many more inequalities, i.e., more constraints on the sum rates, in region ${\mathscr R} (X_{\cal K}, {\cal K})$ (with $R_j^{\text s} = 0$) compared with ${\mathscr R} (X_{\cal K}, {\cal K} \setminus \{j\})$.
    In this sense, case two performs better than case one. 
    Therefore, when $R_j^{\text s} = 0$, there is a trade-off between cases $j \in {\cal K}'$ and $j \in {\overline {{\cal K}'}}$.
    Since Theorem~\ref{lemma_DM_exten} considers all possible partitions of ${\cal K}$, it improves the results of Lemma~\ref{theorem_region_DM}.
\end{remark}

As done for Lemma~\ref{theorem_region_DM}, we also consider two special cases with $R_k^{\text o} = 0, \forall k \in {\cal K}$ and $R_k^{\text s} = 0, \forall k \in {\cal K}$ for Theorem~\ref{lemma_DM_exten}.
First, if $R_k^{\text o} = 0, \forall k \in {\cal K}$, (\ref{region_DM_exten}) becomes
\begin{equation}\label{region_DM_exten_1}
\left\{
\begin{array}{ll}
R_k^{\text s} = 0, ~\forall~ k \in {\overline {{\cal K}'}}, \\
\sum\limits_{k \in \cal S} R_k^{\text s} \leq \left[I(X_{\cal S}, X_{\cal T}; Y| X_{\overline {\cal S}}, X_{\overline {\cal T}}) - I(X_{{\cal S}'}; Z| X_{\overline {{\cal K}'}}) \right]^+, ~\forall~ {\cal S} \subseteq {\cal K}',~ {\cal S}' \subseteq {\cal S},~ {\cal T} \subseteq {\overline {{\cal K}'}},
\end{array} \right.
\end{equation}
which can be simplified as (\ref{region_secrecy_only2}) since for any ${\cal S} \subseteq {\cal K}'$,
\begin{align}
& I(X_{\cal S}; Y| X_{\overline {\cal S}}, X_{\overline {{\cal K}'}}) \leq I(X_{\cal S}, X_{\cal T}; Y| X_{\overline {\cal S}}, X_{\overline {\cal T}}), \nonumber\\
& I(X_{\cal S}; Z| X_{\overline {{\cal K}'}}) \geq I(X_{{\cal S}'}; Z| X_{\overline {{\cal K}'}}), ~\forall~ {\cal S}' \subseteq {\cal S},~ {\cal T} \subseteq {\overline {{\cal K}'}}.
\end{align}
We provide another achievable region for the DM MAC-WT channel with only confidential information in the following lemma and show that it improves the region given in Lemma~\ref{achi_secrecy_only}.
\begin{lemma}\label{achi_secrecy_only2}
	Assume that $R_k^{\text o} = 0, \forall k \in {\cal K}$ and $(X_{\cal K}, Y, Z) \sim \prod_{k=1}^K p(x_k) p(y,z| x_{\cal K})$. 
	For each given $ {\cal K}' \subseteq {\cal K}$, any secrecy rate tuple $(R_1^{\text s}, \cdots, R_K^{\text s})$ satisfying 
	\begin{equation}\label{region_secrecy_only2}
	\left\{\!\!\!
	\begin{array}{ll}
	R_k^{\text s} = 0, ~\forall~ k \in {\overline {{\cal K}'}}, \\
	\sum\limits_{k \in \cal S} R_k^{\text s} \leq \left[I(X_{\cal S}; Y| X_{\overline {\cal S}}, X_{\overline {{\cal K}'}}) - I(X_{\cal S}; Z| X_{\overline {{\cal K}'}}) \right]^+, ~\forall~ {\cal S} \subseteq {\cal K}',
	\end{array} \right.
	\end{equation}
	is achievable.
	Let ${\mathscr R}^{\text s} (X_{\cal K}, {\cal K}')$ denote the set of rate tuples satisfying (\ref{region_secrecy_only2}).
	Then, the convex hull of the union of ${\mathscr R}^{\text s} (X_{\cal K}, {\cal K}')$ over all $\prod_{k=1}^K p(x_k)$ and $ {\cal K}' \subseteq {\cal K}$ is an achievable secrecy rate region of the DM MAC-WT channel with only secret messages.
\end{lemma}
It can be easily found by setting ${\cal K}' = {\cal K}$ that the achievable region given in Lemma~\ref{achi_secrecy_only} is contained in that provided by Lemma~\ref{achi_secrecy_only2}.
In this sense, Lemma~\ref{achi_secrecy_only2} not only improves the result in Lemma~\ref{achi_secrecy_only}, but also those in \cite{tekin2008gaussian, ekrem2008secrecy}, and \cite{tekin2008general}.
We further show this by giving a specific example with two users in Appendix~\ref{K1_2}.

On the other hand, if $R_k^{\text s} = 0, \forall k \in {\cal K}$, for a given ${\cal K}' \subseteq {\cal K}$, (\ref{region_DM_exten}) can be rewritten as
\begin{align}\label{region_DM_exten_4}
\sum_{k \in {\cal S} \setminus {\cal S}'}\! R_k^{\text o} \!+\! \sum_{k \in {\cal T}} R_k^{\text o} & \!\leq\! \left[I(X_{\cal S}, X_{\cal T}; Y| X_{\overline {\cal S}}, X_{\overline {\cal T}}) \!-\! I(X_{{\cal S}'}; Z| X_{\overline {{\cal K}'}}) \right]^+\!\!, \forall {\cal S} \subseteq {\cal K}', {\cal S}' \subseteq {\cal S}, {\cal T} \subseteq {\overline {{\cal K}'}}.
\end{align}
If ${\cal K}' = \phi$, we have ${\overline {{\cal K}'}} = {\cal K}$ and ${\cal S} = {\cal S}' = \phi$.
(\ref{region_DM_exten_4}) in this case becomes
\begin{align}\label{region_DM_exten_7}
\sum\limits_{k \in {\cal T}} R_k^{\text o} \leq I(X_{\cal T}; Y| X_{\overline {\cal T}}), ~\forall~ {\cal T} \subseteq {\cal K},
\end{align}
which over all possible distributions $\prod_{k=1}^K p(x_k)$ constructs the capacity region of a conventional MAC channel with $K$ users and no wiretapping.
If ${\cal K}' \neq \phi$, (\ref{region_DM_exten_4}) can be divided into two parts
\begin{subequations}\label{region_DM_exten_56}
	\begin{align}
	\sum_{k \in {\cal S}} R_k^{\text o} + \sum_{k \in {\cal T}} R_k^{\text o} & \leq I(X_{\cal S}, X_{\cal T}; Y| X_{\overline {\cal S}}, X_{\overline {\cal T}}),  ~\forall~ {\cal S} \subseteq {\cal K}',~ {\cal T} \subseteq {\overline {{\cal K}'}},\label{region_DM_exten_5}\\
	\sum_{k \in {\cal S} \setminus {\cal S}'}\!\!\!\! R_k^{\text o} \!+\! \sum_{k \in {\cal T}}\!\! R_k^{\text o} & \!\leq\! \left[I(X_{\cal S}, X_{\cal T}; Y| X_{\overline {\cal S}}, X_{\overline {\cal T}}) \!-\! I(X_{{\cal S}'}; Z| X_{\overline {{\cal K}'}}) \right]^+\!\!, \forall {\cal S} \subseteq {\cal K}', {\cal S}' \subseteq {\cal S}, {{\cal S}'} \neq \phi, {\cal T} \subseteq {\overline {{\cal K}'}}. \label{region_DM_exten_6}
	\end{align}
\end{subequations}
It is obvious that (\ref{region_DM_exten_5}) is equivalent to (\ref{region_DM_exten_7}).
However, due to (\ref{region_DM_exten_6}), for any ${\cal K}' \neq \phi$, the open message rate region jointly defined by (\ref{region_DM_exten_5}) and (\ref{region_DM_exten_6}) is included in that defined by (\ref{region_DM_exten_7}).
Hence, when $R_k^{\text s} = 0, \forall k \in {\cal K}$, instead of taking into account (\ref{region_DM_exten_4}) for all ${\cal K}' \subseteq {\cal K}$, the region union of ${\mathscr R} (X_{\cal K}, {\cal K}')$ over all ${\cal K}' \subseteq {\cal K}$ can be easily characterized by (\ref{region_DM_exten_7}).
In this sense, the open message rate region defined by (\ref{rate_region0_3}) is improved.

Since this paper considers MAC-WT channels, we are especially concerned about the maximum achievable sum secrecy rate of the system.
In addition, since all users have open messages intended for Bob, then, an interesting question is if all users encode their confidential messages at the maximum sum secrecy rate, what is the maximum sum rate at which they could encode their open messages.
We give the answer in the following Theorem.
\begin{theorem}\label{max_R_s_joint}
	For the considered DM MAC-WT channel, with a particular input distribution $\prod_{k=1}^K p(x_k)$, the maximum achievable sum secrecy rate $\sum_{k \in {\cal K}} R_k^{\text s}$ is
	\begin{equation}\label{R_s_joint_DM}
	R^{\text s} (X_{\cal K}) = \mathop {\max }\limits_{{\cal K}' \subseteq {\cal K}} \left\{ \left[I(X_{{\cal K}'}; Y| X_{\overline {{\cal K}'}}) - I(X_{{\cal K}'}; Z| X_{\overline {{\cal K}'}})\right]^+ \right\}.
	\end{equation}
	Let ${\cal K}'^*$ denote the subset in ${\cal K}$ which achieves (\ref{R_s_joint_DM}) and $ {\overline {{\cal K}'^*}} = {\cal K} \setminus {\cal K}'^*$.
	If $R^{\text s} (X_{\cal K}) > 0$ and all users transmit their confidential messages at sum rate $R^{\text s} (X_{\cal K})$~\footnote{Here we assume $R^{\text s} (X_{\cal K}) > 0$ since otherwise we have $R_k^{\text s} = 0, \forall k \in {\cal K}$, i.e., the system reduces to a conventional MAC channel with only open messages.}, the maximum achievable sum rate at which users in ${\cal K}$ could send their open messages is given by
	\begin{equation} \label{R_o_K1_DM}
	R^{\text o} (X_{\cal K}) = I(X_{\overline {{\cal K}'^*}}; Y) + I(X_{{\cal K}'^*}; Z| X_{\overline {{\cal K}'^*}}).
	\end{equation}
\end{theorem}
\itshape \textbf{Proof:} \upshape
See Appendix \ref{Prove_max_R_s_joint}.
\hfill $\Box$

Theorem~\ref{max_R_s_joint} shows that the channel can support a non-trivial additional open sum rate even if the coding scheme is designed to maximize the sum secrecy rate.

\section{Conclusions}
\label{conclusion}

In this paper, we studied the information-theoretic secrecy of MAC-WT channels where each user had both secret and open messages for the intended receiver.
We provided new achievable regions that enlarge previously known results and gave the general proof for any number of users.

\appendices

\section{Proof of Lemma~\ref{theorem_FM}}
\label{Prove_theorem_FM}

As stated in Remark~\ref{remark_FM}, it is impossible to prove Lemma~\ref{theorem_FM} by directly using the Fourier-Motzkin procedure to eliminate all $R_k^{\text g}$ in (\ref{region_FM1}), not only because of its huge complexity but also due to the fact that the elimination strategy works only if $K$ is given.
Hence, we adopt mathematical induction in the following to prove Lemma~\ref{theorem_FM}.

We first consider the base case with $K = 1$.
By eliminating $R_1^{\text g}$ in (\ref{region_FM1}) using the Fourier-Motzkin procedure \cite[Appendix D]{el2011network}, it can be easily proven that (\ref{rate_region0}) is the projection of (\ref{region_FM1}) onto the hyperplane $\{ R_1^{\text g} = 0\}$.
Lemma~\ref{theorem_FM} can thus be proven for this simple case.

Next, we consider the induction step.
Assume that for any given positive integer $K$, (\ref{rate_region0}) is the projection of (\ref{region_FM1}) onto the hyperplane $\{ R_k^{\text g} = 0, \forall k \in {\cal K}\}$.
Then, by this assumption, it is possible to obtain (\ref{rate_region0}) by eliminating the variables $R_k^{\text g}, \forall k \in {\cal K}$ using the Fourier-Motzkin procedure. 
For convenience, in the following we refer to this assumption as the {\em induction assumption}. 
Under the induction assumption, we shall prove that the statement of Lemma 1 holds  for $K+1$ users.
With $K+1$ users, (\ref{rate_region0}) and (\ref{region_FM1}) become
\begin{align}\label{rate_region_K+1}
\sum_{k \in \cal S} R_k^{\text s} + \sum_{k \in {\cal S}\setminus {\cal S}'} R_k^{\text o} \leq I(X_{\cal S}; Y| X_{\overline {\cal S}}) - I(X_{{\cal S}'}; Z), ~\forall~ {\cal S} \subseteq {\cal K} \cup \{ K + 1 \},~ {\cal S}' \subseteq \cal S,
\end{align}
and
\begin{equation}\label{region_FM_K+1}
\left\{
\begin{array}{ll}
R_k^{\text g} \geq 0, ~\forall~ k \in {\cal K} \cup \{ K + 1 \}, \\
\sum\limits_{k \in {\cal S}} (R_k^{\text s} + R_k^{\text o} + R_k^{\text g}) \leq I(X_{\cal S}; Y| X_{\overline {\cal S}}), ~\forall~ {\cal S} \subseteq {\cal K} \cup \{ K + 1 \}, \\
\sum\limits_{k \in {\cal S}} (R_k^{\text o} + R_k^{\text g}) \geq I(X_{\cal S}; Z), ~\forall~ {\cal S} \subseteq {\cal K} \cup \{ K + 1 \}.
\end{array} \right.
\end{equation}
We need to show that (\ref{rate_region_K+1}) is the projection of (\ref{region_FM_K+1}) onto the hyperplane $\{ R_k^{\text g} = 0, \forall k \in {\cal K} \cup \{ K + 1 \}\}$, i.e., (\ref{rate_region_K+1}) can be obtained by eliminating $R_k^{\text g}, \forall k \in {\cal K}$ as well as $R_{K + 1}^{\text g}$ in (\ref{region_FM_K+1}).
For this purpose, by separating user $K + 1$ from users in set ${\cal K}$, we rewrite (\ref{rate_region_K+1}) equivalently as
\begin{subequations}\label{rate_region_K+1_2}
	\begin{align}
	& \sum_{k \in \cal S} R_k^{\text s} + \sum_{k \in {\cal S}\setminus {\cal S}'} R_k^{\text o} \leq I(X_{\cal S}; Y| X_{\overline {\cal S}}, X_{K + 1}) - I(X_{{\cal S}'}; Z), ~\forall~ {\cal S} \subseteq {\cal K},~ {\cal S}' \subseteq \cal S,\label{rate_region_K+1_2a}\\
	& \sum_{k \in \cal S} R_k^{\text s} \!+\! R_{K + 1}^{\text s} \!+\! \sum_{k \in {\cal S}\setminus {\cal S}'} R_k^{\text o} \!+\! R_{K + 1}^{\text o} \!\leq\! I(X_{\cal S}, X_{K + 1}; Y| X_{\overline {\cal S}}) \!-\! I(X_{{\cal S}'}; Z), \forall {\cal S} \subseteq {\cal K}, {\cal S}' \subseteq \cal S,\label{rate_region_K+1_2b}\\
	& \sum_{k \in \cal S} R_k^{\text s} \!+\! R_{K + 1}^{\text s} \!+\! \sum_{k \in {\cal S}\setminus {\cal S}'} \!R_k^{\text o} \leq I(X_{\cal S}, X_{K + 1}; Y| X_{\overline {\cal S}}) \!-\! I(X_{{\cal S}'}, X_{K + 1}; Z), \forall {\cal S} \subseteq {\cal K},~ {\cal S}' \subseteq \cal S,\label{rate_region_K+1_2c}
	\end{align}
\end{subequations}
and (\ref{region_FM_K+1}) as
\begin{subequations}\label{region_FM_K+1_2}
	\begin{align}
	& R_k^{\text g} \geq 0, ~\forall~ k \in {\cal K}, \label{region_FM_K+1_2a}\\
	& \sum\limits_{k \in {\cal S}} (R_k^{\text s} + R_k^{\text o} + R_k^{\text g}) \leq I(X_{\cal S}; Y| X_{\overline {\cal S}}, X_{K + 1}), ~\forall~ {\cal S} \subseteq {\cal K}, \label{region_FM_K+1_2b}\\
	& \sum\limits_{k \in {\cal S}} (R_k^{\text o} + R_k^{\text g}) \geq I(X_{\cal S}; Z), ~\forall~ {\cal S} \subseteq {\cal K}, \label{region_FM_K+1_2c}\\
	& \sum\limits_{k \in {\cal S}} (R_k^{\text s} \!+\! R_k^{\text o} \!+\! R_k^{\text g}) \leq I(X_{\cal S}, X_{K + 1}; Y| X_{\overline {\cal S}}) \!-\! (R_{K+1}^{\text s} \!+\! R_{K+1}^{\text o} \!+\! R_{K+1}^{\text g}), ~\forall~ {\cal S} \subseteq {\cal K},~ {\cal S} \neq \phi, \label{region_FM_K+1_2d}\\
	& \sum\limits_{k \in {\cal S}} (R_k^{\text o} + R_k^{\text g}) \geq I(X_{\cal S}, X_{K + 1}; Z) - (R_{K+1}^{\text o} + R_{K+1}^{\text g}), ~\forall~ {\cal S} \subseteq {\cal K},~ {\cal S} \neq \phi,\label{region_FM_K+1_2e}\\
	& R_{K+1}^{\text g} \geq 0, \label{region_FM_K+1_2f}\\
	& R_{K+1}^{\text s} + R_{K+1}^{\text o} + R_{K+1}^{\text g} \leq I(X_{K+1}; Y| X_{{\cal K}}), \label{region_FM_K+1_2g}\\
	& R_{K+1}^{\text o} + R_{K+1}^{\text g} \geq I(X_{K+1}; Z). \label{region_FM_K+1_2h}
	\end{align}
\end{subequations}
Note that in (\ref{region_FM_K+1_2d}) and (\ref{region_FM_K+1_2e}) we let ${\cal S} \neq \phi$ since otherwise they will reduce to (\ref{region_FM_K+1_2g}) and (\ref{region_FM_K+1_2h}), which do not contain $R_k^{\text g}, \forall k \in {\cal K}$.
In the following, we eliminate first $R_k^{\text g}, \forall k \in {\cal K}$ and then $R_{K+1}^{\text g}$ in (\ref{region_FM_K+1_2}).

\subsection{Elimination of $R_k^{\text g}, \forall k \in {\cal K}$}

To eliminate $R_k^{\text g}, \forall k \in {\cal K}$ in (\ref{region_FM_K+1_2}), we focus on (\ref{region_FM_K+1_2a})~$\sim$ (\ref{region_FM_K+1_2e}) since only these inequalities contain $R_k^{\text g}, \forall k \in {\cal K}$ while (\ref{region_FM_K+1_2f})~$\sim$ (\ref{region_FM_K+1_2h}) do not.
Since there are $K$ different $R_k^{\text g}$, as stated above, it is impractical to eliminate $R_k^{\text g}$ one by one.
Hence, instead of eliminating $R_k^{\text g}, \forall k \in {\cal K}$ directly from (\ref{region_FM_K+1_2a})~$\sim$ (\ref{region_FM_K+1_2e}), we divide these inequalities into $4$ categories, which together consider all possible upper and lower bound pairs on $R_k^{\text g}, \forall k \in {\cal K}$, and eliminate $R_k^{\text g}, \forall k \in {\cal K}$ in each category using the induction assumption.

\subsubsection{Category~$1$}

We include inequalities (\ref{region_FM_K+1_2a}), (\ref{region_FM_K+1_2b}), and (\ref{region_FM_K+1_2c}) in Category~$1$, and rewrite them as follows for clarity
\begin{equation}\label{region_FM_K+1_2_abc}
\left\{
\begin{array}{ll}
R_k^{\text g} \geq 0, ~\forall~ k \in {\cal K}, \\
\sum\limits_{k \in {\cal S}} (R_k^{\text s} + R_k^{\text o} + R_k^{\text g}) \leq I(X_{\cal S}; Y| X_{\overline {\cal S}}, X_{K + 1}), ~\forall~ {\cal S} \subseteq {\cal K}, \\
\sum\limits_{k \in {\cal S}} (R_k^{\text o} + R_k^{\text g}) \geq I(X_{\cal S}; Z), \forall {\cal S} \subseteq {\cal K}.
\end{array} \right.
\end{equation}
Obviously, (\ref{region_FM_K+1_2_abc}) has a similar formulation as (\ref{region_FM1}).
Hence, from the induction assumption it is known that the projection of (\ref{region_FM_K+1_2_abc}) onto the hyperplane $\{ R_k^{\text g} = 0, \forall k \in {\cal K} \}$ is
\begin{align}\label{R_g_abc}
\sum_{k \in \cal S} R_k^{\text s} + \sum_{k \in {\cal S}\setminus {\cal S}'} R_k^{\text o} \leq I(X_{\cal S}; Y| X_{\overline {\cal S}}, X_{K + 1}) - I(X_{{\cal S}'}; Z), ~\forall~ {\cal S} \subseteq {\cal K},~ {\cal S}' \subseteq \cal S,
\end{align}
which is the same as (\ref{rate_region_K+1_2a}).

\subsubsection{Category~$2$}

In this category we include inequalities (\ref{region_FM_K+1_2a}), (\ref{region_FM_K+1_2d}), and (\ref{region_FM_K+1_2c}), and rewrite them as
\begin{equation}\label{region_FM_K+1_2_adc}
\left\{\!\!\!
\begin{array}{ll}
R_k^{\text g} \geq 0, ~\forall~ k \in {\cal K}, \\
\sum\limits_{k \in {\cal S}} (R_k^{\text s} \!+\! R_k^{\text o} \!+\! R_k^{\text g}) \leq I(X_{\cal S}, X_{K + 1}; Y| X_{\overline {\cal S}}) \!-\! (R_{K+1}^{\text s} \!+\! R_{K+1}^{\text o} \!+\! R_{K+1}^{\text g}), ~\forall {\cal S} \subseteq {\cal K}, {\cal S} \neq \phi, \\
\sum\limits_{k \in {\cal S}} (R_k^{\text o} + R_k^{\text g}) \geq I(X_{\cal S}; Z), ~\forall~ {\cal S} \subseteq {\cal K}.
\end{array} \right.
\end{equation}
Note that though we let ${\cal S} \neq \phi$ in (\ref{region_FM_K+1_2d}), (\ref{region_FM_K+1_2_adc}) still has a similar expression as (\ref{region_FM1}).
It can be easily checked that if ${\cal S} = \phi$, the second inequality of (\ref{region_FM1}) gives $0 \leq 0$, which can be omitted.
Hence, we may also let ${\cal S} \neq \phi$ in (\ref{region_FM1}) without changing its formulation.
The induction assumption can thus be used to eliminate $R_k^{\text g}, \forall k \in {\cal K}$ in (\ref{region_FM_K+1_2_adc}) and obtain
\begin{align}\label{R_g_adc}
& \sum_{k \in \cal S} R_k^{\text s} + \sum_{k \in {\cal S}\setminus {\cal S}'} R_k^{\text o} \leq I(X_{\cal S}, X_{K + 1}; Y| X_{\overline {\cal S}}) - (R_{K+1}^{\text s} + R_{K+1}^{\text o} + R_{K+1}^{\text g}) - I(X_{{\cal S}'}; Z),\nonumber\\
& \quad\quad\quad\quad\quad\quad\quad\quad\; \forall~ {\cal S} \subseteq {\cal K},~ {\cal S} \neq \phi,~ {\cal S}' \subseteq \cal S,
\end{align}
which contains $R_{K+1}^{\text g}$ and thus has to be considered in the next step of eliminating $R_{K+1}^{\text g}$.

\subsubsection{Category~$3$}

In Category~$3$ we include (\ref{region_FM_K+1_2a}), (\ref{region_FM_K+1_2d}), and (\ref{region_FM_K+1_2e}), and rewrite them as follows
\begin{equation}\label{region_FM_K+1_2_ade}
\left\{\!\!\!
\begin{array}{ll}
R_k^{\text g} \geq 0, ~\forall~ k \in {\cal K}, \\
\sum\limits_{k \in {\cal S}} (R_k^{\text s} \!+\! R_k^{\text o} \!+\! R_k^{\text g}) \leq I(X_{\cal S}, X_{K + 1}; Y| X_{\overline {\cal S}}) \!-\! (R_{K+1}^{\text s} \!+\! R_{K+1}^{\text o} \!+\! R_{K+1}^{\text g}), ~\forall~ {\cal S} \subseteq {\cal K}, {\cal S} \neq \phi, \\
\sum\limits_{k \in {\cal S}} (R_k^{\text o} + R_k^{\text g}) \geq I(X_{\cal S}, X_{K + 1}; Z) - (R_{K+1}^{\text o} + R_{K+1}^{\text g}), ~\forall~ {\cal S} \subseteq {\cal K},~ {\cal S} \neq \phi.
\end{array} \right.
\end{equation}
Using the induction assumption to eliminate $R_k^{\text g}, \forall k \in {\cal K}$, we have
\begin{align}\label{R_g_ade}
& \sum_{k \in \cal S} R_k^{\text s} + \sum_{k \in {\cal S}\setminus {\cal S}'} R_k^{\text o} \nonumber\\
\leq & I(X_{\cal S}, X_{K + 1}; Y| X_{\overline {\cal S}}) - (R_{K+1}^{\text s} + R_{K+1}^{\text o} + R_{K+1}^{\text g}) - \left[ I(X_{{\cal S}'}, X_{K + 1}; Z) - (R_{K+1}^{\text o} + R_{K+1}^{\text g}) \right],\nonumber\\
= & I(X_{\cal S}, X_{K + 1}; Y| X_{\overline {\cal S}}) - I(X_{{\cal S}'}, X_{K + 1}; Z) - R_{K+1}^{\text s},~ \forall~ {\cal S} \subseteq {\cal K},~ {\cal S} \neq \phi,~ {\cal S}' \subseteq {\cal S},~ {\cal S}' \neq \phi.
\end{align}
By comparing (\ref{R_g_ade}) with (\ref{rate_region_K+1_2c}), it is known that (\ref{R_g_ade}) consists of partial inequalities in (\ref{rate_region_K+1_2c}) with ${\cal S} \subseteq {\cal K},~ {\cal S} \neq \phi,~ {\cal S}' \subseteq {\cal S},~ {\cal S}' \neq \phi$.

\subsubsection{Category~$4$}

We include inequalities (\ref{region_FM_K+1_2a}), (\ref{region_FM_K+1_2b}), and (\ref{region_FM_K+1_2e}) in Category~$4$, and rewrite them as follows
\begin{equation}\label{region_FM_K+1_2_abe}
\left\{
\begin{array}{ll}
R_k^{\text g} \geq 0, ~\forall~ k \in {\cal K}, \\
\sum\limits_{k \in {\cal S}} (R_k^{\text s} + R_k^{\text o} + R_k^{\text g}) \leq I(X_{\cal S}; Y| X_{\overline {\cal S}}, X_{K + 1}), \forall {\cal S} \subseteq {\cal K}, \\
\sum\limits_{k \in {\cal S}} (R_k^{\text o} + R_k^{\text g}) \geq I(X_{\cal S}, X_{K + 1}; Z) - (R_{K+1}^{\text o} + R_{K+1}^{\text g}), ~\forall~ {\cal S} \subseteq {\cal K},~ {\cal S} \neq \phi.
\end{array} \right.
\end{equation}
The following projection of (\ref{region_FM_K+1_2_abe}) can then be obtained from the induction assumption
\begin{align}\label{R_g_abe}
& \sum_{k \in \cal S} R_k^{\text s} + \sum_{k \in {\cal S}\setminus {\cal S}'} R_k^{\text o} \leq I(X_{\cal S}; Y| X_{\overline {\cal S}}, X_{K + 1}) - I(X_{{\cal S}'}, X_{K + 1}; Z) + R_{K+1}^{\text o} + R_{K+1}^{\text g},\nonumber\\
& \quad\quad\quad\quad\quad\quad\quad\quad\; \forall~ {\cal S} \subseteq {\cal K},~ {\cal S} \neq \phi,~ {\cal S}' \subseteq {\cal S},~ {\cal S}' \neq \phi,
\end{align}
which also contains $R_{K+1}^{\text g}$ and has to be considered in the next step of eliminating $R_{K+1}^{\text g}$.

Combining (\ref{region_FM_K+1_2f})~$\sim$ (\ref{region_FM_K+1_2h}), (\ref{R_g_abc}), (\ref{R_g_adc}), (\ref{R_g_ade}), and (\ref{R_g_abe}), we get a projection of (\ref{region_FM_K+1_2}) onto the hyperplane $\{ R_k^{\text g} = 0, \forall k \in {\cal K} \}$.
To further get a projection of (\ref{region_FM_K+1_2}) onto the hyperplane $\{ R_k^{\text g} = 0, \forall k \in {\cal K} \cup \{ K + 1 \}\}$, we have to eliminate $R_{K + 1}^{\text g}$.

\subsection{Elimination of $R_{K+1}^{\text g}$}

From (\ref{region_FM_K+1_2g}) and (\ref{R_g_adc}), we get the following upper bounds on $R_{K+1}^{\text g}$
\begin{subequations}\label{R_g_ub}
	\begin{align}
	R_{K+1}^{\text g} & \leq I(X_{K+1}; Y| X_{{\cal K}}) - (R_{K+1}^{\text s} + R_{K+1}^{\text o}), \label{R_g_ub1}\\
	R_{K+1}^{\text g} & \leq I(X_{\cal S}, X_{K + 1}; Y| X_{\overline {\cal S}}) - I(X_{{\cal S}'}; Z) - \Big( \sum_{k \in \cal S} R_k^{\text s} + \sum_{k \in {\cal S}\setminus {\cal S}'} R_k^{\text o} + R_{K+1}^{\text s} + R_{K+1}^{\text o} \Big), \nonumber\\
	& \; \forall~ {\cal S} \subseteq {\cal K},~ {\cal S} \neq \phi,~ {\cal S}' \subseteq \cal S.\label{R_g_ub2}
	\end{align}
\end{subequations}
Moreover, the following lower bounds on $R_{K+1}^{\text g}$ can be obtained from (\ref{region_FM_K+1_2f}), (\ref{region_FM_K+1_2h}), and (\ref{R_g_abe})
\begin{subequations}\label{R_g_lb}
	\begin{align}
	R_{K+1}^{\text g} & \geq 0, \label{R_g_lb1}\\
	R_{K+1}^{\text g} & \geq I(X_{K+1}; Z) - R_{K+1}^{\text o}, \label{R_g_lb2}\\
	R_{K+1}^{\text g} & \geq - I(X_{\cal S}; Y| X_{\overline {\cal S}}, X_{K + 1}) + I(X_{{\cal S}'}, X_{K + 1}; Z) + \sum_{k \in \cal S} R_k^{\text s} + \sum_{k \in {\cal S}\setminus {\cal S}'} R_k^{\text o} - R_{K+1}^{\text o}, \nonumber\\
	& \; \forall~ {\cal S} \subseteq {\cal K},~ {\cal S} \neq \phi,~ {\cal S}' \subseteq {\cal S},~ {\cal S}' \neq \phi.\label{R_g_lb3}
	\end{align}
\end{subequations}
Comparing these upper and lower bounds, we can eliminate $R_{K+1}^{\text g}$.

Firstly, we compare (\ref{R_g_ub}) with (\ref{R_g_lb1}), and get
\begin{subequations}\label{R_g_ub_lb1}
	\begin{align}
	& R_{K+1}^{\text s} + R_{K+1}^{\text o} \leq I(X_{K+1}; Y| X_{{\cal K}}), \label{R_g_ub1_lb1}\\
	& \sum_{k \in \cal S} R_k^{\text s} + \sum_{k \in {\cal S}\setminus {\cal S}'} R_k^{\text o} + R_{K+1}^{\text s} + R_{K+1}^{\text o} \leq I(X_{\cal S}, X_{K + 1}; Y| X_{\overline {\cal S}}) - I(X_{{\cal S}'}; Z), \nonumber\\
	& \quad\quad\quad\quad\quad\quad\quad\quad\quad\quad\quad\quad\quad\quad\quad\; \forall~ {\cal S} \subseteq {\cal K},~ {\cal S} \neq \phi,~ {\cal S}' \subseteq \cal S.\label{R_g_ub2_lb1}
	\end{align}
\end{subequations}
Obviously, inequalities (\ref{R_g_ub1_lb1}) and (\ref{R_g_ub2_lb1}) can be integrated into one formula as follows
\begin{align}\label{R_g_ub1_ub2_lb1}
\sum_{k \in \cal S} R_k^{\text s} \!+\! \sum_{k \in {\cal S}\setminus {\cal S}'} R_k^{\text o} \!+\! R_{K+1}^{\text s} \!+\! R_{K+1}^{\text o} \leq I(X_{\cal S}, X_{K + 1}; Y| X_{\overline {\cal S}}) \!-\! I(X_{{\cal S}'}; Z), \forall {\cal S} \subseteq {\cal K}, {{\cal S}'} \subseteq \cal S,
\end{align}
which is the same as (\ref{rate_region_K+1_2b}).

Secondly, we compare (\ref{R_g_ub}) with (\ref{R_g_lb2}), and get
\begin{subequations}\label{R_g_ub_lb2}
	\begin{align}
	& R_{K+1}^{\text s} \leq I(X_{K+1}; Y| X_{{\cal K}}) - I(X_{K+1}; Z), \label{R_g_ub1_lb2}\\
	& \sum_{k \in \cal S} R_k^{\text s} + \sum_{k \in {\cal S}\setminus {\cal S}'} R_k^{\text o} + R_{K+1}^{\text s} \leq I(X_{\cal S}, X_{K + 1}; Y| X_{\overline {\cal S}}) - I(X_{{\cal S}'}; Z) - I(X_{K+1}; Z), \nonumber\\
	& \quad\quad\quad\quad\quad\quad\quad\quad\quad\quad\quad\quad \forall~ {\cal S} \subseteq {\cal K},~ {\cal S} \neq \phi,~ {\cal S}' \subseteq \cal S.\label{R_g_ub2_lb2}
	\end{align}
\end{subequations}
By considering different ${\cal S}'$, i.e., ${\cal S}' = \phi$ and ${\cal S}' \neq \phi$, we may divide (\ref{R_g_ub2_lb2}) into two formulas as below
\begin{subequations}\label{R_g_ub_lb2_2}
	\begin{align}
	& \sum_{k \in \cal S} (R_k^{\text s} + R_k^{\text o}) + R_{K+1}^{\text s} \leq I(X_{\cal S}, X_{K + 1}; Y| X_{\overline {\cal S}}) - I(X_{K+1}; Z), \forall {\cal S} \subseteq {\cal K}, {\cal S} \neq \phi, {\cal S}' = \phi, \label{R_g_ub2_lb2_1}\\
	& \sum_{k \in \cal S} R_k^{\text s} + \sum_{k \in {\cal S}\setminus {\cal S}'} R_k^{\text o} + R_{K+1}^{\text s} \leq I(X_{\cal S}, X_{K + 1}; Y| X_{\overline {\cal S}}) - I(X_{{\cal S}'}; Z) - I(X_{K+1}; Z), \nonumber\\
	& \quad\quad\quad\quad\quad\quad\quad\quad\quad\quad\quad\quad \forall~ {\cal S} \subseteq {\cal K},~ {\cal S} \neq \phi,~ {\cal S}' \subseteq {\cal S},~ {\cal S}' \neq \phi.\label{R_g_ub2_lb2_2}
	\end{align}
\end{subequations}
From (\ref{R_g_ade}), (\ref{R_g_ub1_lb2}), and (\ref{R_g_ub2_lb2_1}), it can be found that these inequalities can be integrated into one formula as follows
\begin{align}\label{rate_region_K+1_2c2}
\sum_{k \in \cal S} R_k^{\text s} \!+\! \sum_{k \in {\cal S}\setminus {\cal S}'} R_k^{\text o} \!+\! R_{K + 1}^{\text s} \!\leq\! I(X_{\cal S}, X_{K + 1}; Y| X_{\overline {\cal S}}) \!-\! I(X_{{\cal S}'}, X_{K + 1}; Z), ~\forall~ {\cal S} \subseteq {\cal K}, {\cal S}' \subseteq \cal S,
\end{align}
which is the same as (\ref{rate_region_K+1_2c}).
Combining (\ref{R_g_abc}), (\ref{R_g_ub1_ub2_lb1}), and (\ref{rate_region_K+1_2c2}), it is known that (\ref{rate_region_K+1_2}) or (\ref{rate_region_K+1}) has already been obtained.
All the other inequalities resulted from the elimination procedure should be redundant if Lemma~\ref{theorem_FM} is true.
Hence, we have to prove that (\ref{R_g_ub2_lb2_2}) is redundant.
Since $X_{K + 1}$ is independent of $X_{{\cal S}'}$, (\ref{R_g_ade}) can be rewritten and relaxed as follows
\begin{align}\label{R_g_ade_relax}
& \sum_{k \in \cal S} R_k^{\text s} + \sum_{k \in {\cal S}\setminus {\cal S}'} R_k^{\text o} + R_{K + 1}^{\text s} \leq I(X_{\cal S}, X_{K + 1}; Y| X_{\overline {\cal S}}) - I(X_{{\cal S}'}, X_{K + 1}; Z)\nonumber\\
& \leq I(X_{\cal S}, X_{K + 1}; Y| X_{\overline {\cal S}}) - I(X_{{\cal S}'}; Z) - I(X_{K+1}; Z), ~\forall~ {\cal S} \subseteq {\cal K},~ {\cal S} \neq \phi,~ {\cal S}' \subseteq {\cal S},~ {\cal S}' \neq \phi,
\end{align}
where the second inequality is the upper bound in (\ref{R_g_ub2_lb2_2}).
Since (\ref{R_g_ade}) has been included in (\ref{rate_region_K+1_2c2}), (\ref{R_g_ub2_lb2_2}) is thus redundant.

Finally, we compare (\ref{R_g_ub}) with (\ref{R_g_lb3}), which results in
\begin{subequations}\label{R_g_ub_lb3}
	\begin{align}
	& \sum_{k \in \cal S} R_k^{\text s} + \sum_{k \in {\cal S}\setminus {\cal S}'} R_k^{\text o} + R_{K+1}^{\text s} \leq I(X_{\cal S}; Y| X_{\overline {\cal S}}, X_{K + 1}) + I(X_{K+1}; Y| X_{{\cal K}}) - I(X_{{\cal S}'}, X_{K+1}; Z), \nonumber\\ 
	& \quad\quad\quad\quad\quad\quad\quad\quad\quad\quad\quad\quad \forall~ {\cal S} \subseteq {\cal K},~ {\cal S} \neq \phi,~ {\cal S}' \subseteq {\cal S},~ {\cal S}' \neq \phi,\vspace{0.5em} \label{R_g_ub1_lb3}\\
	& \sum_{k \in \cal S} R_k^{\text s} \!+\! \sum_{k \in {\cal S}\setminus {\cal S}'}\! R_k^{\text o} \!+\! \sum_{k \in {\cal S}_1} R_k^{\text s} \!+\! \sum_{k \in {\cal S}_1\setminus {\cal S}_1'}\! R_k^{\text o} \!+\! R_{K+1}^{\text s} \leq I(X_{\cal S}; Y| X_{\overline {\cal S}}, X_{K + 1}) - I(X_{{\cal S}'}, X_{K + 1}; Z) \nonumber\\
	& + I(X_{{\cal S}_1}, X_{K + 1}; Y| X_{\overline {{\cal S}_1}}) - I(X_{{\cal S}_1'}; Z), \forall {\cal S}, {\cal S}_1 \subseteq {\cal K}, {\cal S}, {\cal S}_1 \neq \phi, {\cal S}' \subseteq {\cal S}, {\cal S}' \neq \phi, {\cal S}_1' \subseteq {\cal S}_1.\label{R_g_ub2_lb3}
	\end{align}
\end{subequations}
Note that when comparing (\ref{R_g_ub2}) with (\ref{R_g_lb3}), which gives (\ref{R_g_ub2_lb3}), we replace notations ${\cal S}$ and ${\cal S}'$ in (\ref{R_g_ub2}) with ${\cal S}_1$ and ${\cal S}_1'$, respectively, to avoid ambiguity.
As stated after after (\ref{rate_region_K+1_2c2}), (\ref{R_g_ub1_lb3}) and (\ref{R_g_ub2_lb3}) should be redundant if Lemma~\ref{theorem_FM} is true.
We prove the redundancy in the following.

We first prove that (\ref{R_g_ub1_lb3}) is redundant.
Since $X_k, \forall k \in {\cal K} \cup \{ K + 1 \}$ are independent of each other and ${\overline {\cal S}} \subseteq {\cal K}$, we have
\begin{align}\label{R_g_ade_relax2}
& I(X_{\cal S}, X_{K + 1}; Y| X_{\overline {\cal S}}) \!-\! I(X_{{\cal S}'}, X_{K + 1}; Z) \!=\! I(X_{\cal S}; Y| X_{\overline {\cal S}}) \!+\! I(X_{K+1}; Y| X_{{\cal K}}) \!-\! I(X_{{\cal S}'}, X_{K+1}; Z) \nonumber\\
& \leq\! I(X_{\cal S}; Y| X_{\overline {\cal S}}, X_{K \!+\! 1}) \!+\! I(X_{K \!+\! 1}; Y| X_{{\cal K}}) \!-\! I(X_{{\cal S}'}, X_{K \!+\! 1}; Z), \forall {\cal S} \!\subseteq\! {\cal K}, {\cal S} \!\neq\! \phi, {{\cal S}'} \!\subseteq\! {\cal S}, {\cal S}' \!\neq\! \phi.
\end{align}
By replacing the corresponding terms in (\ref{R_g_ade_relax}) with (\ref{R_g_ade_relax2}), the redundancy of (\ref{R_g_ub1_lb3}) can be similarly proven as that of (\ref{R_g_ub2_lb2_2}).
Since the redundancy proof of (\ref{R_g_ub2_lb3}) is not as easy as that of (\ref{R_g_ub2_lb2_2}) or (\ref{R_g_ub1_lb3}), for the sake of clarity, we give the proof in the following separate subsection.

\subsection{Redundancy Proof of (\ref{R_g_ub2_lb3})}

To prove that (\ref{R_g_ub2_lb3}) is redundant, we first rewrite its left-hand side term as follows
\begin{equation}\label{term_ab}
\underbrace { \sum_{k \in \cal S} R_k^{\text s} + \sum_{k \in {\cal S}\setminus {\cal S}'} R_k^{\text o} }_{{\text {Term}}~a} + \underbrace { \sum_{k \in {\cal S}_1} R_k^{\text s} + \sum_{k \in {\cal S}_1\setminus {\cal S}_1'} R_k^{\text o} + R_{K+1}^{\text s} }_{{\text {Term}}~b},
\end{equation}
which is divided into two parts, i.e., Term~$a$ and Term~$b$.
In the following, we exchange the secret message rates $R_k^{\text s}$ and open message rates $R_k^{\text o}$ in Term~$a$ and Term~$b$, and obtain Term~$a'$ as well as Term~$b'$.
When performing this operation, the following criteria have to be met.
\begin{itemize}
	\item Criterion~$1$. Term~$b'$ contains as many secret message rates $R_k^{\text s}$ as possible;
	\item Criterion~$2$. Term~$a'$ contains as many open message rates $R_k^{\text o}$ as possible;
	\item Criterion~$3$. Criterion~$1$ has a higher priority than Criterion~$2$;
	\item Criterion~$4$. For any $k$, there could be only one $R_k^{\text s}$ or $R_k^{\text o}$ in Term~$a'$ or Term~$b'$;
	\item Criterion~$5$. For any $k$, if $R_k^{\text s}$ is not in Term~$a'$ or Term~$b'$, $R_k^{\text o}$ cannot be in this term.
\end{itemize}

\begin{figure}
	\begin{minipage}[t]{0.49\linewidth}
		\centering
		\includegraphics[width=3in]{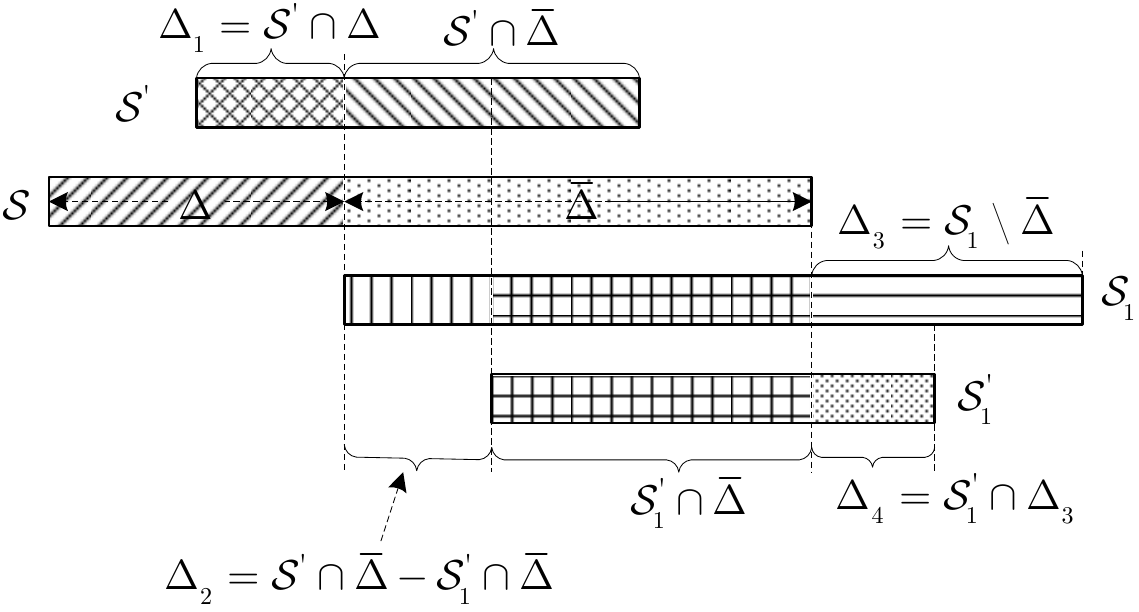}
		\caption{Sets ${\cal S}$, ${\cal S}'$, ${\cal S}_1$, ${\cal S}_1'$, and their divisions.}
		\label{Sets_S}
	\end{minipage}
	\hskip 1ex
	\begin{minipage}[t]{0.49\linewidth}
		\centering
		\includegraphics[width=3in]{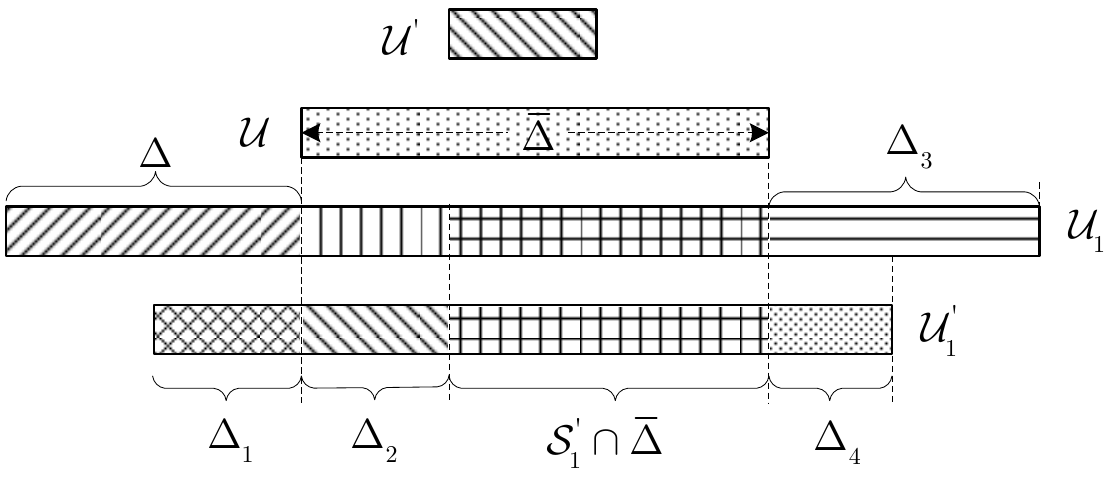}
		\caption{Sets ${\cal U}$, ${\cal U}'$, ${\cal U}_1$, ${\cal U}_1'$ obtained from ${\cal S}$, ${\cal S}'$, ${\cal S}_1$, ${\cal S}_1'$ based on Criteria $1 \sim 5$.}
		\label{Sets_U}
	\end{minipage}
\end{figure}

For ease of understanding, we give an example system with ${\cal K} = \{ 1, 2, 3, 4\}$, ${\cal S} = \{ 1, 3, 4\}$, ${\cal S}' = \{ 4\}$, ${\cal S}_1 = \{ 1, 2, 4\}$, and ${\cal S}_1' = \phi$.
In this case, (\ref{term_ab}) takes on form
\begin{equation}\label{term_ab_eg}
\underbrace { R_1^{\text s} + R_1^{\text o} + R_3^{\text s} + R_3^{\text o} + R_4^{\text s} }_{{\text {Term}}~a} + \underbrace { R_1^{\text s} + R_1^{\text o} + R_2^{\text s} + R_2^{\text o} + R_4^{\text s} + R_4^{\text o} + R_5^{\text s} }_{{\text {Term}}~b}.
\end{equation}
Since $R_3^{\text s}$ is not included in Term~$b$, according to Criterion~$1$ and Criterion~$3$, $R_3^{\text s}$ should be moved into Term~$b$.
Note that $R_3^{\text o}$ has also to be moved into Term~$b$ since otherwise Criterion~$5$ is violated.
Since both Term~$a$ and Term~$b$ contain $R_4^{\text s}$, while only Term~$b$ contains $R_4^{\text o}$, according to Criterion~$2$, $R_4^{\text o}$ should be moved into Term~$a$.
Note that we may not move the $R_1^{\text o}$ in Term~$b$ into Term~$a$ since otherwise Criterion~$4$ is violated, and may not move $R_2^{\text o}$ into Term~$a$ since otherwise Criterion~$5$ is violated.
With these operations, (\ref{term_ab_eg}) becomes
\begin{equation}\label{term_ab_eg2}
\underbrace { R_1^{\text s} + R_1^{\text o} + R_4^{\text s} + R_4^{\text o} }_{{\text {Term}}~a'} + \underbrace { R_1^{\text s} + R_1^{\text o} + R_2^{\text s} + R_2^{\text o} + R_3^{\text s} + R_3^{\text o} + R_4^{\text s} + R_5^{\text s} }_{{\text {Term}}~b'}.
\end{equation}

In the following, we describe these operations mathematically and prove the redundancy of (\ref{R_g_ub2_lb3}).
For convenience, we give an example of sets ${\cal S}$, ${\cal S}'$, ${\cal S}_1$, ${\cal S}_1'$, and their divisions in Fig.~\ref{Sets_S}.
Let ${\overline \Delta}$ denote the set of user indexes which are in both $\cal S$ and ${\cal S}_1$, and $\Delta$ denote the set of indexes which are in $\cal S$ but not in ${\cal S}_1$, i.e.,
\begin{align}\label{Delta}
{\overline \Delta} & = {\cal S} \cap {\cal S}_1, \nonumber\\
\Delta & = {\cal S} - {\cal S}_1.
\end{align}
Let $\Delta_1$ denote the intersection of ${\cal S}'$ and $\Delta$, and $\Delta_2$ denote the set of user indexes which are in ${\cal S}' \cap {\overline \Delta}$ but not in ${\cal S}_1' \cap {\overline \Delta}$, i.e.,
\begin{align}\label{Delta12}
\Delta_1 & = {\cal S}' \cap \Delta, \nonumber\\
\Delta_2 & = {\cal S}' \cap {\overline \Delta} - {\cal S}_1' \cap {\overline \Delta}. 
\end{align}
With the operations described above, let ${\cal U}$, ${\cal U}'$, ${\cal U}_1$, and ${\cal U}_1'$, which respectively correspond to ${\cal S}$, ${\cal S}'$, ${\cal S}_1$, and ${\cal S}_1'$ in (\ref{term_ab}), denote the user indexes in Term~$a'$ and Term~$b'$.
Then, according to Criterion~$1$,
\begin{align}\label{U_U1}
{\cal U} & = {\cal S} \setminus \Delta, \nonumber\\
{\cal U}_1 & = {\cal S}_1 \cup \Delta. 
\end{align}
The complementary sets of ${\cal U}$ and ${\cal U}_1$ are 
\begin{align} \label{U_U1_complement}
{\overline {\cal U}} & = {\cal K} \setminus ({\cal S} \setminus \Delta) = ({\cal K} \setminus {\cal S}) \cup \Delta = {\overline {\cal S}} \cup \Delta, \nonumber\\
{\overline {{\cal U}_1}} & = {\cal K} \setminus ({\cal S}_1 \cup \Delta) = ({\cal K} \setminus {\cal S}_1) \setminus \Delta = {\overline {{\cal S}_1}} \setminus \Delta.
\end{align}
Moreover, according to Criterion~$2$, Criterion~$4$, and Criterion~$5$, we have
\begin{align}\label{U_U1_prime}
{\cal U}' & = {\cal S}' \setminus (\Delta_1 \cup \Delta_2), \nonumber\\
{\cal U}_1' & = {\cal S}_1' \cup (\Delta_1 \cup \Delta_2).
\end{align}
Fig.~\ref{Sets_U} depicts the sets ${\cal U}$, ${\cal U}'$, ${\cal U}_1$, ${\cal U}_1'$ obtained from ${\cal S}$, ${\cal S}'$, ${\cal S}_1$, ${\cal S}_1'$ in Fig.~\ref{Sets_S}.
We thus have 
\begin{align}\label{R_g_ub2_lb3_oper}
& \underbrace { \sum_{k \in \cal S} R_k^{\text s} + \sum_{k \in {\cal S}\setminus {\cal S}'} R_k^{\text o} }_{{\text {Term}}~a} + \underbrace { \sum_{k \in {\cal S}_1} R_k^{\text s} + \sum_{k \in {\cal S}_1\setminus {\cal S}_1'} R_k^{\text o} + R_{K+1}^{\text s} }_{{\text {Term}}~b} \nonumber\\
= & \underbrace { \sum_{k \in \cal U} R_k^{\text s} + \sum_{k \in {\cal U}\setminus {\cal U}'} R_k^{\text o} }_{{\text {Term}}~a'} + \underbrace { \sum_{k \in {\cal U}_1} R_k^{\text s} + \sum_{k \in {\cal U}_1\setminus {\cal U}_1'} R_k^{\text o} + R_{K+1}^{\text s} }_{{\text {Term}}~b'} \nonumber\\
\leq & I(X_{\cal U}; Y| X_{\overline {\cal U}}, X_{K + 1}) - I(X_{{\cal U}'}; Z) + I(X_{{\cal U}_1}, X_{K + 1}; Y| X_{\overline {{\cal U}_1}}) - I(X_{{\cal U}_1'}, X_{K + 1}; Z) \nonumber\\
= & I(X_{\cal S} \setminus X_\Delta; Y| X_{\overline {\cal S}}, X_\Delta, X_{K + 1}) - I(X_{{\cal S}'} \setminus X_{\Delta_1 \cup \Delta_2}; Z) \nonumber\\
+ & I(X_{{\cal S}_1}, X_\Delta, X_{K + 1}; Y| X_{\overline {{\cal S}_1}} \setminus X_\Delta) - I(X_{{\cal S}_1'}, X_{\Delta_1}, X_{\Delta_2}, X_{K + 1}; Z),\nonumber\\
& \forall~ {\cal S},~ {\cal S}_1 \subseteq {\cal K},~ {\cal S},~ {\cal S}_1 \neq \phi,~ {\cal S}' \subseteq {\cal S},~ {\cal S}' \neq \phi,~ {\cal S}_1' \subseteq {\cal S}_1,
\end{align}
where the inequality results from (\ref{R_g_abc}) and (\ref{rate_region_K+1_2c2}).
On the other hand, based on the definitions of $\Delta$, $\Delta_1$, and $\Delta_2$, and the chain rule of mutual information, (\ref{R_g_ub2_lb3}) can be rewritten as follows
\begin{align}\label{R_g_ub2_lb3_re}
& \sum_{k \in \cal S} R_k^{\text s} + \sum_{k \in {\cal S}\setminus {\cal S}'} R_k^{\text o} + \sum_{k \in {\cal S}_1} R_k^{\text s} + \sum_{k \in {\cal S}_1\setminus {\cal S}_1'} R_k^{\text o} + R_{K+1}^{\text s} \nonumber\\
\leq & I(X_{\cal S}; Y| X_{\overline {\cal S}}, X_{K + 1}) - I(X_{{\cal S}'}, X_{K + 1}; Z) + I(X_{{\cal S}_1}, X_{K + 1}; Y| X_{\overline {{\cal S}_1}}) - I(X_{{\cal S}_1'}; Z) \nonumber\\
= & I(X_{\cal S} \setminus X_\Delta, X_\Delta; Y| X_{\overline {\cal S}}, X_{K + 1}) - I(X_{{\cal S}'} \setminus X_{\Delta_1 \cup \Delta_2}, X_{\Delta_1 \cup \Delta_2}, X_{K + 1}; Z) \nonumber\\
+ & I(X_{{\cal S}_1}, X_{K + 1}; Y| X_{\overline {{\cal S}_1}}) - I(X_{{\cal S}_1'}; Z) \nonumber\\
= & I(X_{\cal S} \setminus X_\Delta; Y| X_{\overline {\cal S}}, X_\Delta, X_{K + 1}) + I(X_\Delta; Y| X_{\overline {\cal S}}, X_{K + 1}) - I(X_{{\cal S}'} \setminus X_{\Delta_1 \cup \Delta_2}; Z)\nonumber\\
- & I( X_{\Delta_1}, X_{\Delta_2}, X_{K + 1}; Z| X_{{\cal S}'} \setminus X_{\Delta_1 \cup \Delta_2}) + I(X_{{\cal S}_1}, X_{K + 1}; Y| X_{\overline {{\cal S}_1}}) - I(X_{{\cal S}_1'}; Z), \nonumber\\
& \forall~ {\cal S},~ {\cal S}_1 \subseteq {\cal K},~ {\cal S},~ {\cal S}_1 \neq \phi,~ {\cal S}' \subseteq {\cal S},~ {\cal S}' \neq \phi,~ {\cal S}_1' \subseteq {\cal S}_1.
\end{align}
In the following, we show that the upper bound in (\ref{R_g_ub2_lb3_oper}) is no larger and is thus tighter than that in (\ref{R_g_ub2_lb3_re}).
Then, (\ref{R_g_ub2_lb3_re}) is redundant.
Neglecting the common terms $I(X_{\cal S} \setminus X_\Delta; Y| X_{\overline {\cal S}}, X_\Delta, X_{K + 1})$ and $I(X_{{\cal S}'} \setminus X_{\Delta_1 \cup \Delta_2}; Z)$ in (\ref{R_g_ub2_lb3_oper}) and (\ref{R_g_ub2_lb3_re}), we prove
\begin{align}\label{relation1}
I(X_{{\cal S}_1}, X_\Delta, X_{K + 1}; Y| X_{\overline {{\cal S}_1}} \setminus X_\Delta) \leq I(X_\Delta; Y| X_{\overline {\cal S}}, X_{K + 1}) + I(X_{{\cal S}_1}, X_{K + 1}; Y| X_{\overline {{\cal S}_1}}),
\end{align}
and
\begin{align}\label{relation2}
I(X_{{\cal S}_1'}, X_{\Delta_1}, X_{\Delta_2}, X_{K + 1}; Z) \geq I( X_{\Delta_1}, X_{\Delta_2}, X_{K + 1}; Z| X_{{\cal S}'} \setminus X_{\Delta_1 \cup \Delta_2}) + I(X_{{\cal S}_1'}; Z).
\end{align}

From the definitions of $\overline \Delta$ and $\Delta$ in (\ref{Delta}), it is known that $\Delta \cap {\overline \Delta} = \phi$ and ${\cal S} = \Delta \cup {\overline \Delta}$.
Hence,
\begin{align}\label{S_comple}
{\overline {\cal S}} = {\cal K} \setminus {\cal S} = ({\cal K} \setminus {\overline \Delta}) \setminus \Delta.
\end{align}
Moreover, since ${\overline \Delta} \subseteq {\cal S}_1$,
\begin{align}\label{S_prime_comple}
{\overline {{\cal S}_1}} = {\cal K} \setminus {\cal S}_1 \subseteq {\cal K} \setminus {\overline \Delta}.
\end{align}
Based on (\ref{S_comple}) and (\ref{S_prime_comple}), we have
\begin{align}\label{S_prime_S_comple}
{\overline {{\cal S}_1}} \setminus \Delta \subseteq ({\cal K} \setminus {\overline \Delta}) \setminus \Delta = {\overline {\cal S}}.
\end{align}
Using the chain rule of mutual information, (\ref{S_prime_S_comple}), and the fact that $X_k, \forall k \in {\cal K} \cup \{ K + 1 \}$ are independent of each other, we have
\begin{align}\label{relation1_2}
I(X_{{\cal S}_1}, X_\Delta, X_{K + 1}; Y| X_{\overline {{\cal S}_1}} \setminus X_\Delta) & = I(X_\Delta; Y| X_{\overline {{\cal S}_1}} \setminus X_\Delta) + I(X_{{\cal S}_1}, X_{K + 1}; Y| X_{\overline {{\cal S}_1}}) \nonumber\\
& \leq I(X_\Delta; Y| X_{\overline {\cal S}}, X_{K + 1}) + I(X_{{\cal S}_1}, X_{K + 1}; Y| X_{\overline {{\cal S}_1}}).
\end{align}
The inequation (\ref{relation1}) is thus true.
On the other hand, since ${\cal S} = \Delta \cup {\overline \Delta}$, $\Delta \cap {\overline \Delta} = \phi$, and ${\cal S}' \subseteq {\cal S}$, as shown in Fig.~\ref{Sets_S}, ${\cal S}'$ can be divided into two disjoint parts as follows
\begin{align}\label{S_1_disjoint}
{\cal S}' = ({\cal S}' \cap \Delta) \cup ({\cal S}' \cap {\overline \Delta})  = \Delta_1 \cup ({\cal S}' \cap {\overline \Delta}),
\end{align}
where we used the definition of $\Delta_1$ in (\ref{Delta12}).
Hence,
\begin{equation}\label{S_1_Delta}
{\cal S}' \cap {\overline \Delta} = {\cal S}' \setminus \Delta_1.
\end{equation}
Let $\Delta_3$ denote the set of user indexes which are in ${\cal S}_1$ but not in $\cal S$, and $\Delta_4$ denote the intersection of ${\cal S}_1'$ and $\Delta_3$, i.e., 
\begin{align}\label{Delta34}
\Delta_3 & = {\cal S}_1 - {\cal S} = {\cal S}_1 \setminus {\overline \Delta}, \nonumber\\
\Delta_4 & = {\cal S}_1' \cap \Delta_3. 
\end{align}
It can then be similarly proven as (\ref{S_1_Delta}) that
\begin{equation}\label{S_1_Delta_prime}
{\cal S}_1' \cap {\overline \Delta} = {\cal S}_1' \setminus \Delta_4,
\end{equation}
which can also be found from Fig.~\ref{Sets_S}.
From (\ref{S_1_Delta}), (\ref{S_1_Delta_prime}), and the definition of $\Delta_2$ in (\ref{Delta12}), ${\cal S}'$ in (\ref{S_1_disjoint}) can be further divided into three disjoint parts as follows
\begin{align}\label{S_1_disjoint_2}
{\cal S}' & = \Delta_1 \cup ({\cal S}' \cap {\overline \Delta}) 
~ = \Delta_1 \cup \Delta_2 \cup \left[ ({\cal S}' \cap {\overline \Delta}) \cap ({\cal S}_1' \cap {\overline \Delta}) \right] \nonumber\\
& = \Delta_1 \cup \Delta_2 \cup \left[ ({\cal S}' \setminus \Delta_1) \cap ({\cal S}_1' \setminus \Delta_4) \right] 
~ = \Delta_1 \cup \Delta_2 \cup ({\cal S}' \cap {\cal S}_1'), 
\end{align}
where the last step holds since $\Delta_1 \cap \Delta_4 = \phi$.
Accordingly, we have
\begin{align}\label{S1_S1_prime}
{\cal S}' \setminus (\Delta_1 \cup \Delta_2) = {\cal S}' \cap {\cal S}_1' \subseteq {\cal S}_1'.
\end{align}
Then, using the chain rule of mutual information, (\ref{S1_S1_prime}), and the fact that $X_k, \forall k \in {\cal K} \cup \{ K + 1 \}$ are independent of each other, we have
\begin{align}\label{relation2_2}
I(X_{{\cal S}_1'}, X_{\Delta_1}, X_{\Delta_2}, X_{K + 1}; Z) & = I( X_{\Delta_1}, X_{\Delta_2}, X_{K + 1}; Z| X_{{\cal S}_1'}) + I(X_{{\cal S}_1'}; Z)\nonumber\\
& \geq I( X_{\Delta_1}, X_{\Delta_2}, X_{K + 1}; Z| X_{{\cal S}'} \setminus X_{\Delta_1 \cup \Delta_2}) + I(X_{{\cal S}_1'}; Z),
\end{align}
i.e., (\ref{relation2}) is true.
Combining (\ref{R_g_ub2_lb3_oper}), (\ref{R_g_ub2_lb3_re}), (\ref{relation1_2}), and (\ref{relation2_2}), it is known that (\ref{R_g_ub2_lb3}) is redundant.

So far we have shown that (\ref{rate_region_K+1}) (or (\ref{rate_region_K+1_2})) can be obtained by eliminating $R_k^{\text g}, \forall k \in {\cal K} \cup \{ K + 1 \}$ in (\ref{region_FM_K+1}) (or (\ref{region_FM_K+1_2})), and all the other inequalities resulted from the elimination procedure, i.e., (\ref{R_g_ub2_lb2_2}), (\ref{R_g_ub1_lb3}), and (\ref{R_g_ub2_lb3}), are redundant.
As a result, (\ref{rate_region_K+1}) is the projection of (\ref{region_FM_K+1}) onto the hyperplane $\{ R_k^{\text g} = 0, \forall k \in {\cal K} \cup \{ K + 1 \}\}$.
Lemma~\ref{theorem_FM} is thus proven.

\section{Proof of Theorem~\ref{lemma_FM_gene_K2}}
\label{Prove_lemma_FM_gene_K2}

Since ${\overline {{\cal K}'}}$ has $2^{\left| {\overline {{\cal K}'}} \right| }$ subsets, we may divide the inequality system (\ref{region_FM2}) into $2^{\left| {\overline {{\cal K}'}} \right|}$ subsystems with each one corresponding to a subset ${\cal T} \subseteq {\overline {{\cal K}'}}$.
For any ${\cal T} \subseteq {\overline {{\cal K}'}}$, the inequality subsystem is 
\begin{equation}\label{region_FM2_sub}
\left\{
\begin{array}{ll}
R_k^{\text g} \geq 0, ~\forall~ k \in {\cal K}', \\
\sum\limits_{k \in {\cal S}} (R_k^{\text s} + R_k^{\text o} + R_k^{\text g}) + \sum\limits_{k \in {\cal T}} R_k^{\text o} \leq I(X_{\cal S}, X_{\cal T}; Y| X_{\overline {\cal S}}, X_{\overline {\cal T}}), ~\forall~ {\cal S} \subseteq {\cal K}', \\
\sum\limits_{k \in {\cal S}} (R_k^{\text o} + R_k^{\text g}) \geq I(X_{\cal S}; Z| X_{\overline {{\cal K}'}}), ~\forall~ {\cal S} \subseteq {\cal K}'.
\end{array} \right.
\end{equation}
It is obvious that eliminating $R_k^{\text g}, \forall k \in {\cal K}'$ in (\ref{region_FM2}) is equivalent to eliminating $R_k^{\text g}, \forall k \in {\cal K}'$ in (\ref{region_FM2_sub}) for all ${\cal T} \subseteq {\overline {{\cal K}'}}$.
Due to the assumption $I(X_{\cal S}; Y| X_{\overline {\cal S}}, X_{\overline {{\cal K}'}}) \geq I(X_{\cal S}; Z| X_{\overline {{\cal K}'}}), \forall {\cal S} \subseteq {\cal K}'$ made in Theorem~\ref{lemma_FM_gene_K2}, for a given ${\cal T} \subseteq {\overline {{\cal K}'}}$, we have
\begin{align}
I(X_{\cal S}, X_{\cal T}; Y| X_{\overline {\cal S}}, X_{\overline {\cal T}})  \geq I(X_{\cal S}; Y| X_{\overline {\cal S}}, X_{\overline {{\cal K}'}}) 
~ \geq I(X_{\cal S}; Z| X_{\overline {{\cal K}'}}), ~\forall~ {\cal S} \subseteq {\cal K}'.
\end{align}
Then, by replacing $I(X_{\cal S}; Y| X_{\overline {\cal S}})$ in (\ref{region_FM1}) with $I(X_{\cal S}, X_{\cal T}; Y| X_{\overline {\cal S}}, X_{\overline {\cal T}}) - \sum_{k \in {\cal T}} R_k^{\text o}$ and $I(X_{\cal S}; Z)$ with $I(X_{\cal S}; Z| X_{\overline {{\cal K}'}})$, we can eliminate $R_k^{\text g}, \forall k \in {\cal K}'$ in (\ref{region_FM2_sub}) based on Lemma~\ref{theorem_FM} and get
\begin{align}\label{rate_region0_sub}
\sum_{k \in \cal S} R_k^{\text s} \!+\! \sum_{k \in {\cal S} \setminus {\cal S}'} R_k^{\text o} \!\leq\! I(X_{\cal S}, X_{\cal T}; Y| X_{\overline {\cal S}}, X_{\overline {\cal T}}) \!-\! \sum_{k \in {\cal T}} R_k^{\text o} \!-\! I(X_{{\cal S}'}; Z| X_{\overline {{\cal K}'}}), \forall {\cal S} \subseteq {\cal K}', {\cal S}' \subseteq {\cal S}.
\end{align}
Combining the inequalities (\ref{rate_region0_sub}) for all ${\cal T} \subseteq {\overline {{\cal K}'}}$, (\ref{region_DM0}) can be obtained, and Theorem~\ref{lemma_FM_gene_K2} is thus proven.

\section{Proof of Theorem~\ref{lemma_DM_exten}}
\label{prove_lemma_DM_exten}

We start the proof from a special case with ${\cal K}' = \phi$, i.e., ${\overline {{\cal K}'}} = {\cal K}$.
In this case, (\ref{region_DM_exten}) becomes
\begin{equation}
\left\{
\begin{array}{ll}
R_k^{\text s} = 0, ~\forall~ k \in {\cal K}, \\
\sum_{k \in {\cal T}} R_k^{\text o} \leq I (X_{\cal T}; Y| X_{\overline {\cal T}}), ~\forall~ {\cal T} \subseteq {\cal K},
\end{array} \right.
\end{equation}
indicating that the region ${\mathscr R} (X_{\cal K}, \phi)$ is included in the capacity region of a conventional MAC channel with no wiretapping, the achievability proof of which is well known.

Next, we show that for any ${\cal K}' \neq \phi$, i.e., ${\overline {{\cal K}'}} \subsetneqq {\cal K}$, there exists a $\left( 2^{n R_1^{\text s}}, 2^{n R_1^{\text o}}, \cdots, 2^{n R_K^{\text s}}, 2^{n R_K^{\text o}}, n \right)$ code such that any rate tuple inside region ${\mathscr R} (X_{\cal K}, {\cal K}')$ is achievable.
This, together with the standard time-sharing over coding strategies, suffices to prove the theorem. 
Without loss of generality (w.l.o.g.), we always assume
\begin{equation}\label{assumption}
I(X_{\cal S}; Y| X_{\overline {\cal S}}) > 0, ~\forall~ {\cal S} \subseteq {\cal K},~ {\cal S} \neq \phi,
\end{equation}
since otherwise users in ${\cal S}$ cannot communicate with Bob.
Moreover, for convenience, we assume
\begin{align}\label{assumption_1}
I(X_{\cal S}, X_{\cal T}; Y| X_{\overline {\cal S}}, X_{\overline {\cal T}}) - I(X_{{\cal S}'}; Z| X_{\overline {{\cal K}'}}) > 0, ~\forall~ {\cal S} \subseteq {\cal K}',~ {\cal S}' \subseteq {\cal S},~ {\cal T} \subseteq {\overline {{\cal K}'}},~ {\cal S} \neq \phi.
\end{align}
If (\ref{assumption_1}) is not satisfied, we show later that Theorem~\ref{lemma_DM_exten} could be proven by simply modifying the following proof steps.
Since $X_k, \forall k \in {\cal K}$ are independent of each other, using the chain rule and non-negativity of mutual information, it is known that
\begin{align}\label{two_ineq}
& I(X_{\cal S}; Y| X_{\overline {\cal S}}, X_{\overline {{\cal K}'}}) \leq I(X_{\cal S}, X_{\cal T}; Y| X_{\overline {\cal S}}, X_{\overline {\cal T}}), \nonumber\\
& I(X_{\cal S}; Z| X_{\overline {{\cal K}'}}) \geq I(X_{{\cal S}'}; Z| X_{\overline {{\cal K}'}}), ~\forall~ {\cal S} \subseteq {\cal K}',~ {\cal S}' \subseteq {\cal S},~ {\cal T} \subseteq {\overline {{\cal K}'}}.
\end{align}
Hence, (\ref{assumption_1}) can be simplified as
\begin{equation}\label{assumption_0}
I(X_{\cal S}; Y| X_{\overline {\cal S}}, X_{\overline {{\cal K}'}}) - I(X_{\cal S}; Z| X_{\overline {{\cal K}'}}) > 0, ~\forall~ {\cal S} \subseteq {\cal K}',~ {\cal S} \neq \phi.
\end{equation}
With (\ref{assumption_0}), Theorem~\ref{lemma_FM_gene_K2} can be applied for the achievability proof in the following.

\subsection{Proof of Theorem~\ref{lemma_DM_exten} When Assumption (\ref{assumption_1}) is True}
\label{assump_true}

If assumption (\ref{assumption_1}) or (\ref{assumption_0}) is true, the rate tuples inside region ${\mathscr R} (X_{\cal K}, {\cal K}')$ satisfy
\begin{equation}\label{rate_region1}
\left\{
\begin{array}{ll}
R_k^{\text s} = 0, ~\forall~ k \in {\overline {{\cal K}'}}, \\
\sum\limits_{k \in \cal S} R_k^{\text s} + \sum\limits_{k \in {\cal S} \setminus {\cal S}'} R_k^{\text o} + \sum\limits_{k \in {\cal T}} R_k^{\text o} & < I(X_{\cal S}, X_{\cal T}; Y| X_{\overline {\cal S}}, X_{\overline {\cal T}}) - I(X_{{\cal S}'}; Z| X_{\overline {{\cal K}'}}) - \epsilon,\\
& \forall~ {\cal S} \subseteq {\cal K}',~ {\cal S}' \subseteq {\cal S},~ {\cal T} \subseteq {\overline {{\cal K}'}},~ {\cal S} \cup {\cal T} \neq \phi,
\end{array} \right.
\end{equation}
where $\epsilon$ is an arbitrarily small positive number. 
Using Theorem~\ref{lemma_FM_gene_K2}, it is known that for any rate tuple $\left(R_1^{\text s}, R_1^{\text o},\cdots, R_K^{\text s}, R_K^{\text o}\right)$ satisfying (\ref{rate_region1}), there exist $R_k^{\text g}, \forall k \in {\cal K}'$ such that
\begin{equation}\label{region_FM3}
\left\{\!\!
\begin{array}{ll}
R_k^{\text g} \geq 0, ~\forall~ k \in {\cal K}', \\
\sum\limits_{k \in {\cal S}} (R_k^{\text s} \!+\! R_k^{\text o} \!+\! R_k^{\text g}) \!+\! \sum\limits_{k \in {\cal T}} R_k^{\text o} \!<\! I(X_{\cal S}, X_{\cal T}; Y| X_{\overline {\cal S}}, X_{\overline {\cal T}}) \!-\! \epsilon, \forall {\cal S} \!\subseteq\! {\cal K}', {\cal T} \!\subseteq\! {\overline {{\cal K}'}}, {\cal S} \cup {\cal T} \!\neq\! \phi, \\
\sum\limits_{k \in {\cal S}} (R_k^{\text o} + R_k^{\text g}) > I(X_{\cal S}; Z| X_{\overline {{\cal K}'}}), ~\forall~ {\cal S} \subseteq {\cal K}',~ {\cal S} \neq \phi,
\end{array} \right.
\end{equation}
and $R_k^{\text g}, \forall k \in {\cal K}'$ can be found by applying Dantzig's simplex algorithm \cite{gass2003linear}.

In (\ref{region_FM3}), we introduce a `garbage' message to each user in ${\cal K}'$ to interfere with the decoding of Eve, i.e., besides the secret and open messages, each user $k \in {\cal K}'$ also has to transmit a `garbage' message at rate $R_k^{\text g}$ though it is not necessary for Bob.
The rate of `garbage' messages $R_k^{\text g}, \forall k \in {\cal K}'$ satisfies (\ref{region_FM3}), which is the key point for proving Theorem~\ref{lemma_DM_exten}.
In particular, the second inequation of (\ref{region_FM3}) ensures that the messages (even after adding `garbage' messages) of all users can be perfectly decoded at Bob. With the third inequation in (\ref{region_FM3}), on one hand, Eve obviously cannot decode the messages (including all secret, open, and garbage messages) of users in ${\cal K}'$ using the normal MAC decoding scheme.
On the other hand, we show that Eve cannot extract any confidential information.

In the following, we prove Theorem~\ref{lemma_DM_exten} by first providing a random coding scheme, and then showing that (\ref{RE}) can be satisfied, i.e., all users can communicate with Bob with arbitrarily small probability of error, while the confidential information leaked to Eve tends to zero.

\subsubsection{Coding Scheme}
\label{Prove_theorem1_A}

For a given rate tuple $\left(R_1^{\text s}, R_1^{\text o},\cdots, R_K^{\text s}, R_K^{\text o}\right)$ inside region ${\mathscr R} (X_{\cal K}, {\cal K}')$, choose $R_k^{\text g}, \forall k \in {\cal K}'$ satisfying (\ref{region_FM3}).
Assume w.l.o.g. that $2^{n R_k^{\text s}}$, $2^{n (R_k^{\text o} + R_k^{\text g})}$, $2^{n R_k^{\text g}}, \forall k \in {\cal K}'$, and $2^{n R_k^{\text o}}, \forall k \in {\overline {{\cal K}'}}$, are integers.
Denote
\begin{align}\label{L2}
& {\cal L}_{k, m_k^{\text s}, m_k^{\text o}} = \left[ (m_k^{\text s} - 1) 2^{n (R_k^{\text o} + R_k^{\text g})} + (m_k^{\text o} - 1) 2^{n R_k^{\text g}} + 1 : (m_k^{\text s} - 1) 2^{n (R_k^{\text o} + R_k^{\text g})} + m_k^{\text o} 2^{n R_k^{\text g}} \right],\nonumber\\
& \quad\quad\quad\quad \forall~ k \in {\cal K}',~ m_k^{\text s} \in {\cal M}_k^{\text s},~ m_k^{\text o} \in {\cal M}_k^{\text o},\nonumber\\
& {\cal L}_{k,m_k^{\text s}} = \left[ (m_k^{\text s}-1) 2^{n (R_k^{\text o} + R_k^{\text g})} + 1 : m_k^{\text s} 2^{n (R_k^{\text o} + R_k^{\text g})} \right], ~\forall~ k \in {\cal K}',~ m_k^{\text s} \in {\cal M}_k^{\text s},\nonumber\\
& {\cal L}_k = \left\{ {\cal L}_{k,m_k^{\text s}}, \forall m_k^{\text s} \in {\cal M}_k^{\text s} \right\} 
~ = \left[ 1:2^{n (R_k^{\text s} + R_k^{\text o} + R_k^{\text g})} \right], ~\forall~ k \in {\cal K}'.
\end{align}
Then, a coding scheme is provided below.

\begin{figure*}[ht]
	\centering
	\includegraphics[scale=0.85]{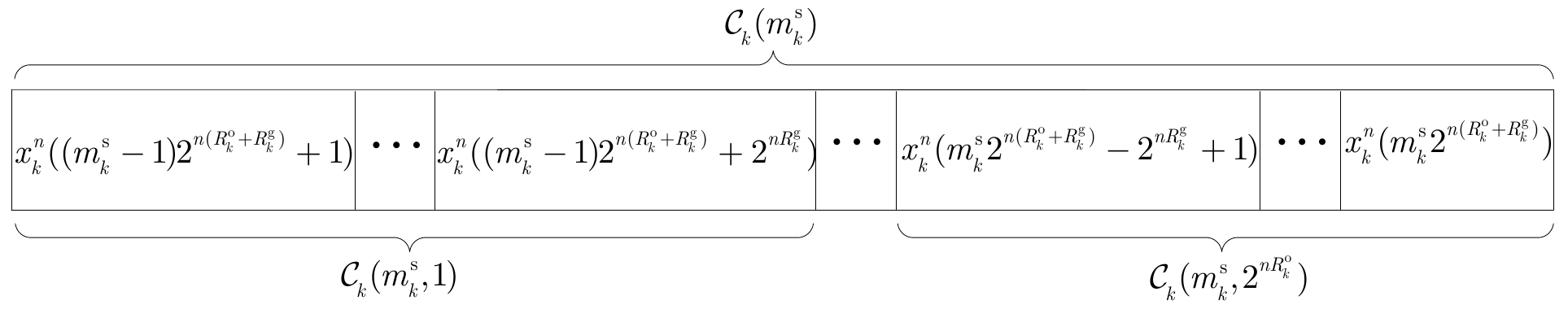}
	\vspace{-3 em}
	\caption{A division of subcodebook ${\cal C}_k (m_k^{\text s})$ of user $k \in {\cal K}'$.}
	\label{Subcodebook}
\end{figure*}
\begin{figure}
	\centering
	\includegraphics[scale=0.80]{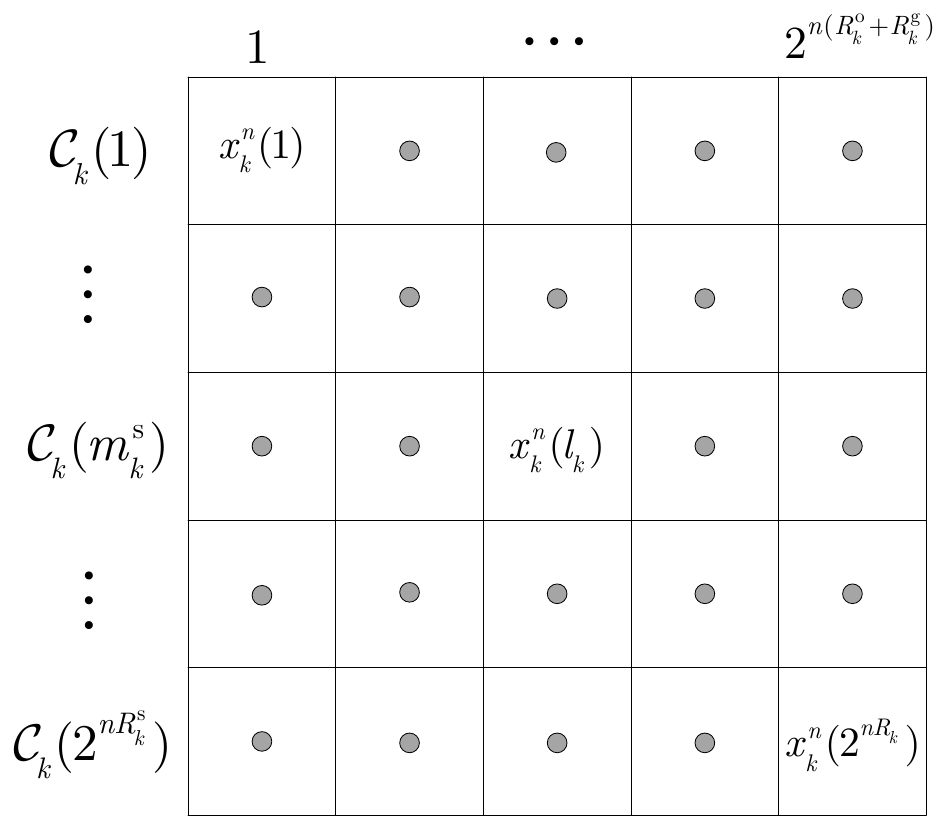}
	\vspace{-1.5 em}
	\caption{Codebook ${\cal C}_k$ of user $k \in {\cal K}'$, where $R_k = R_k^{\text s} + R_k^{\text o} + R_k^{\text g}$.}
	\vspace{-2 em}
	\label{Codebook}
\end{figure}

{\textbf {Codebook generation.}}
For each message pair $(m_k^{\text s}, m_k^{\text o}) \in {\cal M}_k^{\text s} \times {\cal M}_k^{\text o}$ of user $k \in {\cal K}'$, generate a sub-subcodebook ${\cal C}_k (m_k^{\text s}, m_k^{\text o})$ by randomly and independently generating $2^{n R_k^{\text g}}$ sequences $x_k^n(l_k), \forall l_k \in {\cal L}_{k, m_k^{\text s}, m_k^{\text o}}$, each according to $\prod_{i=1}^n p (x_{ki})$. 
For a given secret message $m_k^{\text s}$, the sub-subcodebooks for all open messages constitute subcodebook ${\cal C}_k (m_k^{\text s})$, i.e., ${\cal C}_k (m_k^{\text s}) = \left\{ {\cal C}_k (m_k^{\text s}, m_k^{\text o}), \forall m_k^{\text o} \in {\cal M}_k^{\text o} \right\}$.
Fig.~\ref{Subcodebook} gives an example of subcodebook ${\cal C}_k (m_k^{\text s})$.
Then, as shown in Fig. \ref{Codebook}, these subcodebooks constitute the codebook of user $k$, i.e., ${\cal C}_k = \left\{ {\cal C}_k (m_k^{\text s}), \forall m_k^{\text s} \in {\cal M}_k^{\text s} \right\}$.
For each user $k$ in ${\overline {{\cal K}'}}$, we apply the random coding scheme used in the conventional MAC channel with no wiretapping.
In particular, generate its codebook ${\cal C}_k$ by randomly and independently generating $2^{n R_k^{\text o}}$ sequences $x_k^n(m_k^{\text o}), \forall m_k^{\text o} \in {\cal M}_k^{\text o}$, each according to $\prod_{i=1}^n p (x_{ki})$.
The codebooks of all users are then revealed to all transmitters and receivers, including Eve.

{\textbf {Encoding.}} 
To send message pair $(m_k^{\text s}, m_k^{\text o}) \in {\cal M}_k^{\text s} \times {\cal M}_k^{\text o}$, encoder $k \in {\cal K}'$ uniformly chooses a codeword (with index $l_k$) from ${\cal C}_k (m_k^{\text s},m_k^{\text o})$ and then transmits $x_k^n (l_k)$.
To send message $ m_k^{\text o} \in {\cal M}_k^{\text o}$, encoder $k\in {\overline {{\cal K}'}}$ transmits $x_k^n (m_k^{\text o})$.

{\textbf {Decoding.}} 
The decoder at Bob uses joint typicality decoding to find an estimate of the messages and declares that $( \{ {\hat m}_k^{\text s}, {\hat m}_k^{\text o}, \forall k \in {\cal K}' \}, \{ {\hat m}_k^{\text o}, \forall k \in {\overline {{\cal K}'}} \} )$ is sent if it is the unique message tuple such that 
$( \{ x_k^n (l_k), \forall k \in {\cal K}' \}, \{ x_k^n ({\hat m}_k^{\text o}), \forall k \in {\overline {{\cal K}'}} \}, y^n ) \in {\cal T}_\epsilon^{(n)}( X_{\cal K}, Y)$, for some $l_k$ such that $x_k^n (l_k) \in {\cal C}_k ({\hat m}_k^{\text s},{\hat m}_k^{\text o}), \forall k \in {\cal K}'$. 

\subsubsection{Analysis of the Probability of Error}

Since
\begin{align}
\sum\limits_{k \in {\cal S}} (R_k^{\text s} \!+\! R_k^{\text o} \!+\! R_k^{\text g}) \!+\! \sum\limits_{k \in {\cal T}} R_k^{\text o} < I(X_{\cal S}, X_{\cal T}; Y| X_{\overline {\cal S}}, X_{\overline {\cal T}}) \!-\! \epsilon, \forall {\cal S} \subseteq {\cal K}', {\cal T} \subseteq {\overline {{\cal K}'}}, {\cal S} \cup {\cal T} \neq \phi,
\end{align}
it can be proven by using the law of large numbers (LLN) and the packing lemma that the probability of error averaged over the random codebook and encoding tends to zero as $n \rightarrow \infty$.
The proof follows exactly the same steps used in \cite[Subsection 4.5.1]{el2011network}.
Hence, $\lim_{n \rightarrow \infty} P_{\text e} \leq \delta$.

\subsubsection{Analysis of the Information Leakage Rate}
\label{A-C}

Based on the coding scheme provided above, it is known that for a given codebook ${\cal C}_k$ of user $k \in {\cal K}'$, the secret message $M_k^{\text s}$ is a function of the codeword index $L_k$.
As for user $k \in {\overline {{\cal K}'}}$, its open message $M_k^{\text o}$ is the index of its codeword $X_k^n$.
Note that the messages of all users are independent.
Hence, we have
\begin{align}\label{IM1M2Z}
I(M_{{\cal K}'}^{\text s}; Z^n| M_{\overline {{\cal K}'}}^{\text o}) & = H(M_{{\cal K}'}^{\text s}| M_{\overline {{\cal K}'}}^{\text o}) - H(M_{{\cal K}'}^{\text s}| M_{\overline {{\cal K}'}}^{\text o}, Z^n) \nonumber\\
& = \sum\limits_{k \in {\cal K}'} H(M_k^{\text s}) - H(M_{{\cal K}'}^{\text s}, L_{{\cal K}'}| M_{\overline {{\cal K}'}}^{\text o}, Z^n) + H(L_{{\cal K}'}| M_{{\cal K}'}^{\text s}, M_{\overline {{\cal K}'}}^{\text o}, Z^n ) \nonumber\\
& = \sum\limits_{k \in {\cal K}'} n R_k^{\text s} - H(L_{{\cal K}'}| M_{\overline {{\cal K}'}}^{\text o}, Z^n) + H(L_{{\cal K}'}| M_{{\cal K}'}^{\text s}, M_{\overline {{\cal K}'}}^{\text o}, Z^n) \nonumber\\
& = \sum\limits_{k \in {\cal K}'} n R_k^{\text s} - H(L_{{\cal K}'}| X_{\overline {{\cal K}'}}^n, Z^n) + H(L_{{\cal K}'}| M_{{\cal K}'}^{\text s}, X_{\overline {{\cal K}'}}^n, Z^n).
\end{align}
To measure the information leakage rate, we separately evaluate entropies $H(L_{{\cal K}'}| X_{\overline {{\cal K}'}}^n, Z^n)$ and $H(L_{{\cal K}'}| M_{{\cal K}'}^{\text s}, X_{\overline {{\cal K}'}}^n, Z^n)$.
First, $H(L_{{\cal K}'}| X_{\overline {{\cal K}'}}^n, Z^n)$ can be transformed to
\begin{align}\label{HL1L2Z}
& H(L_{{\cal K}'}| X_{\overline {{\cal K}'}}^n, Z^n) = H(L_{{\cal K}'}| X_{\overline {{\cal K}'}}^n) - I(L_{{\cal K}'}; Z^n| X_{\overline {{\cal K}'}}^n) 
\overset{(a)}{=} H(L_{{\cal K}'}) - I(L_{{\cal K}'}, X_{{\cal K}'}^n; Z^n| X_{\overline {{\cal K}'}}^n)\nonumber\\
& \overset{(b)}{=} \sum\limits_{k \in {\cal K}'} n (R_k^{\text s} \!+\! R_k^{\text o} \!+\! R_k^{\text g} ) \!-\! I(X_{{\cal K}'}^n; Z^n| X_{\overline {{\cal K}'}}^n)~
\overset{(c)}{=} \sum\limits_{k \in {\cal K}'} n (R_k^{\text s} \!+\! R_k^{\text o} \!+\! R_k^{\text g} ) \!-\! n I(X_{{\cal K}'}; Z| X_{\overline {{\cal K}'}}),
\end{align}
where $(a)$ holds since the messages and encoding of different users are independent and for any $k \in {\cal K}'$, $L_k$ is the index of the codeword $X_k^n$, $(b)$ holds since $L_{{\cal K}'} \rightarrow X_{{\cal K}'}^n \rightarrow Z^n$ forms a Markov chain, and $(c)$ follows since $p(x_{{\cal K}'}^n, z^n| x_{\overline {{\cal K}'}}^n) = \prod_{i=1}^n p (x_{{\cal K}' i}, z_i| x_{{\overline {{\cal K}'}} i})$.
Then, we provide an upper bound to $\frac{1}{n} H (L_{{\cal K}'}| M_{{\cal K}'}^{\text s}, X_{\overline {{\cal K}'}}^n, Z^n)$ in the following theorem.
\begin{lemma}\label{entropy_ub}
	Using the coding scheme provided above, we have the following inequation
	\begin{align}\label{upper_bound}
	\lim_{n \rightarrow \infty} \frac{1}{n} H (L_{{\cal K}'}| M_{{\cal K}'}^{\text s}, X_{\overline {{\cal K}'}}^n, Z^n) \leq \sum\limits_{k \in {\cal K}'} ( R_k^{\text o} + R_k^{\text g} ) - I(X_{{\cal K}'}; Z| X_{\overline {{\cal K}'}}) + \delta.
	\end{align}
\end{lemma}
\itshape \textbf{Proof:}  \upshape
See Appendix~\ref{prove_entropy_ub}.
\hfill $\Box$

Combining (\ref{IM1M2Z}), (\ref{HL1L2Z}), and (\ref{upper_bound}), we have
\begin{equation}\label{I_leq_delta}
\lim_{n \rightarrow \infty} R_{{\text E}, {\cal K}'} = \lim_{n \rightarrow \infty} \frac{1}{n} I(M_{{\cal K}'}^{\text s}; Z^n| M_{\overline {{\cal K}'}}^{\text o}) \leq \delta.
\end{equation}
Lemma~\ref{theorem_region_DM} is thus proven if assumption (\ref{assumption_1}) is true.

\subsection{Proof of Theorem~\ref{lemma_DM_exten} When Assumption (\ref{assumption_1}) is not True}

If (\ref{assumption_1}) is not true, i.e., there exist ${\cal S} \subseteq {\cal K}'$, ${\cal S} \neq \phi$, ${\cal S}' \subseteq {\cal S}$, ${\cal S}' \neq \phi$, and ${\cal T} \subseteq {\overline {{\cal K}'}}$ such that
\begin{equation}\label{assumption_2}
I(X_{\cal S}, X_{\cal T}; Y| X_{\overline {\cal S}}, X_{\overline {\cal T}}) - I(X_{{\cal S}'}; Z| X_{\overline {{\cal K}'}}) \leq 0.
\end{equation}
From (\ref{two_ineq}), it is known that (\ref{assumption_2}) ensures
\begin{equation}\label{assumption_6}
I(X_{\cal S}; Y| X_{\overline {\cal S}}, X_{\overline {{\cal K}'}}) - I(X_{\cal S}; Z| X_{\overline {{\cal K}'}}) \leq 0.
\end{equation}
With (\ref{assumption_6}), there exist two possible cases, i.e., 
\begin{equation}\label{assumption_60}
I(X_{{\cal K}'}; Y| X_{\overline {{\cal K}'}}) - I(X_{{\cal K}'}; Z| X_{\overline {{\cal K}'}}) \leq 0,
\end{equation}
and
\begin{equation}\label{assumption_63}
I(X_{{\cal K}'}; Y| X_{\overline {{\cal K}'}}) - I(X_{{\cal K}'}; Z| X_{\overline {{\cal K}'}}) > 0.
\end{equation}
In the following, we prove that Theorem~\ref{lemma_DM_exten} is true when either (\ref{assumption_60}) or (\ref{assumption_63}) holds.

In the first case, i.e., when (\ref{assumption_60}) holds, we have $R_k^{\text s} = 0, \forall k \in {\cal K}$.
(\ref{region_DM_exten}) thus becomes
\begin{equation}\label{region_case1}
\left\{
\begin{array}{ll}
R_k^{\text s} = 0, ~\forall~ k \in {\cal K}, \\
\sum\limits_{k \in {\cal S} \setminus {\cal S}'} R_k^{\text o} + \sum\limits_{k \in {\cal T}} R_k^{\text o}	& \leq \left[ I(X_{\cal S}, X_{\cal T}; Y| X_{\overline {\cal S}}, X_{\overline {\cal T}}) - I(X_{{\cal S}'}; Z| X_{\overline {{\cal K}'}}) \right]^+, \\
& \forall~ {\cal S} \subseteq {\cal K}',~ {\cal S}' \subseteq {\cal S},~ {\cal T} \subseteq {\overline {{\cal K}'}}.
\end{array} \right.
\end{equation}
By always letting ${\cal S}' = \phi$ in (\ref{region_case1}), we get
\begin{equation}\label{region_case1_2}
\left\{
\begin{array}{ll}
R_k^{\text s} = 0, ~\forall~ k \in {\cal K}, \\
\sum\limits_{k \in {\cal S}} R_k^{\text o} + \sum\limits_{k \in {\cal T}} R_k^{\text o}\leq I(X_{\cal S}, X_{\cal T}; Y| X_{\overline {\cal S}}, X_{\overline {\cal T}}),  ~\forall~ {\cal S} \subseteq {\cal K}',~ {\cal T} \subseteq {\overline {{\cal K}'}}.
\end{array} \right.
\end{equation}
Since (\ref{region_case1_2}) considers only partial inequalities in (\ref{region_case1}), the region defined by (\ref{region_case1}) is included in that defined by (\ref{region_case1_2}).
Moreover, (\ref{region_case1_2}) also shows that due to (\ref{assumption_60}), the region ${\mathscr R} (X_{\cal K}, {\cal K}')$ is included in the capacity region of a conventional MAC channel with no wiretapping and its achievability proof is well known and is thus omitted here.

In the second case, i.e., when (\ref{assumption_63}) holds, due to (\ref{assumption_6}), there must exist at least one subset ${\cal K}_0 \subsetneqq {\cal K}'$ such that
\begin{equation}\label{assumption_61}
I(X_{{\cal K}_0}; Y| X_{{\cal K}' \setminus {\cal K}_0}, X_{\overline {{\cal K}'}}) - I(X_{{\cal K}_0}; Z| X_{\overline {{\cal K}'}}) \leq 0,
\end{equation}
and 
\begin{align}\label{assumption_62}
I(X_{{\cal K}_0 \cup {\cal S}}; Y| X_{{\cal K}' \setminus ({\cal K}_0 \cup {\cal S})}, X_{\overline {{\cal K}'}}) - I(X_{{\cal K}_0 \cup {\cal S}}; Z| X_{\overline {{\cal K}'}}) > 0, ~\forall~ {\cal S} \subseteq {\cal K}' \setminus {\cal K}_0,~ {\cal S} \neq \phi.
\end{align}
The inequalities (\ref{assumption_61}) and (\ref{assumption_62}) indicate that ${\cal K}_0$ is the largest set in ${\cal K}'$ which includes all users in ${\cal K}_0$ and ensures (\ref{assumption_61}).
Adding any other users in ${\cal K}' \setminus {\cal K}_0$ to ${\cal K}_0$ results in (\ref{assumption_62}).
Note that if there are multiple subsets in ${\cal K}'$ making (\ref{assumption_61}) and (\ref{assumption_62}) hold, we let ${\cal K}_0$ be any of them.
Let
\begin{align}\label{K_prime}
{\cal K}'' & = {\cal K}' \setminus {\cal K}_0 = {\cal K} \setminus ({\overline {{\cal K}'}} \cup {\cal K}_0), \nonumber\\
{\overline {{\cal K}''}} & = {\cal K} \setminus {\cal K}'' = {\overline {{\cal K}'}} \cup {\cal K}_0.
\end{align}
Then, we give the following lemma.
\begin{lemma}\label{lemma_polytope}
	Let $(X_{\cal K}, Y, Z) \sim \prod_{k=1}^K p(x_k) p(y,z| x_{\cal K})$.
	With ${\cal K}_0$, ${\cal K}''$, and ${\overline {{\cal K}''}}$ defined above, we have
	\begin{align}\label{assumption_4}
	I(X_{\cal S}, X_{\cal T}; Y| X_{\overline {\cal S}}, X_{\overline {\cal T}}) \!-\! I(X_{{\cal S}'}; Z| X_{\overline {{\cal K}''}}) \!>\! 0, ~\forall~ {\cal S} \subseteq {\cal K}'', {\cal S}' \subseteq {\cal S}, {\cal T} \subseteq {\overline {{\cal K}''}}, {\cal S} \cup {\cal T} \neq \phi,
	\end{align}
	where ${\overline {\cal S}} = {\cal K}'' \setminus {\cal S}$ and ${\overline {\cal T}} = {\overline {{\cal K}''}} \setminus {\cal T}$.
	In addition, if a rate tuple $(R_1^{\text s}, R_1^{\text o},\cdots, R_K^{\text s}, R_K^{\text o})$ is in region ${\mathscr R} (X_{\cal K}, {\cal K}')$ defined by Theorem~\ref{lemma_DM_exten} and has (\ref{assumption_61}) as well as (\ref{assumption_62}) met, then, it is also in region ${\mathscr R} (X_{\cal K}, {\cal K}'')$, i.e., it satisfies
	\begin{equation}\label{region_DM1}
	\left\{
	\begin{array}{ll}
	R_k^{\text s} = 0, ~\forall~ k \in {\overline {{\cal K}''}}, \\
	\sum\limits_{k \in \cal S} R_k^{\text s} + \sum\limits_{k \in {\cal S} \setminus {\cal S}'} R_k^{\text o} + \sum\limits_{k \in {\cal T}} R_k^{\text o} & \leq I(X_{\cal S}, X_{\cal T}; Y| X_{\overline {\cal S}}, X_{\overline {\cal T}}) - I(X_{{\cal S}'}; Z| X_{\overline {{\cal K}''}}), \\
	& \forall~ {\cal S} \subseteq {\cal K}'',~ {\cal S}' \subseteq {\cal S},~ {\cal T} \subseteq {\overline {{\cal K}''}}.
	\end{array} \right.
	\end{equation}
\end{lemma}
\itshape \textbf{Proof:} \upshape
See Appendix \ref{prove_lemma_polytope}.
\hfill $\Box$

Based on Lemma~\ref{lemma_polytope} it is known that if a rate tuple $(R_1^{\text s}, R_1^{\text o},\cdots, R_K^{\text s}, R_K^{\text o})$ is inside region ${\mathscr R} (X_{\cal K}, {\cal K}')$ and has (\ref{assumption_61}) and (\ref{assumption_62}) met, it is also inside region ${\mathscr R} (X_{\cal K}, {\cal K}'')$ and thus satisfies
\begin{equation}
\left\{
\begin{array}{ll}
R_k^{\text s} = 0, ~\forall~ k \in {\overline {{\cal K}''}}, \\
\sum\limits_{k \in \cal S} R_k^{\text s} + \sum\limits_{k \in {\cal S} \setminus {\cal S}'} R_k^{\text o} + \sum\limits_{k \in {\cal T}} R_k^{\text o} & < I(X_{\cal S}, X_{\cal T}; Y| X_{\overline {\cal S}}, X_{\overline {\cal T}})- I(X_{{\cal S}'}; Z| X_{\overline {{\cal K}''}}) - \varepsilon, \\
& \forall~ {\cal S} \subseteq {\cal K}'',~ {\cal S}' \subseteq {\cal S},~ {\cal T} \subseteq {\overline {{\cal K}''}},~ {\cal S} \cup {\cal T} \neq \phi.
\end{array} \right.
\end{equation}
Moreover, thanks to (\ref{assumption_4}), for any rate tuple inside ${\mathscr R} (X_{\cal K}, {\cal K}'')$, Theorem~\ref{lemma_FM_gene_K2} and Dantzig's simplex algorithm can be applied to get $R_k^{\text g}, \forall k \in {\cal K}''$ such that
\begin{equation}
\left\{\!\!\!
\begin{array}{ll}
R_k^{\text g} \geq 0, ~\forall~ k \in {\cal K}'', \\
\sum\limits_{k \in {\cal S}} (R_k^{\text s} \!+\! R_k^{\text o} \!+\! R_k^{\text g}) \!+\! \sum\limits_{k \in {\cal T}} R_k^{\text o} \!<\! I(X_{\cal S}, X_{\cal T}; Y| X_{\overline {\cal S}}, X_{\overline {\cal T}}) \!-\! \epsilon, \forall {\cal S} \!\subseteq\! {\cal K}'', {\cal T} \!\subseteq\! {\overline {{\cal K}''}}, {\cal S} \cup {\cal T} \!\neq\! \phi, \\
\sum\limits_{k \in {\cal S}} (R_k^{\text o} + R_k^{\text g}) > I(X_{\cal S}; Z| X_{\overline {{\cal K}''}}), ~\forall~ {\cal S} \subseteq {\cal K}'',~ {\cal S} \neq \phi.
\end{array} \right.
\end{equation}
The techniques provided in the previous Subsection~\ref{assump_true} can then be applied to prove the achievability of the rate tuple.
The proof in this and the previous subsections, together with the standard time-sharing over coding strategies, suffices to prove Theorem~\ref{lemma_DM_exten}. 

\section{Proof of Lemma~\ref{entropy_ub}}
\label{prove_entropy_ub}

For given $n$-th order product distribution on ${\cal X}^n_1 \times \cdots \times {\cal X}^n_K \times {\cal Z}^n$, 
recall the definition of conditional $\epsilon$-typical sets
\begin{align}\label{T_conditional}
{\cal T}_\epsilon^{(n)} (X_{\cal S}| x_{\overline {\cal S}}^n, x_{\overline {{\cal K}'}}^n, z^n) & = \left\{ x_{\cal S}^n| (x_{\cal S}^n, x_{\overline {\cal S}}^n, x_{\overline {{\cal K}'}}^n, z^n) \in {\cal T}_\epsilon^{(n)} (X_{\cal S}, X_{\overline {\cal S}}, X_{\overline {{\cal K}'}}, Z), ~\forall~ x_{\cal S}^n \in \prod_{ k \in {\cal K}' } {\cal X}_k^n \right\}, \nonumber\\
& \;\; \forall~ {\cal S} \subseteq {\cal K}',~ (x_{\overline {\cal S}}^n, x_{\overline {{\cal K}'}}^n, z^n) \in {\cal T}_\epsilon^{(n)} (X_{\overline {\cal S}}, X_{\overline {{\cal K}'}}, Z).
\end{align}
Before proving Lemma~\ref{entropy_ub}, we first define some auxiliary functions.
For given codeword set $x_{\overline {{\cal K}'}}^n$ and received signal $z^n$ at the eavesdropper, assume that they are jointly typical, i.e., $(x_{\overline {{\cal K}'}}^n, z^n) \in {\cal T}_\epsilon^{(n)} (X_{\overline {{\cal K}'}}, Z)$, and define functions
\begin{equation}\label{D}
{\cal D} (m_{{\cal K}'}^{\text s}| x_{\overline {{\cal K}'}}^n, z^n) = \left\{ l_{{\cal K}'}| \left( x_k^n (l_k), \forall k \in {\cal K}' \right) \in {\cal T}_\epsilon^{(n)} (X_{{\cal K}'}| x_{\overline {{\cal K}'}}^n, z^n), ~\forall~ l_{{\cal K}'} \in \prod_{ k \in {\cal K}' } {\cal L}_{k,m_k^{\text s}} \right\},
\end{equation}
and
\begin{equation}\label{Q}
Q (m_{{\cal K}'}^{\text s}| x_{\overline {{\cal K}'}}^n, z^n) = \left|{\cal D} (m_{{\cal K}'}^{\text s}| x_{\overline {{\cal K}'}}^n, z^n) \right|.
\end{equation}
Obviously, ${\cal D} (m_{{\cal K}'}^{\text s}| x_{\overline {{\cal K}'}}^n, z^n)$ records all possible index sets $l_{{\cal K}'} \in \prod_{ k \in {\cal K}' } {\cal L}_{k,m_k^{\text s}}$ with each one ensuring that $x_k^n (l_k), \forall k \in {\cal K}'$ and $(x_{\overline {{\cal K}'}}^n, z^n)$ are jointly typical, and $Q (m_{{\cal K}'}^{\text s}| x_{\overline {{\cal K}'}}^n, z^n)$ denotes the number of these index sets.
By introducing an indicator variable
\begin{equation}\label{indicator_E0}
O' (l_{{\cal K}'}| x_{\overline {{\cal K}'}}^n, z^n) \!=\! \left\{\!\!\!
\begin{array}{ll}
1,&  \!\!\!{\text {if}} ~ \left\{ x_k^n (l_k), \forall k \in {\cal K}' \right\} \!\in {\cal T}_\epsilon^{(n)} (X_{{\cal K}'}| x_{\overline {{\cal K}'}}^n, z^n),\\
0,&  \!\!\!{\text {otherwise}},\\
\end{array} \right.\!\!\!\!
\end{equation} 
where $ l_{{\cal K}'} \in \prod_{ k \in {\cal K}' } {\cal L}_{k,m_k^{\text s}} $, $Q (m_{{\cal K}'}^{\text s}| x_{\overline {{\cal K}'}}^n, z^n)$ can also be represented as
\begin{equation}\label{N0}
Q (m_{{\cal K}'}^{\text s}| x_{\overline {{\cal K}'}}^n, z^n) = \sum_{ l_{{\cal K}'} \in \prod_{ k \in {\cal K}' } {\cal L}_{k,m_k^{\text s}} } O' (l_{{\cal K}'}| x_{\overline {{\cal K}'}}^n, z^n).
\end{equation}
Using the above definitions, we give in the following theorem upper bounds to the expectation and variance of $Q (m_{{\cal K}'}^{\text s}| x_{\overline {{\cal K}'}}^n, z^n)$.
\begin{theorem}\label{exp_var_Q}
	For any $(x_{\overline {{\cal K}'}}^n, z^n) \in {\cal T}_\epsilon^{(n)} (X_{\overline {{\cal K}'}}, Z)$, the expectation and variance of $Q (m_{{\cal K}'}^{\text s}| x_{\overline {{\cal K}'}}^n, z^n)$ can be bounded as
	\begin{align}
	{\mathbb E} \left[ Q (m_{{\cal K}'}^{\text s}| x_{\overline {{\cal K}'}}^n, z^n) \right] & \leq 2^{n \left[ \varOmega_{{\cal K}'} + (K + 1) \epsilon \right] }, \label{EN}\\
	{\text {Var}} \left[ Q (m_{{\cal K}'}^{\text s}| x_{\overline {{\cal K}'}}^n, z^n) \right] & \leq \sum\limits_{ {\cal S} \subsetneqq {\cal K}' } 2^{n \left[ 2 \varOmega_{{\cal K}'} - \varOmega_{\overline {\cal S}} + 2 (K + 1) \epsilon \right]}, \label{VarN}
	\end{align}
	where
	\begin{align}\label{delta}
	\varOmega_{\cal S} & = \sum\limits_{k \in {\cal S}} n (R_k^{\text o} + R_k^{\text g} ) - I(X_{\cal S}; Z| X_{\overline {{\cal K}'}}), \nonumber\\
	\varOmega_{\overline {\cal S}} & = \sum\limits_{k \in {\overline {\cal S}}} n (R_k^{\text o} + R_k^{\text g} ) - I(X_{\overline {\cal S}}; Z| X_{\overline {{\cal K}'}}), ~\forall~ {\cal S} \subseteq {\cal K}',~ {\cal S} \neq \phi.
	\end{align}
\end{theorem}
\itshape \textbf{Proof:}  \upshape
See Appendix~\ref{prove_exp_var_Q}.
\hfill $\Box$

Define event 
\begin{equation}
{\cal E} (m_{{\cal K}'}^{\text s}| x_{\overline {{\cal K}'}}^n, z^n) = \left\{  Q (m_{{\cal K}'}^{\text s}| x_{\overline {{\cal K}'}}^n, z^n) \geq 2^{n \left[ \varOmega_{{\cal K}'} + (K + 1) \epsilon \right] + 1} \right\}.
\end{equation}
We have
\begin{align}\label{p_event}
{\text {Pr}} \left\{{\cal E} (m_{{\cal K}'}^{\text s}| x_{\overline {{\cal K}'}}^n, z^n) \right\} & = {\text {Pr}} \left\{  Q (m_{{\cal K}'}^{\text s}| x_{\overline {{\cal K}'}}^n, z^n) \geq 2^{n \left[ \varOmega_{{\cal K}'} + (K + 1) \epsilon \right] + 1} \right\}\nonumber\\
& \leq {\text {Pr}} \left\{  Q (m_{{\cal K}'}^{\text s}| x_{\overline {{\cal K}'}}^n, z^n) \geq {\mathbb E} \left[ Q (m_{{\cal K}'}^{\text s}| x_{\overline {{\cal K}'}}^n, z^n) \right] + 2^{n \left[ \varOmega_{{\cal K}'} + (K + 1) \epsilon \right]} \right\}\nonumber\\
& \leq {\text {Pr}} \left\{ \left| Q (m_{{\cal K}'}^{\text s}| x_{\overline {{\cal K}'}}^n, z^n) \!-\! {\mathbb E} \left[ Q (m_{{\cal K}'}^{\text s}| x_{\overline {{\cal K}'}}^n, z^n) \right] \right| \!\geq\! 2^{n \left[ \varOmega_{{\cal K}'} + (K + 1) \epsilon \right]} \right\}\nonumber\\
& \overset{(a)}{\leq} \frac{{\text {Var}} \left[ Q (m_{{\cal K}'}^{\text s}| x_{\overline {{\cal K}'}}^n, z^n) \right]}{2^{2n \left[ \varOmega_{{\cal K}'} + (K + 1) \epsilon \right]}} \overset{(b)}{\leq} \sum\limits_{ {\cal S} \subsetneqq {\cal K}' } 2^{ - n \varOmega_{\overline {\cal S}} },
\end{align}
where step $(a)$ follows by applying the Chebyshev inequality, and $(b)$ follows by (\ref{VarN}).
Due to (\ref{region_FM3}), we have $\varOmega_{\overline {\cal S}} > 0, \forall {\cal S} \subsetneqq {\cal K}'$.
Then, it is obvious that ${\text {Pr}} \left\{{\cal E} (m_{{\cal K}'}^{\text s}| x_{\overline {{\cal K}'}}^n, z^n) \right\} \rightarrow 0$ as $n \rightarrow \infty$.
For any $(x_{\overline {{\cal K}'}}^n, z^n) \in {\cal T}_\epsilon^{(n)} (X_{\overline {{\cal K}'}}, Z)$, define indicator variable
\begin{equation}\label{indicator_E}
O (m_{{\cal K}'}^{\text s}| x_{\overline {{\cal K}'}}^n, z^n) \!=\! \left\{\!\!\!
\begin{array}{ll}
1,& \!\! {\text {if}}~ {\cal E} (m_{{\cal K}'}^{\text s}| x_{\overline {{\cal K}'}}^n, z^n) ~{\text {occurs}},\\
0,& \!\! {\text {otherwise}}.\\
\end{array} \right.
\end{equation} 
Then, ${\text {Pr}} \left\{ O (m_{{\cal K}'}^{\text s}| x_{\overline {{\cal K}'}}^n, z^n) = 1 \right\} \rightarrow 0$ as $n \rightarrow \infty$.

Since there are $2^{n(R_k^{\text o} + R_k^{\text g})}$ codewords in each subcodebook ${\cal C}_k (m_k^{\text s}), \forall k \in \cal K$, we have
\begin{align}\label{HL1L2Zm1m21}
& H (L_{{\cal K}'}^{\text s}| m_{{\cal K}'}^{\text s}, x_{\overline {{\cal K}'}}^n, z^n) \leq H (L_{{\cal K}'}^{\text s}) = \sum_{k \in {\cal K}'} H (L_k^{\text s}) \nonumber\\
& = \sum_{k \in {\cal K}'} n ( R_k^{\text o} + R_k^{\text g} ), ~\forall~ (m_{{\cal K}'}^{\text s}, x_{\overline {{\cal K}'}}^n, z^n) \in \prod_{ k \in {\cal K}' } {\cal M}_k^{\text s} \times \prod_{ j \in {\overline {{\cal K}'}} } {\cal X}_j^n \times {\cal Z}^n,
\end{align}
and 
\begin{align}\label{HL1L2Zm1m23}
H (L_{{\cal K}'}^{\text s}| m_{{\cal K}'}^{\text s}, X_{\overline {{\cal K}'}}^n, Z^n) & = \sum_{ (x_{\overline {{\cal K}'}}^n, z^n) \in \prod_{ k \in {\overline {{\cal K}'}} } {\cal X}_k^n \times {\cal Z}^n } p(x_{\overline {{\cal K}'}}^n, z^n) H (L_{{\cal K}'}^{\text s}| m_{{\cal K}'}^{\text s}, x_{\overline {{\cal K}'}}^n, z^n) \nonumber\\
& \leq \sum_{k \in {\cal K}'} n ( R_k^{\text o} + R_k^{\text g} ), ~\forall~ m_{{\cal K}'}^{\text s} \in \prod_{ k \in {\cal K}' } {\cal M}_k^{\text s}.
\end{align}
Moreover, based on the definitions of $Q (m_{{\cal K}'}^{\text s}| x_{\overline {{\cal K}'}}^n, z^n)$ and $O (m_{{\cal K}'}^{\text s}| x_{\overline {{\cal K}'}}^n, z^n)$ in (\ref{Q}) and (\ref{indicator_E}),
\begin{align}\label{HL1L2Zm1m22}
& H(L_{{\cal K}'}^{\text s}| m_{{\cal K}'}^{\text s}, x_{\overline {{\cal K}'}}^n, z^n, O (m_{{\cal K}'}^{\text s}| x_{\overline {{\cal K}'}}^n, z^n) = 0) \leq \log(Q (m_{{\cal K}'}^{\text s}| x_{\overline {{\cal K}'}}^n, z^n) ) \nonumber\\
& \leq n \left[ \varOmega_{{\cal K}'} + (K + 1) \epsilon \right] + 1, ~\forall~ m_{{\cal K}'}^{\text s} \in \prod_{ k \in {\cal K}' } {\cal M}_k^{\text s},~ (x_{\overline {{\cal K}'}}^n, z^n) \in {\cal T}_\epsilon^{(n)} (X_{\overline {{\cal K}'}}, Z),
\end{align}
where the last step holds since when $O (m_{{\cal K}'}^{\text s}| x_{\overline {{\cal K}'}}^n, z^n) = 0$, $Q (m_{{\cal K}'}^{\text s}| x_{\overline {{\cal K}'}}^n, z^n) < 2^{n \left[ \varOmega_{{\cal K}'} + (K + 1) \epsilon \right] + 1}$.
Then, a tighter upper bound on $H (L_{{\cal K}'}| m_{{\cal K}'}^{\text s}, X_{\overline {{\cal K}'}}^n, Z^n)$ (in contrast to (\ref{HL1L2Zm1m23})) can be obtained as
\begin{align}\label{HL1L2Zm1m2}
& H (L_{{\cal K}'}| m_{{\cal K}'}^{\text s}, X_{\overline {{\cal K}'}}^n, Z^n) \nonumber\\
= & {\text {Pr}} \left\{ (X_{\overline {{\cal K}'}}^n, Z^n) \in {\cal T}_\epsilon^{(n)} (X_{\overline {{\cal K}'}}, Z) \right\} H ( L_{{\cal K}'}| m_{{\cal K}'}^{\text s}, X_{\overline {{\cal K}'}}^n, Z^n, (X_{\overline {{\cal K}'}}^n, Z^n) \in {\cal T}_\epsilon^{(n)} (X_{\overline {{\cal K}'}}, Z) ) \nonumber\\
+ & {\text {Pr}} \left\{ (X_{\overline {{\cal K}'}}^n, Z^n) \notin {\cal T}_\epsilon^{(n)} (X_{\overline {{\cal K}'}}, Z) \right\} H ( L_{{\cal K}'}| m_{{\cal K}'}^{\text s}, X_{\overline {{\cal K}'}}^n, Z^n, (X_{\overline {{\cal K}'}}^n, Z^n) \notin {\cal T}_\epsilon^{(n)} (X_{\overline {{\cal K}'}}, Z) ) \nonumber\\
\overset{(a)}{\leq} & \sum_{ (x_{\overline {{\cal K}'}}^n, z^n) \in {\cal T}_\epsilon^{(n)} (X_{\overline {{\cal K}'}}, Z) } p (x_{\overline {{\cal K}'}}^n, z^n) H (L_{{\cal K}'}| m_{{\cal K}'}^{\text s}, x_{\overline {{\cal K}'}}^n, z^n) + n \alpha_1 \nonumber\\
= & \sum_{ (x_{\overline {{\cal K}'}}^n, z^n) \in {\cal T}_\epsilon^{(n)} (X_{\overline {{\cal K}'}}, Z) } \!\!\!p (x_{\overline {{\cal K}'}}^n, z^n) \!\left\{  {\text {Pr}} \left\{ O (m_{{\cal K}'}^{\text s}| x_{\overline {{\cal K}'}}^n, z^n) \!=\! 1 \right\} H (L_{{\cal K}'}| m_{{\cal K}'}^{\text s}, x_{\overline {{\cal K}'}}^n, z^n, O (m_{{\cal K}'}^{\text s}| x_{\overline {{\cal K}'}}^n, z^n) \!=\! 1) \right. \nonumber\\
+ & \left. {\text {Pr}} \left\{ O (m_{{\cal K}'}^{\text s}| x_{\overline {{\cal K}'}}^n, z^n) = 0 \right\} H (L_{{\cal K}'}| m_{{\cal K}'}^{\text s}, x_{\overline {{\cal K}'}}^n, z^n, O (m_{{\cal K}'}^{\text s}| x_{\overline {{\cal K}'}}^n, z^n) = 0)\right\} + n \alpha_1 \nonumber\\
\overset{(b)}{\leq} & \sum_{ (x_{\overline {{\cal K}'}}^n, z^n) \in {\cal T}_\epsilon^{(n)} (X_{\overline {{\cal K}'}}, Z) } p (x_{\overline {{\cal K}'}}^n, z^n) \big \{ \alpha_2 H (L_{{\cal K}'}| m_{{\cal K}'}^{\text s}, x_{\overline {{\cal K}'}}^n, z^n) \nonumber\\
+ & H (L_{{\cal K}'}| m_{{\cal K}'}^{\text s}, x_{\overline {{\cal K}'}}^n, z^n, O (m_{{\cal K}'}^{\text s}| x_{\overline {{\cal K}'}}^n, z^n) = 0) \big \} + n \alpha_1 \nonumber\\
\overset{(c)}{\leq} & n \left[ \varOmega_{{\cal K}'} \!+\! (K \!+\! 1) \epsilon \!+\! \frac{1}{n} \!+\! \alpha_2 \sum_{k \in {\cal K}'} ( R_k^{\text o} + R_k^{\text g} ) \!+\! \alpha_1 \right] \overset{(d)}{\leq} n (\varOmega_{{\cal K}'} \!+\! \delta), ~\forall~ m_{{\cal K}'}^{\text s} \in \prod_{ k \in {\cal K}' } {\cal M}_k^{\text s},
\end{align}
where (a) follows from using (\ref{HL1L2Zm1m23}), (b) holds due to the fact that conditioning reduces entropy and ${\text {Pr}} \left\{ O (m_{{\cal K}'}^{\text s}| x_{\overline {{\cal K}'}}^n, z^n) = 0 \right\} \leq 1$, (c) follows from using (\ref{HL1L2Zm1m21}) and (\ref{HL1L2Zm1m22}), and
\begin{align}\label{del}
& \alpha_1 = {\text {Pr}} \left\{ (X_{\overline {{\cal K}'}}^n, Z^n) \notin {\cal T}_\epsilon^{(n)} (X_{\overline {{\cal K}'}}, Z) \right\} \sum_{k \in {\cal K}'} ( R_k^{\text o} + R_k^{\text g} ),\nonumber\\
& \alpha_2 = \max \left\{ {\text {Pr}} \left\{ O (m_{{\cal K}'}^{\text s}| x_{\overline {{\cal K}'}}^n, z^n) = 1 \right\}, ~\forall~ (x_{\overline {{\cal K}'}}^n, z^n) \in {\cal T}_\epsilon^{(n)} (X_{\overline {{\cal K}'}}, Z) \right\}.
\end{align}
Note that by the LLN, ${\text {Pr}} \left\{ (X_{\overline {{\cal K}'}}^n, Z^n) \notin {\cal T}_\epsilon^{(n)} (X_{\overline {{\cal K}'}}, Z) \right\} \rightarrow 0$ as $n \rightarrow \infty$. 
Hence, $\alpha_1 \rightarrow 0$ as $n \rightarrow \infty$.
In addition, since ${\text {Pr}} \left\{ O (m_{{\cal K}'}^{\text s}| x_{\overline {{\cal K}'}}^n, z^n) = 1 \right\} \rightarrow 0, \forall (x_{\overline {{\cal K}'}}^n, z^n) \in {\cal T}_\epsilon^{(n)} (X_{\overline {{\cal K}'}}, Z)$ as $n \rightarrow \infty$, $\alpha_2 \rightarrow 0$ as $n \rightarrow \infty$. 
$(K + 1) \epsilon + \frac{1}{n} + \alpha_2 \sum_{k \in {\cal K}'} ( R_k^{\text o} + R_k^{\text g} ) + \alpha_1$ in (c) of (\ref{HL1L2Zm1m2}) can thus be arbitrarily small as $n \rightarrow \infty$, making the last step of (\ref{HL1L2Zm1m2}) hold.
Hence,
\begin{align}\label{upper_bound1}
\lim_{n \rightarrow \infty} \frac{1}{n} H (L_{{\cal K}'}| M_{{\cal K}'}^{\text s}, X_{\overline {{\cal K}'}}^n, Z^n) & = \lim_{n \rightarrow \infty} \sum_{ m_{{\cal K}'}^{\text s} \in \prod_{ k \in {\cal K}' } {\cal M}_k^{\text s} } \frac{1}{n} 2^{-n \sum_{k \in {\cal K}'} R_k^{\text s} } H (L_{{\cal K}'}| m_{{\cal K}'}^{\text s}, X_{\overline {{\cal K}'}}^n, Z^n)\nonumber\\
& \leq \varOmega_{{\cal K}'} + \delta.
\end{align}
Lemma~\ref{entropy_ub} is thus proven.

\section{Proof of Theorem \ref{exp_var_Q}}
\label{prove_exp_var_Q}

Using the conditional typicality lemma, for sufficiently large $n$, we have
\begin{equation} \label{T_conditional1}
|{\cal T}_\epsilon^{(n)} (X_{\cal S}| x_{\overline {\cal S}}^n, x_{\overline {{\cal K}'}}^n, z^n)| \!\leq\! 2^{n \left[ H(X_{\cal S}| X_{\overline {\cal S}}, X_{\overline {{\cal K}'}}, Z) + \epsilon \right] }, \forall {\cal S} \!\subseteq\! {\cal K}', (x_{\overline {\cal S}}^n, x_{\overline {{\cal K}'}}^n, z^n) \!\in\! {\cal T}_\epsilon^{(n)} (X_{\overline {\cal S}}, X_{\overline {{\cal K}'}}, Z).
\end{equation}
Denote
\begin{equation}\label{p_S}
p_{\cal S} = {\text {Pr}} \left\{ X_{\cal S}^n \in {\cal T}_\epsilon^{(n)} (X_{\cal S}| x_{\overline {\cal S}}^n, x_{\overline {{\cal K}'}}^n, z^n) \right\}, ~\forall~ {\cal S} \subseteq {\cal K}',~ (x_{\overline {\cal S}}^n, x_{\overline {{\cal K}'}}^n, z^n) \in {\cal T}_\epsilon^{(n)} (X_{\overline {\cal S}}, X_{\overline {{\cal K}'}}, Z).
\end{equation}
Since $X_k, \forall k \in {\cal K}$ are independent, $p_{\cal S}$ can be upper bounded by
\begin{align}\label{p_S_up}
p_{\cal S} & = \sum_{ x_{\cal S}^n \in {\cal T}_\epsilon^{(n)} (X_{\cal S}| x_{\overline {\cal S}}^n, x_{\overline {{\cal K}'}}^n, z^n) } \prod_{ k \in {\cal S} } p(x_k^n) ~ \leq 2^{n \left[ H(X_{\cal S}| X_{\overline {\cal S}}, X_{\overline {{\cal K}'}}, Z) + \epsilon \right] } \prod_{ k \in {\cal S} } 2^{- n \left[ H(X_k) - \epsilon \right] } \nonumber\\
& = 2^{n \left[ H(X_{\cal S}| X_{\overline {\cal S}}, X_{\overline {{\cal K}'}}, Z) + \epsilon \right] } \prod_{ k \in {\cal S} } 2^{- n \left[ H(X_k| X_{\overline {\cal S}}, X_{\overline {{\cal K}'}}) - \epsilon \right] } ~ = 2^{- n \left[ I(X_{\cal S}; Z| X_{\overline {\cal S}}, X_{\overline {{\cal K}'}}) - (|{\cal S}| + 1) \epsilon \right]} \nonumber\\
& \leq 2^{- n \left[ I(X_{\cal S}; Z| X_{\overline {\cal S}}, X_{\overline {{\cal K}'}}) - (K + 1) \epsilon \right] }, ~\forall~ {\cal S} \subseteq {\cal K}',~ (x_{\overline {\cal S}}^n, x_{\overline {{\cal K}'}}^n, z^n) \in {\cal T}_\epsilon^{(n)} (X_{\overline {\cal S}}, X_{\overline {{\cal K}'}}, Z).
\end{align}
For the special ${\cal S} = {\cal K}'$ case, if $(x_{\overline {{\cal K}'}}^n, z^n) \in {\cal T}_\epsilon^{(n)} (X_{\overline {{\cal K}'}}, Z)$, (\ref{p_S_up}) becomes
\begin{align}\label{p_K_up}
p_{{\cal K}'} = {\text {Pr}} \left\{ X_{{\cal K}'}^n \in {\cal T}_\epsilon^{(n)} (X_{{\cal K}'}| x_{\overline {{\cal K}'}}^n, z^n) \right\} ~ \leq 2^{-n \left[ I(X_{{\cal K}'}; Z| X_{\overline {{\cal K}'}}) - (K + 1) \epsilon \right]}.
\end{align}
Then, using $O' (l_{{\cal K}'}| x_{\overline {{\cal K}'}}^n, z^n)$ defined in (\ref{indicator_E0}),
we have 
\begin{align}\label{N_expectation}
{\mathbb E} \left[ Q (m_{{\cal K}'}^{\text s}| x_{\overline {{\cal K}'}}^n, z^n) \right] = & \sum_{ l_{{\cal K}'} \in \prod_{ k \in {\cal K}' } {\cal L}_{k,m_k^{\text s}} } {\mathbb E} \left[ O' (l_{{\cal K}'}| x_{\overline {{\cal K}'}}^n, z^n) \right] ~ = \sum_{ l_{{\cal K}'} \in \prod_{ k \in {\cal K}' } {\cal L}_{k,m_k^{\text s}} } p_{{\cal K}'} \nonumber\\
= & \left| \prod_{ k \in {\cal K}' } {\cal L}_{k,m_k^{\text s}} \right| p_{{\cal K}'} ~ = 2^{ \sum_{k \in {\cal K}'} n ( R_k^{\text o} + R_k^{\text g} ) } p_{{\cal K}'} ~ \leq 2^{n \left[ \varOmega_{{\cal K}'} + (K + 1) \epsilon \right]},
\end{align}
and 
\begin{align}\label{N_square}
& {\mathbb E} \left[( Q (m_{{\cal K}'}^{\text s}| x_{\overline {{\cal K}'}}^n, z^n) )^2\right] = {\mathbb E} \left[ \left( \sum_{ l_{{\cal K}'} \in \prod_{ k \in {\cal K}' } {\cal L}_{k,m_k^{\text s}} } O' (l_{{\cal K}'}| x_{\overline {{\cal K}'}}^n, z^n) \right)^2 \right] \nonumber\\
& = {\mathbb E} \left[ \left( \sum_{ l_{{\cal K}'} \in \prod_{ k \in {\cal K}' } {\cal L}_{k,m_k^{\text s}} } O' (l_{{\cal K}'}| x_{\overline {{\cal K}'}}^n, z^n) \right) \times \left( \sum_{ {\hat l}_{{\cal K}'} \in \prod_{ k \in {\cal K}' } {\cal L}_{k,m_k^{\text s}} } O' ({\hat l}_{{\cal K}'}| x_{\overline {{\cal K}'}}^n, z^n) \right) \right] \nonumber\\
& = \sum_{ l_{{\cal K}'} \in \prod_{ k \in {\cal K}' } {\cal L}_{k,m_k^{\text s}} } {\mathbb E} \left[ O' (l_{{\cal K}'}| x_{\overline {{\cal K}'}}^n, z^n) \times \left( \sum_{ {\hat l}_{{\cal K}'} \in \prod_{ k \in {\cal K}' } {\cal L}_{k,m_k^{\text s}} } O' ({\hat l}_{{\cal K}'}| x_{\overline {{\cal K}'}}^n, z^n) \right) \right] \nonumber\\
& = \sum_{ l_{{\cal K}'} \in \prod_{ k \in {\cal K}' } {\cal L}_{k,m_k^{\text s}} }\!\!\!\!\! {\text {Pr}} \left\{ O' (l_{{\cal K}'}| x_{\overline {{\cal K}'}}^n, z^n) \!=\! 1 \right\} \!\left\{ \sum_{ {\hat l}_{{\cal K}'} \in \prod_{ k \in {\cal K}' } {\cal L}_{k,m_k^{\text s}} }\!\!\!\!\! {\text {Pr}} \left\{\! O' ({\hat l}_{{\cal K}'}| x_{\overline {{\cal K}'}}^n, z^n) \!=\! 1| O' (l_{{\cal K}'}| x_{\overline {{\cal K}'}}^n, z^n) \!=\! 1 \!\right\} \!\!\right\}\nonumber\\
& = \sum_{ l_{{\cal K}'} \in \prod_{ k \in {\cal K}' } {\cal L}_{k,m_k^{\text s}} } {\text {Pr}} \left\{ \left( x_k^n (l_k), \forall k \in {\cal K}' \right) \in {\cal T}_\epsilon^{(n)} (X_{{\cal K}'}| x_{\overline {{\cal K}'}}^n, z^n) \right\} \times \nonumber\\
& \Bigg\{ 1 + \sum\limits_{{\cal S} \subseteq {\cal K}',~ {\cal S} \neq \phi} ~ \sum_{ {\hat l}_{\cal S} \in \prod_{ k \in {\cal S} } {\cal L}_{k,m_k^{\text s}},~ {\hat l}_k \neq l_k,~ \forall k \in {\cal S} }
{\text {Pr}} \left\{ \left( x_k^n ({\hat l}_k), \forall k \in {\cal S} \right) \in {\cal T}_\epsilon^{(n)} (X_{\cal S}| x_{\overline {\cal S}}^n, x_{\overline {{\cal K}'}}^n, z^n) \right\} \Bigg\}\nonumber\\
& = 2^{ \sum_{k \in {\cal K}'} n ( R_k^{\text o} + R_k^{\text g} ) } p_{{\cal K}'} \left[ 1 + \sum\limits_{{\cal S} \subseteq {\cal K}',~ {\cal S} \neq \phi}~ \prod\limits_{k \in {\cal S}} (2^{n (R_k^{\text o} + R_k^{\text g})} - 1) p_{\cal S} \right]\nonumber\\
& \leq 2^{ \sum_{k \in {\cal K}'} n ( R_k^{\text o} + R_k^{\text g} ) } p_{{\cal K}'} \left[ 1 + \sum\limits_{{\cal S} \subseteq {\cal K}',~ {\cal S} \neq \phi}~ 2^{ \sum_{k \in {\cal S}} n ( R_k^{\text o} + R_k^{\text g} ) } p_{\cal S} \right]\nonumber\\
& = 2^{ \sum_{k \in {\cal K}'} n ( R_k^{\text o} + R_k^{\text g} ) } p_{{\cal K}'} \left[ 1 + \sum\limits_{ {\cal S} \subsetneqq {\cal K}',~ {\cal S} \neq \phi }~ 2^{ \sum_{k \in {\cal S}} n ( R_k^{\text o} + R_k^{\text g} ) } p_{\cal S} + 2^{ \sum_{k \in {\cal K}'} n ( R_k^{\text o} + R_k^{\text g} ) } p_{{\cal K}'} \right]\nonumber\\
& \leq 2^{n \left[ \varOmega_{{\cal K}'} + (K + 1) \epsilon \right]} + \sum\limits_{ {\cal S} \subsetneqq {\cal K}',~ {\cal S} \neq \phi } 2^{n \left[ 2 \varOmega_{{\cal K}'} - \varOmega_{\overline {\cal S}} + 2 (K + 1) \epsilon \right]} + \left\{ {\mathbb E} [ Q (m_{{\cal K}'}^{\text s}| x_{\overline {{\cal K}'}}^n, z^n) ] \right\}^2,
\end{align}
where the last step follows from (\ref{N_expectation}) and using (\ref{p_S_up}) as well as (\ref{p_K_up}) to get
\begin{align}
& 2^{ \sum_{k \in {\cal K}'} n ( R_k^{\text o} + R_k^{\text g} ) } p_{{\cal K}'} 2^{ \sum_{k \in {\cal S}} n ( R_k^{\text o} + R_k^{\text g} ) } p_{\cal S} \leq 2^{n \left[ \varOmega_{{\cal K}'} + (K + 1) \epsilon \right]} 2^{ n \left[ \sum_{k \in {\cal S}} ( R_k^{\text o} + R_k^{\text g} ) - I(X_{\cal S}; Z| X_{\overline {\cal S}}, X_{\overline {{\cal K}'}}) + (K + 1) \epsilon \right] } \nonumber\\
& = 2^{n \left[ \varOmega_{{\cal K}'} + (K + 1) \epsilon \right]} 2^{ n \left[ \sum_{k \in {\cal K}'} ( R_k^{\text o} + R_k^{\text g} ) - I(X_{{\cal K}'}; Z| X_{\overline {{\cal K}'}}) - \sum_{k \in {\overline {\cal S}}} ( R_k^{\text o} + R_k^{\text g} ) + I(X_{\overline {\cal S}}; Z| X_{\overline {{\cal K}'}}) + (K + 1) \epsilon \right] } \nonumber\\
& = 2^{n \left[ 2 \varOmega_{{\cal K}'} - \varOmega_{\overline {\cal S}} + 2 (K + 1) \epsilon \right]}, ~\forall~ {\cal S} \subsetneqq {\cal K}',~ {\cal S} \neq \phi.
\end{align}
According to (\ref{N_expectation}) and (\ref{N_square}),
\begin{align}\label{VarN1}
& {\text {Var}} \left[ Q (m_{{\cal K}'}^{\text s}| x_{\overline {{\cal K}'}}^n, z^n) \right] = {\mathbb E} \left[( Q (m_{{\cal K}'}^{\text s}| x_{\overline {{\cal K}'}}^n, z^n) )^2\right] - \left\{{\mathbb E} [ Q (m_{{\cal K}'}^{\text s}| x_{\overline {{\cal K}'}}^n, z^n) ]\right\}^2 \nonumber\\
& \leq 2^{n \left[ \varOmega_{{\cal K}'} + (K + 1) \epsilon \right]} + \sum\limits_{ {\cal S} \subsetneqq {\cal K}',~ {\cal S} \neq \phi } 2^{n \left[ 2 \varOmega_{{\cal K}'} - \varOmega_{\overline {\cal S}} + 2 (K + 1) \epsilon \right]} 
~ \leq \sum\limits_{ {\cal S} \subsetneqq {\cal K}' } 2^{n \left[ 2 \varOmega_{{\cal K}'} - \varOmega_{\overline {\cal S}} + 2 (K + 1) \epsilon \right]}.
\end{align}
Theorem \ref{exp_var_Q} is thus proven.

\section{Proof of Lemma~\ref{lemma_polytope}}
\label{prove_lemma_polytope}

We first prove (\ref{assumption_4}).
Using the definitions of ${\cal K}''$ and ${\overline {{\cal K}''}}$ given in (\ref{K_prime}) and the chain rule of mutual information, the inequality (\ref{assumption_62}) can be rewritten as 
\begin{align}\label{assumption_62_rewrite}
& I(X_{{\cal K}_0 \cup {\cal S}}; Y| X_{{\cal K}' \setminus ({\cal K}_0 \cup {\cal S})}, X_{\overline {{\cal K}'}}) - I(X_{{\cal K}_0 \cup {\cal S}}; Z| X_{\overline {{\cal K}'}}) \nonumber\\
= & I(X_{{\cal K}_0}, X_{\cal S}; Y| X_{\overline {\cal S}}, X_{\overline {{\cal K}'}}) - I(X_{{\cal K}_0}, X_{\cal S}; Z| X_{\overline {{\cal K}'}}) \nonumber\\
= & I(X_{{\cal K}_0}; Y| X_{{\cal K}''}, X_{\overline {{\cal K}'}}) - I(X_{{\cal K}_0}; Z| X_{\overline {{\cal K}'}}) + I(X_{\cal S}; Y| X_{\overline {\cal S}}, X_{\overline {{\cal K}'}}) - I(X_{\cal S}; Z| X_{{\overline {{\cal K}'}} \cup {\cal K}_0}) \nonumber\\
= & I(X_{{\cal K}_0}; Y| X_{{\cal K}' \setminus {\cal K}_0}, X_{\overline {{\cal K}'}}) - I(X_{{\cal K}_0}; Z| X_{\overline {{\cal K}'}}) + I(X_{\cal S}; Y| X_{\overline {\cal S}}, X_{\overline {{\cal K}'}}) - I(X_{\cal S}; Z| X_{\overline {{\cal K}''}}) \nonumber\\
> & 0, ~\forall~ {\cal S} \subseteq {\cal K}'',~ {\cal S} \neq \phi.
\end{align}
Due to (\ref{assumption_61}), it is clear from (\ref{assumption_62_rewrite}) that
\begin{equation}
I(X_{\cal S}; Y| X_{\overline {\cal S}}, X_{\overline {{\cal K}'}}) - I(X_{\cal S}; Z| X_{\overline {{\cal K}''}}) > 0, ~\forall~ {\cal S} \subseteq {\cal K}'',~ {\cal S} \neq \phi,
\end{equation}
based on which we get
\begin{align}\label{assumption_64}
& I(X_{\cal S}, X_{\cal T}; Y| X_{\overline {\cal S}}, X_{\overline {\cal T}}) - I(X_{{\cal S}'}; Z| X_{\overline {{\cal K}''}}) \geq I(X_{\cal S}; Y| X_{\overline {\cal S}}, X_{\overline {{\cal K}''}}) - I(X_{\cal S}; Z| X_{\overline {{\cal K}''}}) \nonumber\\
& \geq I(X_{\cal S}; Y| X_{\overline {\cal S}}, X_{\overline {{\cal K}'}}) - I(X_{\cal S}; Z| X_{\overline {{\cal K}''}})
~ > 0, ~\forall~ {\cal S} \subseteq {\cal K}'',~ {\cal S} \neq \phi,~ {\cal S}' \subseteq {\cal S},~ {\cal T} \subseteq {\overline {{\cal K}''}}.
\end{align}
(\ref{assumption_64}) shows that (\ref{assumption_4}) is true when ${\cal S} \neq \phi$.
If ${\cal S} = \phi$, (\ref{assumption_4}) becomes 
\begin{equation}\label{assumption_65}
I(X_{\cal T}; Y| X_{{\cal K}''}, X_{\overline {\cal T}}) > 0, ~\forall~ {\cal T} \subseteq {\overline {{\cal K}''}},~ {\cal T} \neq \phi,
\end{equation}
which is also true due to assumption (\ref{assumption}).
Combining (\ref{assumption_64}) and (\ref{assumption_65}), we get (\ref{assumption_4}).

Next, we show that for any rate tuple in region ${\mathscr R} (X_{\cal K}, {\cal K}')$ defined by Theorem~\ref{lemma_DM_exten}, if (\ref{assumption_61}) and (\ref{assumption_62}) can be satisfied, it is also in region ${\mathscr R} (X_{\cal K}, {\cal K}'')$.

If $(R_1^{\text s}, R_1^{\text o},\cdots, R_K^{\text s}, R_K^{\text o})$ is in region ${\mathscr R} (X_{\cal K}, {\cal K}')$ and satisfies (\ref{assumption_61}), we have
\begin{equation}\label{Rs_zero_K0}
R_k^{\text s} = 0, ~\forall~ k \in {\overline {{\cal K}''}}.
\end{equation}
To prove that this rate tuple is also in region ${\mathscr R} (X_{\cal K}, {\cal K}'')$, we need to further verify the upper bounds on $\sum_{k \in \cal S} R_k^{\text s} + \sum_{k \in {\cal S} \setminus {\cal S}'} R_k^{\text o} + \sum_{k \in {\cal T}} R_k^{\text o}$ for all possible set choices given in (\ref{region_DM1}).
To that end, we separately prove
\begin{align}\label{part1}
\sum\limits_{k \in \cal S} R_k^{\text s} + \sum\limits_{k \in {\cal S} \setminus {\cal S}'} R_k^{\text o} + \sum\limits_{k \in {\cal T}} R_k^{\text o} & \leq I(X_{\cal S}, X_{\cal T}; Y| X_{\overline {\cal S}}, X_{{\overline {{\cal K}'}} \setminus {\cal T}}, X_{{\cal K}_0}) - I(X_{{\cal S}'}; Z| X_{\overline {{\cal K}''}}), \nonumber\\
& ~\forall~ {\cal S} \subseteq {\cal K}'',~ {\cal S}' \subseteq {\cal S},~ {\cal T} \subseteq {\overline {{\cal K}'}},
\end{align}
and
\begin{align}\label{part2}
\sum\limits_{k \in \cal S} R_k^{\text s} + \sum\limits_{k \in {\cal S} \setminus {\cal S}'} R_k^{\text o} + \sum\limits_{k \in {\cal T}} R_k^{\text o} & \leq I(X_{\cal S}, X_{\cal T}; Y| X_{\overline {\cal S}}, X_{{\overline {{\cal K}''}} \setminus {\cal T}}) - I(X_{{\cal S}'}; Z| X_{\overline {{\cal K}''}}), \nonumber\\
& ~\forall~ {\cal S} \subseteq {\cal K}'',~ {\cal S}' \subseteq {\cal S},~ {\cal T} \subseteq {\overline {{\cal K}''}},~ {\cal T} \cap {\cal K}_0 \neq \phi.
\end{align}
Note that ${\cal T}$ in (\ref{part1}) and (\ref{part2}) belongs to different sets.
To avoid ambiguity, instead of using $\overline {\cal T}$, we use its definition directly in (\ref{part1}) and (\ref{part2}).

We first prove (\ref{part1}).
Using (\ref{Rs_zero_K0}), the left-hand side term of (\ref{part1}) can be rewritten as
\begin{align}\label{prove_part1}
& \sum_{k \in \cal S} R_k^{\text s} + \sum_{k \in {\cal S} \setminus {\cal S}'} R_k^{\text o} + \sum_{k \in {\cal T}} R_k^{\text o} = \sum_{k \in \cal S} R_k^{\text s} + \sum_{k \in {\cal K}_0} R_k^{\text s} + \sum_{k \in {\cal S} \setminus {\cal S}'} R_k^{\text o} + \sum_{k \in {\cal T}} R_k^{\text o} \nonumber\\
& = \sum_{k \in {\cal S} \cup {\cal K}_0} R_k^{\text s} + \sum_{k \in {\cal S} \cup {\cal K}_0 \setminus ({\cal S}' \cup {\cal K}_0)} R_k^{\text o} + \sum_{k \in {\cal T}} R_k^{\text o} \nonumber\\
& \overset{(a)}{\leq} I(X_{{\cal S} \cup {\cal K}_0}, X_{\cal T}; Y| X_{{\cal K}' \setminus ({\cal S} \cup {\cal K}_0)}, X_{{\overline {{\cal K}'}} \setminus {\cal T}}) - I(X_{{\cal S}' \cup {\cal K}_0}; Z| X_{\overline {{\cal K}'}}) \nonumber\\
& = I(X_{{\cal K}_0}; Y| X_{{\cal K}' \setminus {\cal K}_0}, X_{\overline {{\cal K}'}}) - I(X_{{\cal K}_0}; Z| X_{\overline {{\cal K}'}}) \nonumber\\
& + I(X_{\cal S}, X_{\cal T}; Y| X_{{\cal K}' \setminus ({\cal S} \cup {\cal K}_0)}, X_{{\overline {{\cal K}'}} \setminus {\cal T}}) - I(X_{{\cal S}'}; Z| X_{\overline {{\cal K}'}}, X_{{\cal K}_0}) \nonumber\\
& \overset{(b)}{\leq} I(X_{\cal S}, X_{\cal T}; Y| X_{{\cal K}' \setminus ({\cal S} \cup {\cal K}_0)}, X_{{\overline {{\cal K}'}} \setminus {\cal T}}) - I(X_{{\cal S}'}; Z| X_{\overline {{\cal K}'}}, X_{{\cal K}_0}) \nonumber\\
& \overset{(c)}{=} I(X_{\cal S}, X_{\cal T}; Y| X_{\overline {\cal S}}, X_{{\overline {{\cal K}'}} \setminus {\cal T}}) - I(X_{{\cal S}'}; Z| X_{\overline {{\cal K}''}}) \nonumber\\
& \overset{(d)}{\leq} I(X_{\cal S}, X_{\cal T}; Y| X_{\overline {\cal S}}, X_{{\overline {{\cal K}'}} \setminus {\cal T}}, X_{{\cal K}_0}) - I(X_{{\cal S}'}; Z| X_{\overline {{\cal K}''}}), ~\forall~ {\cal S} \subseteq {\cal K}'',~ {\cal S}' \subseteq {\cal S},~ {\cal T} \subseteq {\overline {{\cal K}'}},
\end{align}
where $(a)$ holds since $(R_1^{\text s}, R_1^{\text o},\cdots, R_K^{\text s}, R_K^{\text o})$ is in region ${\mathscr R} (X_{\cal K}, {\cal K}')$ and thus satisfies (\ref{region_DM_exten}), $(b)$ follows from using (\ref{assumption_61}), $(c)$ is true since ${\overline {\cal S}} = {\cal K}'' \setminus {\cal S} = {\cal K}' \setminus ({\cal S} \cup {\cal K}_0)$ and ${\overline {{\cal K}''}} = {\overline {{\cal K}'}} \cup {\cal K}_0$, and $(d)$ holds due to the fact that $X_k, \forall k \in {\cal K}$ are independent of each other.
(\ref{part1}) is thus proven.

Next we prove (\ref{part2}).
If ${\cal T} \subseteq {\overline {{\cal K}''}}$ and ${\cal T} \cap {\cal K}_0 \neq \phi$, ${\cal T}$ can be divided into two disjoint subsets, ${\cal T}_1$ and ${\cal T}_2$, with ${\cal T}_1 \subseteq {\overline {{\cal K}'}}$ and ${\cal T}_2 \subseteq {\cal K}_0$.
As defined in (\ref{assumption_61}), ${\cal K}_0 \subsetneqq {\cal K}'$.
Hence, ${\cal T}_2 \subseteq {\cal K}'$.
Since the rate tuple $(R_1^{\text s}, R_1^{\text o},\cdots, R_K^{\text s}, R_K^{\text o})$ is in region ${\mathscr R} (X_{\cal K}, {\cal K}')$ and thus satisfies (\ref{region_DM_exten}).
By setting ${\cal S} = {\cal T}_2$, ${\cal S}' = \phi$, and ${\cal T} = \phi$ in (\ref{region_DM_exten}), and using (\ref{Rs_zero_K0}), we have
\begin{align}\label{prove_part2_1}
\sum_{k \in {\cal T}_2} R_k^{\text o} = \sum_{k \in {\cal T}_2} R_k^{\text s} + \sum_{k \in {\cal T}_2} R_k^{\text o} 
~ \leq I(X_{{\cal T}_2}; Y| X_{{\cal K}' \setminus {\cal T}_2}, X_{\overline {{\cal K}'}}) 
~ = I(X_{{\cal T}_2}; Y| X_{{\cal K}''}, X_{\overline {{\cal K}'}}, X_{{\cal K}_0 \setminus {\cal T}_2}).
\end{align}
Then,
\begin{align}\label{prove_part2}
& \sum_{k \in \cal S} R_k^{\text s} + \sum_{k \in {\cal S} \setminus {\cal S}'} R_k^{\text o} + \sum_{k \in {\cal T}} R_k^{\text o} = \sum_{k \in \cal S} R_k^{\text s} + \sum_{k \in {\cal S} \setminus {\cal S}'} R_k^{\text o} + \sum_{k \in {\cal T}_1} R_k^{\text o} + \sum_{k \in {\cal T}_2} R_k^{\text o} \nonumber\\
\leq & I(X_{\cal S}, X_{{\cal T}_1}; Y| X_{\overline {\cal S}}, X_{{\overline {{\cal K}'}} \setminus {\cal T}_1}) - I(X_{{\cal S}'}; Z| X_{\overline {{\cal K}''}}) + I(X_{{\cal T}_2}; Y| X_{{\cal K}''}, X_{\overline {{\cal K}'}}, X_{{\cal K}_0 \setminus {\cal T}_2})\nonumber\\
\leq & I(X_{\cal S}, X_{{\cal T}_1}; Y| X_{\overline {\cal S}}, X_{{\overline {{\cal K}'}} \setminus {\cal T}_1}, X_{{\cal K}_0 \setminus {\cal T}_2}) - I(X_{{\cal S}'}; Z| X_{\overline {{\cal K}''}}) + I(X_{{\cal T}_2}; Y| X_{{\cal K}''}, X_{\overline {{\cal K}'}}, X_{{\cal K}_0 \setminus {\cal T}_2})\nonumber\\
= & I(X_{\cal S}, X_{{\cal T}_1}, X_{{\cal T}_2}; Y| X_{\overline {\cal S}}, X_{{\overline {{\cal K}'}} \setminus {\cal T}_1}, X_{{\cal K}_0 \setminus {\cal T}_2}) - I(X_{{\cal S}'}; Z| X_{\overline {{\cal K}''}}) \nonumber\\
= & I(X_{\cal S}, X_{{\cal T}}; Y| X_{\overline {\cal S}}, X_{{\overline {{\cal K}''}} \setminus {\cal T}}) - I(X_{{\cal S}'}; Z| X_{\overline {{\cal K}''}}), ~\forall {\cal S} \subseteq {\cal K}'', {\cal S}' \subseteq {\cal S}, {\cal T} \subseteq {\overline {{\cal K}''}}, {\cal T} \cap {\cal K}_0 \neq \phi,
\end{align}
where the first inequality follows from using (\ref{prove_part1}) and (\ref{prove_part2_1}).
Note that here we use $(c)$ in (\ref{prove_part1}) instead of $(d)$.
(\ref{part2}) is thus proven.
Combining (\ref{Rs_zero_K0}), (\ref{part1}), and (\ref{part2}), it is known that $(R_1^{\text s}, R_1^{\text o},\cdots, R_K^{\text s}, R_K^{\text o})$ satisfies (\ref{region_DM1}) and is thus also in ${\mathscr R} (X_{\cal K}, {\cal K}'')$.
Lemma~\ref{lemma_polytope} is then proven.

\section{}
\label{specific_j}

In this appendix, we consider a specific user $j$ in ${\cal K}$ and show that if $R_j^{\text s} = 0$, there is a trade-off in terms of the resulted achievable regions between cases $j \in {\cal K}'$ and $j \in {\overline {{\cal K}'}}$.
For convenience, assume that all the other users ${\cal K} \setminus \{j\}$ are in ${\cal K}'$.

In the first case with $j \in {\cal K}'$, ${\cal K}' = {\cal K}$ and ${\overline {{\cal K}'}} = \phi$.
For convenience, assume 
\begin{equation}
I(X_{\cal S}; Y| X_{\overline {\cal S}}) \geq I(X_{\cal S}; Z), ~\forall~ {\cal S} \subseteq {\cal K},
\end{equation}
which guarantees
\begin{equation}
I(X_{\cal S}; Y| X_{\overline {\cal S}}) \geq I(X_{{\cal S}'}; Z), ~\forall~ {\cal S} \subseteq {\cal K},~ {\cal S}' \subseteq {\cal S}.
\end{equation}
Then, it is known from Lemma~\ref{theorem_FM} that for any rate tuple $(R_1^{\text s}, R_1^{\text o},\cdots, R_K^{\text s}, R_K^{\text o})$ satisfying
\begin{equation}\label{FM_project1}
\left\{
\begin{array}{ll}
R_j^{\text s} = 0, \\
\sum\limits_{k \in \cal S} R_k^{\text s} + \sum\limits_{k \in {\cal S} \setminus {\cal S}'} R_k^{\text o} \leq I(X_{\cal S}; Y| X_{\overline {\cal S}}) - I(X_{{\cal S}'}; Z), ~\forall~ {\cal S} \subseteq {\cal K},~ {\cal S}' \subseteq {\cal S},
\end{array} \right.
\end{equation}
there exist $R_k^{\text g}, \forall k \in {\cal K}$ such that
\begin{equation}\label{region_FM_j1}
\left\{
\begin{array}{ll}
R_j^{\text s} = 0, \\
R_k^{\text g} \geq 0, ~\forall~ k \in {\cal K}, \\
\sum\limits_{k \in {\cal S}} (R_k^{\text s} + R_k^{\text o} + R_k^{\text g}) \leq I(X_{\cal S}; Y| X_{\overline {\cal S}}), ~\forall~ {\cal S} \subseteq {\cal K}, \\
\sum\limits_{k \in {\cal S}} (R_k^{\text o} + R_k^{\text g}) \geq I(X_{\cal S}; Z), ~\forall~ {\cal S} \subseteq {\cal K}.
\end{array} \right.
\end{equation}
As shown in Theorem~\ref{lemma_DM_exten}, (\ref{FM_project1}) actually constructs the achievable region ${\mathscr R} (X_{\cal K}, {\cal K})$ (with $R_j^{\text s} = 0$).
Note that due to $\sum_{k \in {\cal S}} (R_k^{\text o} + R_k^{\text g}) \geq I(X_{\cal S}; Z), \forall {\cal S} \subseteq {\cal K}$, which considers the combination of all users, including user $j$, Eve cannot decode the open message $M_j^{\text o}$.
Hence, in the achievability proof of (\ref{FM_project1}), we use metric $\frac{1}{n} I (M_{\cal K}^{\text s}; Z^n)$ and hope it vanishes as $n \rightarrow \infty$.

In (\ref{FM_project1}), there are three cases on whether $j$ belongs to ${\cal S}$ and ${\cal S}'$ or not, i.e., 1) $j \notin {\cal S}$, 2) $j \in {\cal S}$ and $j \notin {\cal S}'$, 3) $j \in {\cal S}$ and $j \in {\cal S}'$.
Note that here we mean ${\cal S}$ and ${\cal S}'$ in (\ref{FM_project1}), which may be different from those in somewhere else, e.g., those in (\ref{FM_project2}).
By considering these cases separately, the inequalities in (\ref{FM_project1}) can be divided into three parts as follows
\begin{subequations}\label{FM_project2}
	\begin{align}
	& \sum\limits_{k \in \cal S} R_k^{\text s} + \sum\limits_{k \in {\cal S} \setminus {\cal S}'} R_k^{\text o} \leq I(X_{\cal S}; Y| X_{\overline {\cal S}}, X_j) - I(X_{{\cal S}'}; Z), ~\forall~ {\cal S} \subseteq {\cal K} \setminus \{j\},~ {\cal S}' \subseteq {\cal S}, \label{FM_project2_1} \\
	& \sum\limits_{k \in \cal S} R_k^{\text s} + \sum\limits_{k \in {\cal S} \setminus {\cal S}'} R_k^{\text o} + R_j^{\text o} \leq I(X_{\cal S}, X_j; Y| X_{\overline {\cal S}}) - I(X_{{\cal S}'}; Z), ~\forall~ {\cal S} \subseteq {\cal K} \setminus \{j\},~ {\cal S}' \subseteq {\cal S},~ \label{FM_project2_2}\\
	& \sum\limits_{k \in \cal S} R_k^{\text s} + \sum\limits_{k \in {\cal S} \setminus {\cal S}'} R_k^{\text o} \leq I(X_{\cal S}, X_j; Y| X_{\overline {\cal S}}) - I(X_{{\cal S}'}, X_j; Z), ~\forall~ {\cal S} \subseteq {\cal K} \setminus \{j\},~ {\cal S}' \subseteq {\cal S}. \label{FM_project2_3}
	\end{align}
\end{subequations}
(\ref{FM_project2_1}) and (\ref{FM_project2_3}) show that for each choice of ${\cal S} \subseteq {\cal K} \setminus \{j\}$ and ${\cal S}' \subseteq {\cal S}$, there are two bounds on $\sum_{k \in \cal S} R_k^{\text s} + \sum_{k \in {\cal S} \setminus {\cal S}'} R_k^{\text o}$.
We show in the following that compared with the $j \in {\overline {{\cal K}'}}$ case, $(\ref{FM_project2})$ sets looser but more upper bounds to the rates.

Now we consider the second case with $j \in {\overline {{\cal K}'}}$.
In this case, ${\cal K}' = {\cal K} \setminus \{j\}$ and ${\overline {{\cal K}'}} = \{j\}$.
Again for convenience, assume 
\begin{equation}
I(X_{\cal S}; Y| X_{\overline {\cal S}}, X_j) \geq I(X_{\cal S}; Z| X_j), ~\forall~ {\cal S} \subseteq {\cal K} \setminus \{j\},
\end{equation}
which guarantees
\begin{align}
I(X_{\cal S}, X_{\cal T}; Y| X_{\overline {\cal S}}, X_{\overline {\cal T}}) \geq I(X_{{\cal S}'}; Z| X_j), ~\forall~ {\cal S} \subseteq {\cal K} \setminus \{j\},~ {\cal S}' \subseteq {\cal S},~ {\cal T} \subseteq \{j\}.
\end{align}
It is known from Theorem~\ref{lemma_FM_gene_K2} that for any rate tuple $(R_1^{\text s}, R_1^{\text o},\cdots, R_K^{\text s}, R_K^{\text o})$ satisfying
\begin{equation}\label{FM_project3}
\left\{
\begin{array}{ll}
R_j^{\text s} = 0, \\
\sum\limits_{k \in \cal S} R_k^{\text s} + \sum\limits_{k \in {\cal S} \setminus {\cal S}'} R_k^{\text o} + \sum\limits_{k \in {\cal T}} R_k^{\text o}	& \leq I(X_{\cal S}, X_{\cal T}; Y| X_{\overline {\cal S}}, X_{\overline {\cal T}}) - I(X_{{\cal S}'}; Z| X_j), \\
& \forall~ {\cal S} \subseteq {\cal K} \setminus \{j\},~ {\cal S}' \subseteq {\cal S},~ {\cal T} \subseteq \{j\},
\end{array} \right.
\end{equation}
there exist $R_k^{\text g}, \forall k \in {\cal K} \setminus \{j\}$ such that
\begin{equation}\label{region_FM_j2}
\left\{
\begin{array}{ll}
R_j^{\text s} = 0, \\
R_k^{\text g} \geq 0, ~\forall~ k \in {\cal K} \setminus \{j\}, \\
\sum\limits_{k \in {\cal S}} (R_k^{\text s} + R_k^{\text o} + R_k^{\text g}) + \sum\limits_{k \in {\cal T}} R_k^{\text o} \leq I(X_{\cal S}, X_{\cal T}; Y| X_{\overline {\cal S}}, X_{\overline {\cal T}}), ~\forall~ {\cal S} \subseteq {\cal K} \setminus \{j\},~ {\cal T} \subseteq \{j\}, \\
\sum\limits_{k \in {\cal S}} (R_k^{\text o} + R_k^{\text g}) \geq I(X_{\cal S}; Z| X_j), ~\forall~ {\cal S} \subseteq {\cal K} \setminus \{j\}.
\end{array} \right.
\end{equation}
It is obvious from Theorem~\ref{lemma_DM_exten} that (\ref{FM_project3}) constructs the achievable region ${\mathscr R} (X_{\cal K}, {\cal K} \setminus \{j\})$.
Different from the first case, (\ref{region_FM_j2}) ensures $\sum_{k \in {\cal S}} (R_k^{\text o} + R_k^{\text g}) \geq I(X_{\cal S}; Z| X_j), \forall {\cal S} \subseteq {\cal K} \setminus \{j\}$, which does not take user $j$ into account.
It is thus possible for Eve to decode $M_j^{\text o}$.
Hence, $X_j$ is treated as a known information in (\ref{FM_project3}) and (\ref{region_FM_j2}), and the worse metric $\frac{1}{n} I (M_{{\cal K} \setminus \{j\}}^{\text s}; Z^n| M_j^{\text o})$ ( in contrast to $\frac{1}{n} I (M_{{\cal K} \setminus \{j\}}^{\text s}; Z^n)$) has to be used for the achievability proof of (\ref{FM_project3}).

In (\ref{FM_project3}), there are two cases on whether $j$ is in ${\cal T}$ or not, i.e., 1) $j \notin {\cal T}$, 2) $j \in {\cal T}$.
Considering these cases separately, the inequalities in (\ref{FM_project3}) can be divided into two classes as follows
\begin{subequations}\label{FM_project4}
	\begin{align}
	& \sum\limits_{k \in \cal S} R_k^{\text s} + \sum\limits_{k \in {\cal S} \setminus {\cal S}'} R_k^{\text o} \leq I(X_{\cal S}; Y| X_{\overline {\cal S}}, X_j) - I(X_{{\cal S}'}; Z| X_j), ~\forall {\cal S} \subseteq {\cal K} \setminus \{j\}, {\cal S}' \subseteq {\cal S}, \label{FM_project4_1} \\
	& \sum\limits_{k \in \cal S} R_k^{\text s} + \sum\limits_{k \in {\cal S} \setminus {\cal S}'} R_k^{\text o} + R_j^{\text o} \leq I(X_{\cal S}, X_j; Y| X_{\overline {\cal S}}) - I(X_{{\cal S}'}; Z| X_j), \forall {\cal S} \subseteq {\cal K} \setminus \{j\}, {\cal S}' \subseteq {\cal S}. \label{FM_project4_2}
	\end{align}
\end{subequations}
By comparing (\ref{FM_project4_1}) with (\ref{FM_project2_1}) it can be found that they set upper bounds to the same rate sums, but due to the known information $X_j$, (\ref{FM_project4_1}) gives tighter bounds in contrast to (\ref{FM_project2_1}).
Similar observations can also be made by comparing (\ref{FM_project4_2}) with (\ref{FM_project2_2}).
In this sense, adding $j$ to ${\overline {{\cal K}'}}$ shrinks the achievable region.
However, as we can also see, by including user $j$ in ${\cal K}'$, (\ref{FM_project2}) gives many more upper bounds on the rate sums than (\ref{FM_project4}), i.e., (\ref{FM_project2_3}), which is obviously a disadvantage in determining the achievable region.

In conclusion, when $R_j^{\text s} = 0$, there is a trade-off between $j \in {\cal K}'$ and $j \in {\overline {{\cal K}'}}$ in determining the achievable regions.
To get a better region, all possible cases have to be taken into account.

\section{}
\label{K1_2}

In this appendix, we consider a special case with $K = 2$ and $R_k^{\text o} = 0, \forall k \in {\cal K}$, and show that the achievable secrecy rate region provided in Lemma~\ref{achi_secrecy_only} is part of that provided in Lemma~\ref{achi_secrecy_only2}.

For a given joint distribution $\prod_{k=1}^K p(x_k)$, assume that $ I(X_{\cal S}; Y| X_{\overline {\cal S}}) - I(X_{\cal S}; Z| X_{\overline {{\cal K}'}}) \geq 0, \forall {\cal K}' \subseteq {\cal K}, {\cal S} \subseteq {\cal K}'$, for convenience.
When ${\cal K}' = \{1, 2\}$, the inequation systems (\ref{region_secrecy_only}) and (\ref{region_secrecy_only2}) can be written in detail as
\begin{align}\label{region_K_2_K0_empty}
\left\{
\begin{array}{ll}
R_1^{\text s} \leq I(X_1; Y| X_2) - I(X_1; Z), \\
R_2^{\text s} \leq I(X_2; Y| X_1) - I(X_2; Z), \\
R_1^{\text s} + R_2^{\text s} \leq I(X_1, X_2; Y) - I(X_1, X_2; Z),
\end{array} \right.
\end{align}
which constructs regions ${\mathscr R}^{\text s} (X_{\cal K})$ and ${\mathscr R}^{\text s} (X_{\cal K}, \{1, 2\})$.
When ${\cal K}'$ is $\{1\}$ or $\{2\}$, (\ref{region_secrecy_only2}) can be rewritten as
\begin{align}\label{region_K_2_K0_2}
\left\{
\begin{array}{ll} 
R_2^{\text s} = 0, \\
R_1^{\text s} \leq I(X_1; Y| X_2) - I(X_1; Z| X_2),
\end{array} \right.
\end{align}
and
\begin{align}\label{region_K_2_K0_1}
\left\{
\begin{array}{ll}
R_1^{\text s} = 0, \\
R_2^{\text s} \leq I(X_2; Y| X_1) - I(X_2; Z| X_1),
\end{array} \right.
\end{align}
which respectively construct regions ${\mathscr R}^{\text s} (X_{\cal K}, \{1\})$ and ${\mathscr R}^{\text s} (X_{\cal K}, \{2\})$.
The region ${\mathscr R}^{\text s} (X_{\cal K}, {\cal K}')$ over all ${\cal K}'$ is then ${\mathscr R}^{\text s} (X_{\cal K}, \{1, 2\}) \cup {\mathscr R}^{\text s} (X_{\cal K}, \{1\}) \cup {\mathscr R}^{\text s} (X_{\cal K}, \{2\})$.

\begin{figure}
	\begin{minipage}[t]{0.49\linewidth}
		\centering
		\includegraphics[width=3in]{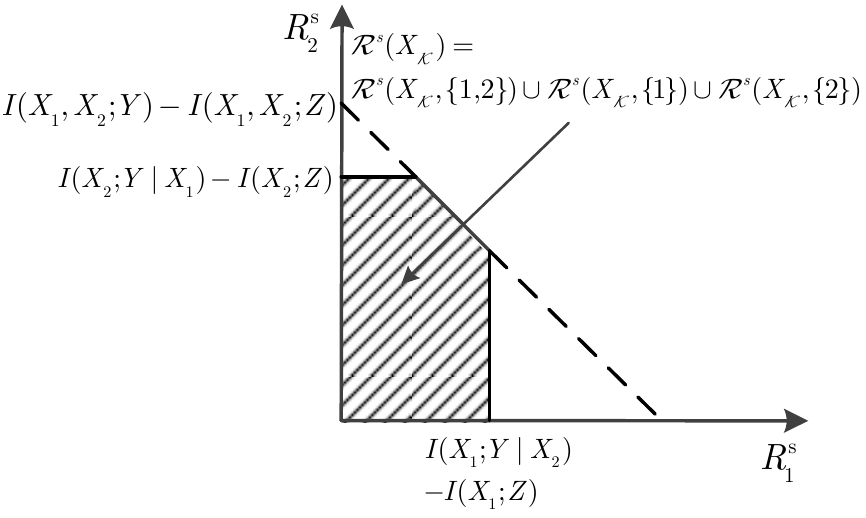}
		\caption{Achievable secrecy rate regions provided by Lemma~\ref{achi_secrecy_only} and Lemma~\ref{achi_secrecy_only2} for a given distribution $\prod_{k=1}^K p(x_k)$ when $K = 2$ and (\ref{leq_min}) holds.}
		\label{rate_region_geq_max}
	\end{minipage}
	\hskip 1ex
	\begin{minipage}[t]{0.49\linewidth}
		\centering
		\includegraphics[width=3in]{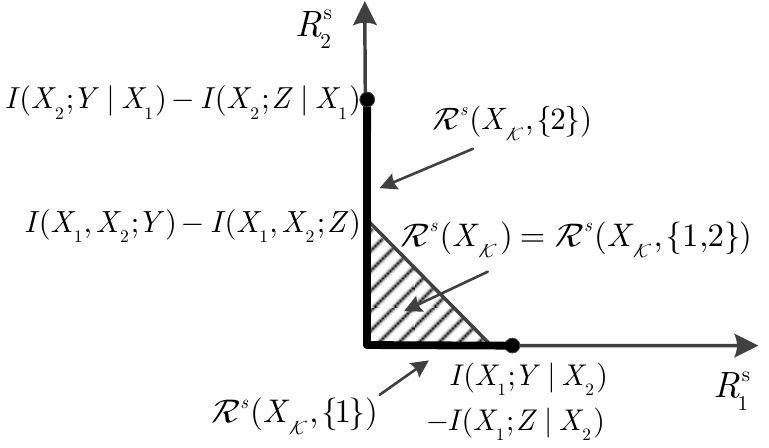}
		\caption{Achievable secrecy rate regions provided by Lemma~\ref{achi_secrecy_only} and Lemma~\ref{achi_secrecy_only2} for a given distribution $\prod_{k=1}^K p(x_k)$ when $K = 2$ and (\ref{leq_min}) holds.}
		\label{rate_region_leq_min}
	\end{minipage}
\end{figure}

If
\begin{align}\label{geq_max}
& I(X_1, X_2; Y) - I(X_1, X_2; Z) \nonumber\\
\geq & \max \left\{ I(X_1; Y| X_2) - I(X_1; Z| X_2), I(X_2; Y| X_1) - I(X_2; Z| X_1) \right\},
\end{align}
it is obvious that the regions ${\mathscr R}^{\text s} (X_{\cal K}, \{1\})$ and ${\mathscr R}^{\text s} (X_{\cal K}, \{2\})$ are included in ${\mathscr R}^{\text s} (X_{\cal K}, \{1, 2\})$.
Hence, as shown by Fig.~\ref{rate_region_geq_max},
\begin{equation}\label{equal}
{\mathscr R}^{\text s} (X_{\cal K}) = {\mathscr R}^{\text s} (X_{\cal K}, \{1, 2\}) \cup {\mathscr R}^{\text s} (X_{\cal K}, \{1\}) \cup {\mathscr R}^{\text s} (X_{\cal K}, \{2\}).
\end{equation}
If
\begin{align}\label{leq_min}
& I(X_1, X_2; Y) - I(X_1, X_2; Z) \nonumber\\
< & \min \left\{ I(X_1; Y| X_2) - I(X_1; Z| X_2), I(X_2; Y| X_1) - I(X_2; Z| X_1) \right\},
\end{align}
it can be seen from Fig.~\ref{rate_region_leq_min} that there exit points in ${\mathscr R}^{\text s} (X_{\cal K}, \{1\})$ and ${\mathscr R}^{\text s} (X_{\cal K}, \{2\})$ which are not included in ${\mathscr R}^{\text s} (X_{\cal K}, \{1, 2\})$.
Hence,
\begin{equation}\label{subsetneqq}
{\mathscr R}^{\text s} (X_{\cal K}) \subsetneqq {\mathscr R}^{\text s} (X_{\cal K}, \{1, 2\}) \cup {\mathscr R}^{\text s} (X_{\cal K}, \{1\}) \cup {\mathscr R}^{\text s} (X_{\cal K}, \{2\}).
\end{equation}
It can be similarly proven that (\ref{subsetneqq}) still holds, if the value of $I(X_1, X_2; Y) - I(X_1, X_2; Z)$ is between those of $I(X_1; Y| X_2) - I(X_1; Z| X_2)$ and $I(X_2; Y| X_1) - I(X_2; Z| X_1)$.
Note that the achievable regions provided by Lemma~\ref{achi_secrecy_only} and Lemma~\ref{achi_secrecy_only2} are respectively the unions of ${\mathscr R}^{\text s} (X_{\cal K})$ and ${\mathscr R}^{\text s} (X_{\cal K}, \{1, 2\}) \cup {\mathscr R}^{\text s} (X_{\cal K}, \{1\}) \cup {\mathscr R}^{\text s} (X_{\cal K}, \{2\})$ over all $\prod_{k=1}^K p(x_k)$.
Hence, the achievable region provided by Lemma~\ref{achi_secrecy_only} is contained in that provided by Lemma~\ref{achi_secrecy_only2}.
In this sense, Lemma~\ref{achi_secrecy_only2} not only improves the result of Lemma~\ref{achi_secrecy_only}, but also that of \cite[Theorem~$1$]{tekin2008gaussian} and \cite[Theorem~$2$]{ekrem2008secrecy}.

\section{Proof of Theorem~\ref{max_R_s_joint}}
\label{Prove_max_R_s_joint}

If $R_k^{\text o} = 0, \forall k \in {\cal K}$, for a given ${\cal K}' \subseteq {\cal K}$, (\ref{region_DM_exten}) becomes (\ref{region_secrecy_only2}), which can be divided into
\begin{equation}\label{region_DM_exten_3}
\left\{
\begin{array}{ll}
R_k^{\text s} = 0, ~\forall~ k \in {\overline {{\cal K}'}}, \\
\sum\limits_{k \in \cal S} R_k^{\text s} \leq \left[I(X_{\cal S}; Y| X_{\overline {\cal S}}, X_{\overline {{\cal K}'}}) - I(X_{\cal S}; Z| X_{\overline {{\cal K}'}}) \right]^+, ~\forall~ {\cal S} \subsetneqq {\cal K}',
\end{array} \right.
\end{equation}
and
\begin{equation}\label{region_DM_exten_8}
\sum\limits_{k \in {\cal K}'} R_k^{\text s} \leq \left[I(X_{{\cal K}'}; Y| X_{\overline {{\cal K}'}}) - I(X_{{\cal K}'}; Z| X_{\overline {{\cal K}'}}) \right]^+.
\end{equation}
As stated in Lemma~\ref{achi_secrecy_only2}, (\ref{region_DM_exten_3}) and (\ref{region_DM_exten_8}) jointly define an achievable secrecy rate region ${\mathscr R}^{\text s} (X_{\cal K}, {\cal K}')$, and (\ref{region_DM_exten_8}) shows that for any secrecy rate tuple in this region, the sum rate $\sum_{k \in {\cal K}'} R_k^{\text s}$ is no larger than $\left[I(X_{{\cal K}'}; Y| X_{\overline {{\cal K}'}}) - I(X_{{\cal K}'}; Z| X_{\overline {{\cal K}'}}) \right]^+$.
We now show that this upper bound is achievable.
To that end, we only need to prove that the inequality (\ref{region_DM_exten_8}) is not redundant, i.e., any linear combination of inequalities in (\ref{region_DM_exten_3}) does not generate (\ref{region_DM_exten_8}) or a tighter upper bound to $\sum_{k \in {\cal K}'} R_k^{\text s}$.
In \cite[Appendix~E]{xu2022achievable}, we have provided the proof for the case with ${\cal K}' = {\cal K}$.
When ${\cal K}' \subsetneqq {\cal K}$, we could complete the proof by following similar steps.
We omit the details here for brevity.
Note that for a given ${\cal K}' \subseteq {\cal K}$, we have $R_k^{\text s} = 0, ~\forall~ k \in {\overline {{\cal K}'}}$.
Hence, the achievable upper bound on $\sum_{k \in {\cal K}} R_k^{\text s}$ given by (\ref{region_DM_exten_3}) and (\ref{region_DM_exten_8}) is $\left[I(X_{{\cal K}'}; Y| X_{\overline {{\cal K}'}}) - I(X_{{\cal K}'}; Z| X_{\overline {{\cal K}'}}) \right]^+$.
Then, considering all possible ${\cal K}' \subseteq {\cal K}$, the maximum achievable sum secrecy rate $\sum_{k \in {\cal K}} R_k^{\text s}$ is
\begin{equation}\label{R_s_joint_DM2}
R^{\text s} (X_{\cal K}) = \mathop {\max }\limits_{{\cal K}' \subseteq {\cal K}} \left\{ \left[ I(X_{{\cal K}'}; Y| X_{\overline {{\cal K}'}}) - I(X_{{\cal K}'}; Z| X_{\overline {{\cal K}'}}) \right]^+ \right\}.
\end{equation}

Let ${\cal K}'^*$ denote the subset in ${\cal K}$ which achieves (\ref{R_s_joint_DM2}) and assume 
\begin{equation}\label{ineq0}
I(X_{{\cal K}'^*}; Y| X_{\overline {{\cal K}'^*}}) - I(X_{{\cal K}'^*}; Z| X_{\overline {{\cal K}'^*}}) > 0,
\end{equation}
since otherwise we have $R_k^{\text s} = 0, \forall k \in {\cal K}$, i.e., the system reduces to a conventional MAC channel with only open messages.
Before proving the second part of Theorem~\ref{max_R_s_joint}, i.e., (\ref{R_o_K1_DM}), we first show that with ${\cal K}'^*$ defined above, we have
\begin{equation}\label{ineq1}
I(X_{\cal S}; Y| X_{\overline {\cal S}}, X_{\overline {{\cal K}'^*}}) - I(X_{\cal S}; Z| X_{\overline {{\cal K}'^*}}) \geq 0, ~\forall~ {\cal S} \subsetneqq {\cal K}'^*.
\end{equation}
(\ref{ineq1}) can be proven by reductio ad absurdum.
If (\ref{ineq1}) is not true, then there exist subsets in ${\cal K}'^*$ such that the corresponding inequalities in (\ref{ineq1}) do not hold.
W.l.o.g., we assume that there exists only one subset ${\cal S}_0$ in ${\cal K}'^*$ such that
\begin{equation}\label{ineq2}
I(X_{{\cal S}_0}; Y| X_{\overline {{\cal S}_0}}, X_{\overline {{\cal K}'^*}}) - I(X_{{\cal S}_0}; Z| X_{\overline {{\cal K}'^*}}) < 0.
\end{equation}
Using the chain rule of mutual information and the fact that $X_k, \forall k \in {\cal K}$ are independent of each other, the left-hand side term of (\ref{ineq0}) is upper bounded by
\begin{align}\label{ineq0_ub}
& I(X_{{\cal K}'^*}; Y| X_{\overline {{\cal K}'^*}}) - I(X_{{\cal K}'^*}; Z| X_{\overline {{\cal K}'^*}}) = I(X_{{\cal K}'^* \setminus {\cal S}_0}, X_{{\cal S}_0}; Y| X_{\overline {{\cal K}'^*}}) - I(X_{{\cal K}'^* \setminus {\cal S}_0}, X_{{\cal S}_0}; Z| X_{\overline {{\cal K}'^*}}) \nonumber\\
& = I(X_{{\cal K}'^* \setminus {\cal S}_0}; Y| X_{{\cal S}_0}, X_{\overline {{\cal K}'^*}}) - I(X_{{\cal K}'^* \setminus {\cal S}_0}; Z| X_{\overline {{\cal K}'^*}}) + I(X_{{\cal S}_0}; Y| X_{\overline {{\cal K}'^*}}) - I(X_{{\cal S}_0}; Z| X_{{\cal K}'^* \setminus {\cal S}_0}, X_{\overline {{\cal K}'^*}}) \nonumber\\
& \leq I(X_{{\cal K}'^* \setminus {\cal S}_0}; Y| X_{{\cal S}_0}, X_{\overline {{\cal K}'^*}}) - I(X_{{\cal K}'^* \setminus {\cal S}_0}; Z| X_{\overline {{\cal K}'^*}}) + I(X_{{\cal S}_0}; Y| X_{\overline {{\cal S}_0}}, X_{\overline {{\cal K}'^*}}) - I(X_{{\cal S}_0}; Z| X_{\overline {{\cal K}'^*}}) \nonumber\\
& < I(X_{{\cal K}'^* \setminus {\cal S}_0}; Y| X_{{\cal S}_0}, X_{\overline {{\cal K}'^*}}) - I(X_{{\cal K}'^* \setminus {\cal S}_0}; Z| X_{\overline {{\cal K}'^*}}),
\end{align}
where the last step holds due to (\ref{ineq2}).
Then, ${\cal K}' = {\cal K}'^* \setminus {\cal S}_0$ and ${\overline {{\cal K}'}} = {\overline {{\cal K}'^*}} \cup {\cal S}_0$ result in a larger value in (\ref{R_s_joint_DM2}) than ${\cal K}'^*$, i.e., $R^{\text s} (X_{\cal K})$ can be increased.
This is contradicted to the assumption that ${\cal K}'^*$ achieves (\ref{R_s_joint_DM2}).
When there are more subsets in ${\cal K}'^*$ such that the inequalities in (\ref{ineq1}) do not hold, we may prove by following similar steps that $R^{\text s} (X_{\cal K})$ can be further increased.
As a result, if ${\cal K}'^* \subseteq {\cal K}$ achieves (\ref{R_s_joint_DM2}) and (\ref{ineq0}) is true, we have (\ref{ineq1}).

Now we show that if users in ${\cal K}'^*$ transmit their confidential messages at sum rate $R^{\text s} (X_{\cal K})$, the maximum achievable sum rate at which users in ${\cal K}$ could send their open messages is given by (\ref{R_o_K1_DM}).
We divide the users in ${\cal K}$ into two classes, i.e., ${\cal K}'^*$ and ${\overline {{\cal K}'^*}}$, and consider their maximum sum open message rate separately below.

First, if $( \{R_k^{\text s}, R_k^{\text o}, \forall k \in {\cal K}'^*\}, \{R_k^{\text s} = 0, \forall k \in {\overline {{\cal K}'^*}}\}, \{R_k^{\text o}, \forall k \in {\overline {{\cal K}'^*}}\} )$ is a rate tuple in region ${\mathscr R} (X_{\cal K}, {\cal K}'^*)$ defined by Theorem~\ref{lemma_DM_exten} and $\sum_{k \in {\cal K}'^*} R_k^{\text s} = R^{\text s} (X_{\cal K})$, which is assumed to be positive, by setting ${\cal S} = {\cal K}'^*$, ${\cal S}' = \phi$, and ${\cal T} = \phi$ in (\ref{region_DM_exten}), we get
\begin{align}\label{R_o_DM_max}
\sum_{k \in {\cal K}'^*} R_k^{\text o} \leq I(X_{{\cal K}'^*}; Y| X_{\overline {{\cal K}'^*}}) - R^{\text s} (X_{\cal K}) 
~ = I(X_{{\cal K}'^*}; Z| X_{\overline {{\cal K}'^*}}),
\end{align}
indicating that the sum rate at which users in ${\cal K}'^*$ can encode their open messages is no larger than $I(X_{{\cal K}'^*}; Z| X_{\overline {{\cal K}'^*}})$.
Then, we prove that this rate is achievable.
Since the rate tuple is in region ${\mathscr R} (X_{\cal K}, {\cal K}'^*)$, with inequalities (\ref{ineq0}) and (\ref{ineq1}), we could use Theorem~\ref{lemma_FM_gene_K2} and find $R_k^{\text g}, \forall k \in {\cal K}'^*$ such that
\begin{equation}\label{Fourier_Motzkin}
\left\{
\begin{array}{ll}
R_k^{\text g} \geq 0, ~\forall~ k \in {\cal K}'^*, \\
\sum\limits_{k \in {\cal S}} (R_k^{\text s} + R_k^{\text o} + R_k^{\text g}) + \sum\limits_{k \in {\cal T}} R_k^{\text o} \leq I(X_{\cal S}, X_{\cal T}; Y| X_{\overline {\cal S}}, X_{\overline {\cal T}}), ~\forall~ {\cal S} \subseteq {\cal K}'^*,~ {\cal T} \subseteq {\overline {{\cal K}'^*}}, \\
\sum\limits_{k \in {\cal S}} (R_k^{\text o} + R_k^{\text g}) \geq I(X_{\cal S}; Z| X_{\overline {{\cal K}'^*}}), ~\forall~ {\cal S} \subseteq {\cal K}'^*.
\end{array} \right.
\end{equation}
It is obvious from (\ref{Fourier_Motzkin}) that if $\sum_{k \in {\cal K}'^*} R_k^{\text o} < I(X_{{\cal K}'^*}; Z| X_{\overline {{\cal K}'^*}})$, we can always split partial rate in $R_k^{\text g}$ to $R_k^{\text o}$, and get ${\hat R}_k^{\text g}$ as well as ${\hat R}_k^{\text o}$, such that
\begin{align}\label{R_o_g_hat}
{\hat R}_k^{\text g} & \geq 0, ~\forall~ k \in {\cal K}'^*, \nonumber\\
{\hat R}_k^{\text o} + {\hat R}_k^{\text g} & = R_k^{\text o} + R_k^{\text g}, ~\forall~ k \in {\cal K}'^*,
\end{align}
and
\begin{equation}\label{max_R_o2}
\sum_{k \in {\cal K}'^*} {\hat R}_k^{\text o} = I(X_{{\cal K}'^*}; Z| X_{\overline {{\cal K}'^*}}).
\end{equation}
With (\ref{R_o_g_hat}), it can be easily verified that rate tuple $( \{R_k^{\text s}, {\hat R}_k^{\text o}, {\hat R}_k^{\text g}, \forall k \in {\cal K}'^*\}, \{R_k^{\text s} = 0, \forall k \in {\overline {{\cal K}'^*}}\}, \{R_k^{\text o}, \forall k \in {\overline {{\cal K}'^*}}\} )$ is in the region defined by (\ref{Fourier_Motzkin}). 
Since the region ${\mathscr R} (X_{\cal K}, {\cal K}'^*)$ can be obtained by projecting (\ref{Fourier_Motzkin}) onto hyperplane $\{ R_k^{\text g} = 0, \forall k \in {\cal K}'^* \}$, then, according to the characteristics of Fourier-Motzkin elimination \cite[Appendix D]{el2011network}, we know that rate tuple $( \{R_k^{\text s}, {\hat R}_k^{\text o}, \forall k \in {\cal K}'^*\}, \{R_k^{\text s} = 0, \forall k \in {\overline {{\cal K}'^*}}\}, \{R_k^{\text o}, \forall k \in {\overline {{\cal K}'^*}}\} )$ is in region ${\mathscr R} (X_{\cal K}, {\cal K}'^*)$.
Due to (\ref{max_R_o2}), with this rate tuple, users in ${\cal K}'^*$ could send their open messages at sum rate $I(X_{{\cal K}'^*}; Z| X_{\overline {{\cal K}'^*}})$.

On the other hand, from the perspective of Bob, we are considering a MAC channel with $K$ users.
Then, it is known from \cite{el2011network} that with a particular input distribution $\prod_{k=1}^K p(x_k)$, an achievable sum rate users in ${\overline {{\cal K}'^*}}$ could send their open messages is $I(X_{\overline {{\cal K}'^*}}; Y)$, which could be achieved by letting Bob jointly decode these users' messages first (before decoding the messages of users in ${\cal K}'^*$).
Hence, the maximum achievable sum rate at which users in ${\overline {{\cal K}'^*}}$ could send their open messages is no less than $I(X_{\overline {{\cal K}'^*}}; Y)$.
In addition, if $\sum_{k \in {\cal K}'^*} R_k^{\text s} = R^{\text s} (X_{\cal K})$, no matter users in ${\cal K}'^*$ transmit open messages or not, the coding scheme provided in Appendix~\ref{prove_lemma_DM_exten} shows that a `garbage' message at rate $R_k^{\text g}$ has to be introduced to each user $k \in {\cal K}'^*$, and
\begin{equation}\label{max_sum_rate_g}
\sum_{k \in {\cal K}'^*} \left( R_k^{\text o} + R_k^{\text g} \right) \geq I(X_{{\cal K}'^*}; Z| X_{\overline {{\cal K}'^*}}),
\end{equation}
has to be guaranteed.
Considering that the maximum achievable sum rate of the MAC link (between all users and Bob) is $I (X_{\cal K}; Y) = I (X_{{\cal K}'^*}, X_{\overline {{\cal K}'^*}}; Y)$, we have
\begin{align}\label{max_sum_rate_o}
\sum_{k \in {\overline {{\cal K}'^*}}} R_k^{\text o} & \leq I (X_{{\cal K}'^*}, X_{\overline {{\cal K}'^*}}; Y) - \sum_{k \in {\cal K}'^*} \left( R_k^{\text s} + R_k^{\text o} + R_k^{\text g} \right) \nonumber\\
& \leq I (X_{{\cal K}'^*}, X_{\overline {{\cal K}'^*}}; Y) - R^{\text s} (X_{\cal K}) - I(X_{{\cal K}'^*}; Z| X_{\overline {{\cal K}'^*}}) ~= I (X_{\overline {{\cal K}'^*}}; Y).
\end{align}
Therefore, the maximum sum open message rate of users in ${\overline {{\cal K}'^*}}$ is $I (X_{\overline {{\cal K}'^*}}; Y)$.

Accordingly, the maximum achievable sum rate at which users in ${\cal K} = {\cal K}'^* \cup {\overline {{\cal K}'^*}}$ could send their open messages is given by (\ref{R_o_K1_DM}).
Theorem~\ref{max_R_s_joint} is thus proven.

\bibliographystyle{IEEEtran}
\bibliography{IEEEabrv,Ref}

\begin{thebibliography}{10}
\providecommand{\url}[1]{#1}
\csname url@samestyle\endcsname
\providecommand{\newblock}{\relax}
\providecommand{\bibinfo}[2]{#2}
\providecommand{\BIBentrySTDinterwordspacing}{\spaceskip=0pt\relax}
\providecommand{\BIBentryALTinterwordstretchfactor}{4}
\providecommand{\BIBentryALTinterwordspacing}{\spaceskip=\fontdimen2\font plus
\BIBentryALTinterwordstretchfactor\fontdimen3\font minus
  \fontdimen4\font\relax}
\providecommand{\BIBforeignlanguage}[2]{{%
\expandafter\ifx\csname l@#1\endcsname\relax
\typeout{** WARNING: IEEEtran.bst: No hyphenation pattern has been}%
\typeout{** loaded for the language `#1'. Using the pattern for}%
\typeout{** the default language instead.}%
\else
\language=\csname l@#1\endcsname
\fi
#2}}
\providecommand{\BIBdecl}{\relax}
\BIBdecl

\bibitem{xu2022achievable}
H.~Xu, T.~Yang, K.-K. Wong, and G.~Caire, ``Achievable regions and precoder
  designs for the multiple access wiretap channels with confidential and open
  messages,'' \emph{IEEE J. Sel. Areas Commun.}, vol.~40, no.~5, pp.
  1407--1427, May 2022.

\bibitem{yang2015safeguarding}
N.~Yang, L.~Wang, G.~Geraci, M.~Elkashlan, J.~Yuan, and M.~Di~Renzo,
  ``Safeguarding {5G} wireless communication networks using physical layer
  security,'' \emph{IEEE Commun. Mag.}, vol.~53, no.~4, pp. 20--27, Apr. 2015.

\bibitem{7762075}
X.~Chen, D.~W.~K. Ng, W.~H. Gerstacker, and H.-H. Chen, ``A survey on
  multiple-antenna techniques for physical layer security,'' \emph{IEEE Commun.
  Surveys Tuts.}, vol.~19, no.~2, pp. 1027--1053, 2nd Quart., 2017.

\bibitem{shannon1949communication}
C.~E. Shannon, ``Communication theory of secrecy systems,'' \emph{Bell Sys.
  Tech. J.}, vol.~28, no.~4, pp. 656--715, Oct. 1949.

\bibitem{wyner1975wire}
A.~D. Wyner, ``The wire-tap channel,'' \emph{Bell Sys. Tech. J.}, vol.~54,
  no.~8, pp. 1355--1387, Oct. 1975.

\bibitem{leung1978gaussian}
S.~Leung-Yan-Cheong and M.~Hellman, ``The gaussian wire-tap channel,''
  \emph{IEEE Trans. Inf. Theory}, vol.~24, no.~4, pp. 451--456, July 1978.

\bibitem{csiszar1978broadcast}
I.~Csisz{\'a}r and J.~K{\"o}rner, ``Broadcast channels with confidential
  messages,'' \emph{IEEE Trans. Inf. Theory}, vol.~24, no.~3, pp. 339--348, May
  1978.

\bibitem{ekrem2008secrecy}
E.~Ekrem and S.~Ulukus, ``On the secrecy of multiple access wiretap channel,''
  in \emph{Proc. 46th Allerton Conf. Commun., Contr., Comput.}, Illinois, USA,
  Sep. 2008, pp. 1014--1021.

\bibitem{4036106}
R.~Liu, I.~Maric, R.~D. Yates, and P.~Spasojevic, ``The discrete memoryless
  multiple access channel with confidential messages,'' in \emph{Proc. IEEE
  Int. Symp. Inf. Theory (ISIT)}, Seattle, WA, USA, July 2006, pp. 957--961.

\bibitem{4455769}
Y.~Liang and H.~V. Poor, ``Multiple-access channels with confidential
  messages,'' \emph{IEEE Trans. Inf. Theory}, vol.~54, no.~3, pp. 976--1002,
  Mar. 2008.

\bibitem{5961828}
R.~Liu, Y.~Liang, and H.~V. Poor, ``Fading cognitive multiple-access channels
  with confidential messages,'' \emph{IEEE Trans. Inf. Theory}, vol.~57, no.~8,
  pp. 4992--5005, Aug. 2011.

\bibitem{nafea2019generalizing}
M.~Nafea and A.~Yener, ``Generalizing multiple access wiretap and wiretap {II}
  channel models: Achievable rates and cost of strong secrecy,'' \emph{IEEE
  Trans. Inf. Theory}, vol.~65, no.~8, pp. 5125--5143, Aug. 2019.

\bibitem{tekin2008gaussian}
E.~Tekin and A.~Yener, ``The gaussian multiple access wire-tap channel,''
  \emph{IEEE Trans. Inf. Theory}, vol.~54, no.~12, pp. 5747--5755, Dec. 2008.

\bibitem{tekin2008general}
------, ``The general gaussian multiple-access and two-way wiretap channels:
  Achievable rates and cooperative jamming,'' \emph{IEEE Trans. Inf. Theory},
  vol.~54, no.~6, pp. 2735--2751, June 2008.

\bibitem{9174164}
H.~Xu, G.~Caire, and C.~Pan, ``An achievable region for the multiple access
  wiretap channels with confidential and open messages,'' in \emph{Proc. IEEE
  Int. Symp. Inf. Theory (ISIT)}, Los Angeles, CA, USA, Jun. 2020, pp.
  949--954.

\bibitem{5730586}
E.~Ekrem and S.~Ulukus, ``The secrecy capacity region of the gaussian {MIMO}
  multi-receiver wiretap channel,'' \emph{IEEE Trans. Inf. Theory}, vol.~57,
  no.~4, pp. 2083--2114, Apr. 2011.

\bibitem{5550390}
R.~Liu, T.~Liu, H.~V. Poor, and S.~Shamai, ``Multiple-input multiple-output
  gaussian broadcast channels with confidential messages,'' \emph{IEEE Trans.
  Inf. Theory}, vol.~56, no.~9, pp. 4215--4227, Sep. 2010.

\bibitem{5605348}
H.~D. Ly, T.~Liu, and Y.~Liang, ``Multiple-input multiple-output gaussian
  broadcast channels with common and confidential messages,'' \emph{IEEE Trans.
  Inf. Theory}, vol.~56, no.~11, pp. 5477--5487, Nov. 2010.

\bibitem{6584931}
S.~A.~A. Fakoorian and A.~L. Swindlehurst, ``On the optimality of linear
  precoding for secrecy in the {MIMO} broadcast channel,'' \emph{IEEE J. Sel.
  Areas Commun.}, vol.~31, no.~9, pp. 1701--1713, Sep. 2013.

\bibitem{9133130}
X.~Yu, D.~Xu, Y.~Sun, D.~W.~K. Ng, and R.~Schober, ``Robust and secure wireless
  communications via intelligent reflecting surfaces,'' \emph{IEEE J. Sel.
  Areas Commun.}, vol.~38, no.~11, pp. 2637--2652, Nov. 2020.

\bibitem{4529283}
R.~Liu, I.~Maric, P.~Spasojevic, and R.~D. Yates, ``Discrete memoryless
  interference and broadcast channels with confidential messages: Secrecy rate
  regions,'' \emph{IEEE Trans. Inf. Theory}, vol.~54, no.~6, pp. 2493--2507,
  May 2008.

\bibitem{4595013}
O.~O. Koyluoglu, H.~El~Gamal, L.~Lai, and H.~V. Poor, ``On the secure degrees
  of freedom in the ${K}$-user gaussian interference channel,'' in \emph{Proc.
  IEEE Int. Symp. Inf. Theory (ISIT)}, Toronto, ON, Canada, July 2008, pp.
  384--388.

\bibitem{5752448}
X.~Tang, R.~Liu, P.~Spasojević, and H.~V. Poor, ``Interference assisted secret
  communication,'' \emph{IEEE Trans. Inf. Theory}, vol.~57, no.~5, pp.
  3153--3167, May 2011.

\bibitem{6006610}
O.~O. Koyluoglu and H.~El~Gamal, ``Cooperative encoding for secrecy in
  interference channels,'' \emph{IEEE Trans. Inf. Theory}, vol.~57, no.~9, pp.
  5682--5694, Sep. 2011.

\bibitem{7060726}
A.~Kalantari, S.~Maleki, G.~Zheng, S.~Chatzinotas, and B.~Ottersten, ``Joint
  power control in wiretap interference channels,'' \emph{IEEE Trans. Wireless
  Commun.}, vol.~14, no.~7, pp. 3810--3823, July 2015.

\bibitem{7313047}
D.~Park, ``Weighted sum rate maximization of {MIMO} broadcast and interference
  channels with confidential messages,'' \emph{IEEE Trans. Wireless Commun.},
  vol.~15, no.~3, pp. 1742--1753, Mar. 2016.

\bibitem{955145}
Y.~Oohama, ``Coding for relay channels with confidential messages,'' in
  \emph{Proc. IEEE Inf. Theory Workshop (ITW)}, Cairns, QLD, Australia, Sep.
  2001, pp. 87--89.

\bibitem{4608977}
L.~Lai and H.~El~Gamal, ``The relay-eavesdropper channel: Cooperation for
  secrecy,'' \emph{IEEE Trans. Inf. Theory}, vol.~54, no.~9, pp. 4005--4019,
  Sep. 2008.

\bibitem{5352243}
L.~Dong, Z.~Han, A.~P. Petropulu, and H.~V. Poor, ``Improving wireless physical
  layer security via cooperating relays,'' \emph{IEEE Trans. Sig. Process.},
  vol.~58, no.~3, pp. 1875--1888, Mar. 2010.

\bibitem{6601774}
Y.~Zou, X.~Wang, and W.~Shen, ``Optimal relay selection for physical-layer
  security in cooperative wireless networks,'' \emph{IEEE J. Sel. Areas
  Commun.}, vol.~31, no.~10, pp. 2099--2111, Oct. 2013.

\bibitem{7105936}
B.~Dai and Z.~Ma, ``Multiple-access relay wiretap channel,'' \emph{IEEE Trans.
  Inf. Forensics Security}, vol.~10, no.~9, pp. 1835--1849, May 2015.

\bibitem{7355564}
X.~Chen, C.~Zhong, C.~Yuen, and H.-H. Chen, ``Multi-antenna relay aided
  wireless physical layer security,'' \emph{IEEE Commun. Mag.}, vol.~53,
  no.~12, pp. 40--46, Dec. 2015.

\bibitem{7551149}
C.~Liu, N.~Yang, R.~Malaney, and J.~Yuan, ``Artificial-noise-aided transmission
  in multi-antenna relay wiretap channels with spatially random
  eavesdroppers,'' \emph{IEEE Trans. Wireless Commun.}, vol.~15, no.~11, pp.
  7444--7456, Nov. 2016.

\bibitem{hao2018resource}
H.~Xu, C.~Pan, W.~Xu, M.~Chen, and W.~Heng, ``Improving wireless physical layer
  security via {D2D} communication,'' in \emph{Proc. IEEE GLOBECOM}, Abu Dhabi,
  UAE, Dec. 2018, pp. 1--7.

\bibitem{8895802}
H.~Xu, G.~Caire, W.~Xu, and M.~Chen, ``Weighted sum secrecy rate maximization
  for {D2D} underlaid cellular networks,'' \emph{IEEE Trans. Commun.}, vol.~68,
  no.~1, pp. 349--362, Jan. 2020.

\bibitem{lee2017precoder}
H.~Lee, C.~Song, J.~Moon, and I.~Lee, ``Precoder designs for {MIMO} gaussian
  multiple access wiretap channels,'' \emph{IEEE Trans. Veh. Tech.}, vol.~66,
  no.~9, pp. 8563--8568, Sep. 2017.

\bibitem{el2011network}
A.~El~Gamal and Y.-H. Kim, \emph{Network information theory}.\hskip 1em plus
  0.5em minus 0.4em\relax {Cambridge University Press}, 2011.

\bibitem{xu2022note}
H.~Xu, K.-K. Wong, and G.~Caire, ``A note on fourier-motzkin elimination with
  three eliminating variables,'' \emph{ReaearchGate, DOI:
  10.13140/RG.2.2.11816.65288}, pp. 1--26, Jul. 2022.

\bibitem{gass2003linear}
S.~I. Gass, \emph{Linear programming: methods and applications}.\hskip 1em plus
  0.5em minus 0.4em\relax {Courier Corporation}, 2003.

\bibitem{7547360}
Y.~Sun, P.~Babu, and D.~P. Palomar, ``Majorization-minimization algorithms in
  signal processing, communications, and machine learning,'' \emph{IEEE Trans.
  Sig. Process.}, vol.~65, no.~3, pp. 794--816, Feb. 2017.

\bibitem{jeon2016joint}
Y.~Jeon, S.-H. Park, C.~Song, J.~Moon, S.~Maeng, and I.~Lee, ``Joint designs of
  fronthaul compression and precoding for full-duplex cloud radio access
  networks,'' \emph{IEEE Wireless Commun. Lett.}, vol.~5, no.~6, pp. 632--635,
  Dec. 2016.

\bibitem{parada2005secrecy}
P.~Parada and R.~Blahut, ``Secrecy capacity of {SIMO} and slow fading
  channels,'' in \emph{Proc. IEEE Int. Symp. Inf. Theory (ISIT)}, Adelaide, SA,
  Australia, Sep. 2005, pp. 2152--2155.

\bibitem{khisti2010secureI}
A.~Khisti and G.~W. Wornell, ``Secure transmission with multiple antennas {I:
  the MISOME} wiretap channel,'' \emph{IEEE Trans. Inf. Theory}, vol.~56,
  no.~7, pp. 3088--3104, July 2010.

\bibitem{shafiee2009towards}
S.~Shafiee, N.~Liu, and S.~Ulukus, ``Towards the secrecy capacity of the
  gaussian {MIMO} wiretap channel: The 2-2-1 channel,'' \emph{IEEE Trans. Inf.
  Theory}, vol.~55, no.~9, pp. 4033--4039, Sep. 2009.

\bibitem{vaezi2017optimal}
M.~Vaezi, W.~Shin, and H.~V. Poor, ``Optimal beamforming for {Gaussian MIMO}
  wiretap channels with two transmit antennas,'' \emph{IEEE Trans. Wireless
  Commun.}, vol.~16, no.~10, pp. 6726--6735, Oct. 2017.

\bibitem{vaezi2017mimo}
M.~Vaezi, W.~Shin, H.~V. Poor, and J.~Lee, ``{MIMO Gaussian} wiretap channels
  with two transmit antennas: Optimal precoding and power allocation,'' in
  \emph{Proc. IEEE Int. Symp. Inf. Theory (ISIT)}, Aachen, Germany, June 2017,
  pp. 1708--1712.

\bibitem{khisti2010secure}
A.~Khisti and G.~W. Wornell, ``Secure transmission with multiple antennas-{Part
  II: The MIMOME} wiretap channel,'' \emph{IEEE Trans. Inf. Theory}, vol.~56,
  no.~11, pp. 5515--5532, Nov. 2010.

\bibitem{fakoorian2012optimal}
S.~A.~A. Fakoorian and A.~L. Swindlehurst, ``Optimal power allocation for
  gsvd-based beamforming in the mimo gaussian wiretap channel,'' in \emph{Proc.
  IEEE Int. Symp. Inf. Theory (ISIT)}, Cambridge, MA, USA, July, 2012, pp.
  2321--2325.

\bibitem{zhang2020rotation}
X.~Zhang, Y.~Qi, and M.~Vaezi, ``A rotation-based method for precoding in
  {Gaussian MIMOME} channels,'' \emph{IEEE Trans. Commun.}, vol.~69, no.~2, pp.
  1189--1200, Feb. 2021.

\bibitem{au1971note}
Y.-H. Au-Yeung, ``A note on some theorems on simultaneous diagonalization of
  two hermitian matrices,'' in \emph{Mathematical Proceedings of the Cambridge
  Philosophical Society}, vol.~70, no.~3.\hskip 1em plus 0.5em minus
  0.4em\relax {Cambridge University Press}, 1971, pp. 383--386.

\bibitem{boyd2004convex}
S.~Boyd and L.~Vandenberghe, \emph{Convex Optimization}.\hskip 1em plus 0.5em
  minus 0.4em\relax {Cambridge University Press}, 2004.

\bibitem{hunger2005floating}
R.~Hunger, \emph{Floating Point Operations in Matrix-vector Calculus}.\hskip
  1em plus 0.5em minus 0.4em\relax Munich University of Technology, Inst. for
  Circuit Theory and Signal Processing, 2005.

\bibitem{access2010further}
E.~U. T.~R. Access, ``Further advancements for {E-UTRA} physical layer
  aspects,'' 3GPP TR 36.814, Tech. Rep., 2010.

\end{thebibliography}

\end{document}